\documentclass[bibyear]{aa}  

\usepackage{natbib}
\usepackage{hyperref}
\hypersetup{colorlinks=true,linkcolor=[rgb]{1.,0.2,0.2},citecolor=[rgb]{0.1,0.4,1.},filecolor=[rgb]{0.7,0.2,0.2},urlcolor=[rgb]{0.7,0.2,0.2}}
\usepackage{xcolor}
\usepackage{orcidlink}
\usepackage{txfonts}

\usepackage{graphicx} 
\usepackage{color}    
\usepackage{float}
\usepackage{multicol}
\usepackage{amsmath}
\usepackage{verbatim}
\usepackage{array}
\usepackage{booktabs}
\usepackage{threeparttable}
\usepackage{amssymb}
\usepackage{array,multirow}
\usepackage{tikz}
\usetikzlibrary{shapes,backgrounds,calc,positioning,arrows.meta}

\newcommand{\stormy}{\emph{stormy}\xspace}
\newcommand{\cloudy}{\emph{cloudy}\xspace}
\newcommand{\rainy}{\emph{rainy}\xspace}
\newcommand{\sunny}{\emph{sunny}\xspace}

\begin{document}

\title{{\bfseries\scshape BlackHoleWeather –} Spin-coupled chaotic cold accretion across the meso scale: {\Large Morphology and thermodynamics}}
   \titlerunning{SMBH spin modeling and regulation in chaotic cold accretion}
   \authorrunning{O.~Piana et al.}

\author{Olmo Piana \thanks{olmopiana@unimore.it}
        \inst{1}\orcidlink{0000-0002-1558-5289},
        Massimo Gaspari
        \inst{1}\orcidlink{0000-0003-2754-9258},
        Filippo Barbani
        \inst{1}\orcidlink{0000-0002-1620-2577},
        Vieri Cammelli
        \inst{1}\orcidlink{0000-0002-2070-9047},
        Giovanni Stel
        \inst{1}\orcidlink{0009-0007-0585-9462},
        Davide M. Brustio
        \inst{1}\orcidlink{0009-0009-7700-1910},
        Valeria Olivares
        \inst{2,3}\orcidlink{0000-0001-6638-4324},
        Francesco Salvestrini
        \inst{4}\orcidlink{0000-0003-4751-7421},
        Ashkbiz Danehkar
        \inst{5}\orcidlink{0000-0003-4552-5997},
        Francesco Tombesi
        \inst{6,7,8}\orcidlink{0000-0002-6562-8654},
        Pasquale Temi
        \inst{9}\orcidlink{0000-0002-8341-342X},
        Filippo M. Maccagni
        \inst{10,11}\orcidlink{0000-0002-9930-1844},
        \and
        Martin Fournier
        \inst{12}\orcidlink{0009-0006-2593-1583}
        }

\institute{
Department of Physics, Informatics and Mathematics, University of Modena and Reggio Emilia, I-41125 Modena, Italy
\and
Department of Physics, Universidad de Santiago de Chile, Santiago, Chile
\and
CIRAS, Universidad de Santiago de Chile, Santiago, Chile
\and
INAF -- Osservatorio Astronomico di Trieste, Via G. Tiepolo 11, I-34143 Trieste, Italy
\and
Science and Technology Institute, Universities Space Research Association, Huntsville, AL 35805, USA
\and
INAF -- Astronomical Observatory of Rome, 00078 Monte Porzio Catone (Rome), Italy
\and
Department of Physics, University of Rome ``Tor Vergata'', 00133 Rome, Italy
\and
INFN -- Rome ``Tor Vergata'' Section, 00133 Rome, Italy
\and
NASA Ames Research Center, MS 245-6, Moffett Field, CA 94035-1000, USA
\and
INAF -- Osservatorio Astronomico di Cagliari, via della Scienza 5, 09047, Selargius (CA), Italy
\and
Wits Centre for Astrophysics, School of Physics, University of the Witwatersrand, 2000, Johannesburg, South Africa
\and
Universit\"{a}t Hamburg, Hamburger Sternwarte, Gojenbergsweg 112, 21029 Hamburg, Germany
}

 
\abstract
{Supermassive black hole (SMBH) spin is a key but poorly constrained ingredient of the feeding-feedback loop. Chaotic cold accretion (CCA) of cold gas clouds delivers rapidly varying three-dimensional torques that drive spin evolution and jet-axis reorientation, and in turn spin regulates jet power.}
{We introduce a time-dependent SMBH spin model linking resolved multiphase feeding at large and meso scales to unresolved relativistic angular-momentum transfer at the innermost stable circular orbit (ISCO).}
{We perform GPU-accelerated hyper-zoom hydrodynamical simulations of a radiatively cooling group atmosphere with jet feedback and SMBH spin evolution, resolving multiphase inflow and angular-momentum direction below parsec scales. We compare fixed-axis, direct, and hybrid prescriptions, with the latter preserving the resolved torque direction while filtering its magnitude through a Kerr ISCO closure. We then apply the hybrid model to low- and high-turbulence group setups.}
{The cold-gas reservoir is nearly independent of whether the jet is fixed, spin-coupled, or rapidly reorienting. The spin prescription instead controls the inner feeding-feedback coupling, modulating central accretion, jet efficiency/power, and feedback geometry. The hybrid model remains bracketed by analytic limits, whereas the direct model overestimates spin variability and jet-axis wandering, showing that an ISCO closure is required. Low-spin SMBHs are easier to reorient because a misaligned torque acts on a smaller angular-momentum reservoir. The decisive quantity is the coherence of the delivered angular momentum: the low-turbulence run preserves longer rainy connections and stronger secular spin evolution, whereas stronger turbulence fragments the inflow and enhances torque cancellation.}
{CCA represents both a stochastic fuel supply and a chaotic torque engine. In the {\sc BlackHoleWeather} framework, turbulence regulates whether the cold reservoir remains connected, how the angular momentum reaches the SMBH, where the next jet points, and how feedback is imprinted onto the halo.}


   \maketitle
%

\section{Introduction} \label{s:intro}
Observations have long established the solidity of several correlations between central supermassive black holes (SMBHs) and the physical properties of host galaxies \citep{kormendy2013}, making the study of these objects central to galaxy evolution. The observed bipolar conical jets emitted by accreting SMBHs are among the most energetic events in the observed Universe and are believed to be able to fundamentally shape the evolution not only of the host galaxy, but also of the large-scale environment (see \citealt{mukherjee2025} for a review). These jets are observed in a great variety of shapes, sizes, and energies, often with misaligned axis, as in X-, S- and Z-shaped jets \citep{krause2019, bruni2021, misra2025}. Also in the case of restarted radio galaxies the bubbles inflated by the jets can be misaligned, indicating multiple cycles of AGN outbursts occurring on a short timescale in different directions \citep[e.g.,][]{ubertosi2023}. These observations point to a picture in which jet precession can be closely linked to the variability of the SMBH spin. Indeed, according to the long-standing Blandford--Znajek model \citep{blandford1977, tchekhovskoy2011}, jets are powered by the rotational energy (spin) of SMBHs, which is extracted during the launch phase by the local magnetic field. Part of this energy is then deposited into the surrounding medium, from the interstellar medium (ISM) to the intragroup and intracluster medium (IGrM/ICM), injecting momentum and turbulence together with stellar feedback \citep[e.g.,][]{barbani2023,barbani2025} and thereby regulating the availability of fuel for SMBH growth. At the same time, the angular momentum of the material falling into the black hole actively impacts its spin \citep{king2008, ricarte2025}. Given this tight accretion--spin--jet connection, studying spin evolution offers a privileged perspective in understanding the processes that regulate the interaction between the central SMBH and its environment \citep[see][for the importance of spin in galaxy evolution]{dimatteo2005, volonteri2013, sesana2014}, as well as providing a coherent explanation for a wide range of observed properties and morphologies of AGN and radiogalaxies \citep{sikora2007, garofalo2010, tchekhovskoy2010, martinez2011, danekhar2024}.

The physical regime in which we aim to study this connection is chaotic cold accretion \citep[CCA;][]{gaspari2012, gaspari2013,gaspari2015,gaspari2017,gaspari2017b}, namely the multiphase, time-variable mode of SMBH fueling expected in hot halos subject to radiative cooling, turbulence, and AGN heating \citep[see][for a review]{gaspari2020}. In this framework, AGN feedback is a self-regulated mechanism within which nonlinear condensation, chaotic fueling, and recurrent jet-regulated feedback emerge as coupled elements of the same multiscale baryon cycle in stratified hot halos. In this environment, nonlinear thermal instability causes cold clouds and filaments to condense out of the hot atmosphere and rain toward the galaxy centre, where repeated collisions, mixing, and fragmentation generate a strongly stochastic supply of mass and angular momentum. This general interpretation is supported by a growing body of multiwavelength observations of multiphase gas, chaotic kinematics, and CCA-regulated feedback in hot halos \citep{mcdonald2018, tremblay2018, maccagni2021, temi2022, olivares2022, olivares2025, reefe2025, romero2025, omoruyi2026}. 
Notably, in this regime the accretion rate becomes highly variable, and the black hole is fed through a sequence of misaligned, three-dimensional torque episodes across recurrent multiphase duty cycles, expected to drive both spin variability and jet reorientation. To study spin evolution, it is then important to check whether the inflow preserves a coherent torque direction over 100 Myr timescales, or is repeatedly reset by the multiphase weather cycle: cold gas condenses, the accretion rate rises, jet feedback reheats and partially clears the nucleus, and condensation resumes in the next cycle. 

Observationally, in addition to the direct connection with the jet power, SMBH spin is strictly linked to jet precession, as it is commonly assumed that jets are launched along the spin axis. The re-orientation of AGN jets can be driven by several physical processes, including warped accretion discs \citep{greenhill2003, nixon2016}, misalignment between the accretion disc and the SMBH spin vector \citep{bardeen1975, lu2005, krolik2015}, misaligned accreting filaments \citep{aalto2015}, binary dynamics of the SMBH \citep{krause2019, gerosa2019}, with different typical precession timescales depending on the underlying physical process. Despite all of the evidence, direct measurements of SMBH spin in the real Universe are still limited, as of now, and mostly delivered by X-ray studies \citep{brenneman2013, reynolds2021, danekhar2024}. Theoretical studies have tried to model the evolution of SMBH spins with semi-analytic models \citep{volonteri2005, king2008, perego2009, dotti2013, volonteri2013, sesana2014}, using analytical results from general relativity (GR) to follow spin evolution in highly-idealized (symmetric) scenarios \citep{bardeen1972}. Numerical simulations, on the other hand, struggle to couple the inherently multi-scale problem of AGN feedback; GRMHD simulations aiming at resolving the inner accretion disc accurately capture jet launching from magnetically charged, spinning black holes \citep{pu2020, narayan2022, lowell2024}, but are often limited in box size, and might miss the contribution of the large-scale gas cycle, which ultimately replenishes the accretion disc. Big-box simulations \citep{dubois2014, cielo2018, beckmann2019, bustamante2019, horton2020, talbot2021, talbot2022, beckmann2024}, instead, are extremely useful to study the spin evolution of the overall AGN population from a cosmological perspective, but do not have the horizon-scale resolution required to accurately track the angular momentum transfer between the accretion disc and the black hole in individual systems, especially in those characterized by high variability. The purpose of this work is to start building a coherent bridge between the micro-scales, which determine the details of SMBH accretion and spin evolution, and the meso- and macro-scales, which regulate the delivery of cold gas toward the centre of the galaxy. We do this by building a grid with multiple refinement levels in a group-size box \citep[e.g.,][]{gaspari2013,gaspari2017,guo2023}, reaching a central resolution of 0.7 pc, and ultimately filling the gap with the horizon scale by applying a GR-informed ISCO (innermost stable circular orbit) model of the accreted mass and angular momentum. 

This work is embedded within the broader {\sc BlackHoleWeather} program \citep{gaspari2020}, whose goal is to build a unified, self-consistent framework for SMBH feeding and feedback throughout the entire baryon cycle, from halo scales down to the immediate black-hole vicinity. Within this perspective, AGN activity is not treated as a one-way heating process, but as a genuinely multiscale weather engine in which cooling, condensation, accretion, and feedback continuously reshape one another. In this broader context, companion studies \cite[][B26a,b]{barbani2026a, barbani2026b} explore the pure feeding side of the problem, isolating how radiative cooling and turbulence alone generate different CCA weather regimes across scales, while \cite[][C26a,b]{cammelli2026a, cammelli2026b} investigate how fixed AGN jets shape the thermodynamics of the multiphase medium. The specific role of the present study is to include in this framework the spin evolution and the derived jet precession in a self-consistent fashion, as opposed to treating it as a free parameter. In particular, we focus on how the multiphase CCA cycle delivers angular momentum to the SMBH in a time-dependent, three-dimensional manner, and on how this stochastic torque cascade affects both the secular evolution of the spin and the direction and power of the associated jets. 

Similarly to B26a,b and C26a,b, the present spin-focused study is divided into two companion articles. In this paper (P26a), we introduce and validate the SMBH spin framework, focusing on how spin coupling affects the morphology and thermodynamic evolution of the multiphase medium through angular-momentum transfer and jet feedback. The companion paper \cite[][P26b]{piana2026b} will instead address complementary kinematic and variability diagnostics, with particular emphasis on torque coherence, jet-axis reorientation, and time-dependent signatures of CCA.

In the following sections, we first describe the numerical setup in \S\ref{s:num}, including the jet prescription and the SMBH spin-evolution model. In \S\ref{s:res1}, we analyse a suite of idealized no-turbulence runs used as control experiments to isolate the role of spin--feedback coupling. These include a \textit{Benchmark} model with a fixed-axis, spin-decoupled jet, and a \textit{Direct} model in which the resolved sink-scale angular momentum is deposited onto the SMBH without an ISCO-based closure. We compare both controls with our fiducial \textit{Hybrid} model, which preserves the resolved, time-dependent direction of the inflowing angular momentum while filtering its magnitude through a GR-motivated ISCO prescription. After testing the self-consistency of this fiducial model, we apply it in \S\ref{s:res2} to higher-resolution turbulent CCA runs, focusing on how spin-coupled feedback modifies the inner feeding--feedback loop, jet morphology, thermodynamic structure, and SMBH spin response. We discuss the broader implications and comparison with previous spin and jet-coupling models in \S\ref{s:disc}, and summarize our main conclusions in \S\ref{s:conc}.


\section{Numerical setup} \label{s:num}

In this work we employ the well-tested numerical simulation \texttt{AthenaPK}, a powerful GPU-accelerated public MHD code, to which we have added several modules to model black hole feeding, feedback, and spin evolution. \texttt{AthenaPK}\footnote{https://github.com/parthenon-hpc-lab/\texttt{AthenaPK}} \citep{grete2023, grete2025, fournier2024, prasad2026} is built on \texttt{Athena++} \citep{stone2020}, extended through the Parthenon and Kokkos \citep{edwards2014} libraries, ensuring both portability and scalability across current and next-generation HPC (high-performance computing) systems. In this design, Parthenon manages adaptive mesh refinement (AMR) and inter-node communication, while Kokkos provides highly efficient on-node data parallelism. Although the details of feeding and feedback in different regimes are presented in the companion papers by B26a and C26a, the focus of this paper is to develop and implement a spin evolution model fully coupled to both feeding and feedback processes. 
In practice, the simulations resolve the multiphase inflow and its time-dependent angular-momentum direction down to the sink scale, whereas the final coupling between the accreted gas and the SMBH spin is modeled through an inner relativistic closure described below. This separation is central to the present work, whose goal is to connect resolved CCA torque delivery to unresolved horizon-scale spin evolution in a physically controlled way.

The mesh structure is a box with a side length of 100 kpc, divided into a root grid of 128 cells per side. The grid is further divided into mesh blocks, each made up of 32 cells per side, which act as computational units. The first level of refinement takes the central cube formed by 32 meshblocks and refines it with a subgrid effectively halving the cell-size, and hence the resolution, within the refined region. We perform this refinement operation until we reach the target resolution of 0.7 pc (for the highest-resolution runs) within the innermost region, which includes the sink, with a radius of 4 cells, and the jet injection cells. Given that the code is non-relativistic, to maintain numerical stability and consistency, we impose velocity and temperature ceilings of respectively $v_\mathrm{ceil} = 0.3c$ and $T_\mathrm{ceil}=5 \times 10^9$ K (although both very rarely reached), making sure that the removed energies account for up to just a few percent of the energy injected with the jet. As per common practice, we also impose density and temperature floors and ceilings to prevent local instabilities caused by the steep gradients between high-resolution cells close to the jet region. The density floors and ceilings correspond to roughly $10^{-6}$ and $10^6 \mathrm{cm^{-3}}$, and the temperature floor to 1 K. The cooling function for $T \geq 10^{4.2}$ K is taken from \cite{schure2009} for solar metallicity, while for $T < 10^{4.2}$ K we use the volumetric cooling rate defined in \cite{gaspari2017} as $n_H^2 \Lambda_{\text{cold}}$, with
\begin{align}\label{lambda}
\Lambda_{\text{cold}} &= 2 \times 10^{-19} 
   \exp[-1.184\times 10^5/(T+10^3)] \notag \\
&\quad + 2.8 \times 10^{-27}\sqrt{T}\,\exp[-92/T].
\end{align}
This includes atomic line cooling, rotovibrational line cooling, and molecular collisions with dust grains, and is necessary to generate atomic and molecular gas.

The simulations start from a gaseous halo in hydrostatic equilibrium within a static gravitational potential. We model a representative intermediate-mass galaxy group, which both reflects the dominant baryonic environment of the present-day cosmic web and enables shorter central cooling times with $t_{\rm cool}\sim10$--$20$ Myr \citep[e.g.][]{osullivan2017}. The potential includes a dark matter halo with mass $M_\mathrm{NFW} = 1.5 \times 10^{13} M_\odot$ described by an NFW profile \citep{navarro1996}, a central dominant galaxy modeled with a Hernquist profile \citep{hernquist1990} and stellar mass $M_\mathrm{\ast} = 1.4 \times 10^{11} M_\odot$, and a point-mass SMBH with $M_\bullet = 2.8 \times 10^{8} M_\odot$, with total acceleration $g_{\rm tot}=g_{\rm NFW}+g_{\rm cD}+g_\bullet$. This SMBH mass is consistent with low-redshift central black holes in galaxy groups, as well as with predictions from semi-analytic models \citep{piana2021, piana2024, cammelli2025}. Following the ACCEPT catalog \citep{cavagnolo2009}, the gas entropy profile follows $K(r)=K_0+K_{100}(r/100\,{\rm kpc})^{\alpha_K}$, rescaled to the galaxy-group regime, and is combined with hydrostatic equilibrium to derive the initial density and pressure profiles. See B26a (Figure 2) and C26a for more details. For the detailed time evolution of the thermodynamic radial profiles, see Appendix \ref{app:A}.

For the runs with driven turbulence presented in Section \ref{s:res2}, turbulence is injected through a continuous stochastic acceleration field \citep[e.g.][]{schmidt2009,gaspari2013_turb,grete2018,grete2025}. The forcing is solenoidal and is injected at a characteristic mode number $n_{\rm peak}=4$, corresponding to a scale of $L_{\rm inj}\simeq25$ kpc. The correlation time is $t_{\rm corr}=30$ Myr.

Throughout our morphological and thermodynamic analysis, we normalize the time to the characteristic raining time 
$\tau \equiv t/t_{\rm rain}$, which denotes the time of first nonlinear cold-gas condensation and is used only as a normalization scale for cross-run comparison. We adopt \(t_{\rm rain}=14\,{\rm Myr}\) for the benchmark runs and \(t_{\rm rain}=18\,{\rm Myr}\) for the turbulent runs.
Furthermore, throughout the analysis we use the same thermodynamic phase bins as adopted in the companion {\sc BlackHoleWeather} papers. We define molecular gas as \(T<2\times10^2\,{\rm K}\), cold atomic gas as \(2\times10^2\leq T<1.6\times10^4\,{\rm K}\), warm gas as \(1.6\times10^4\leq T<1.16\times10^6\,{\rm K}\), hot soft-X gas as \(1.16\times10^6\leq T<5.8\times10^6\,{\rm K}\), and hot hard-X gas as \(T\geq5.8\times10^6\,{\rm K}\).\footnote{These labels are used as thermodynamic analysis bins, not as full synthetic observables.}

\subsection{Jet power and efficiency}

Both the jet and disc radiative efficiency are usually parametrized or assumed in many jet models; however, in this work, we use a formalism based on GR results of the Kerr metric combined with the Blandford-Znajek process, assuming the thin Novikov-Thorne disc model \citep{novikov1973, rezzolla2016}. According to the Blandford-Znajek process \citep{blandford1977}, the magnetic field around the horizon of the accreting SMBH extracts angular momentum from the spinning black hole. We then consider two separate efficiencies: $\epsilon_\mathrm{isco}$ is defined as the specific (dimensionless) total mass-energy fraction a particle carries during its infall towards the ISCO radius. In practice, it represents the fraction of the mass that falls into the sink effectively reaching the ISCO. For a Kerr black hole with dimensionless spin $a$, it is computed as
\begin{equation}
\epsilon_\mathrm{isco}=\frac{r_\mathrm{isco}^2 - 2r_\mathrm{isco} \pm a\sqrt{r_\mathrm{isco}}} {r_\mathrm{isco} \left(r_\mathrm{isco}^2 - 3r_\mathrm{isco} \pm 2a\sqrt{r_\mathrm{isco}}\right)^{1/2}},
\label{eq_eps_isco}
\end{equation}
where the dimensionless radius of the ISCO is calculated as $r_{\text{isco}} = 3 + Z_2 \mp \sqrt{(3 - Z_1)(3 + Z_1 + 2Z_2)}$, with the minus sign indicating prograde orbits and the plus sign retrograde orbits \citep{bardeen1972}. The auxiliary functions $Z_1$ and $Z_2$ are defined as in \citet{bardeen1972}.
Within this picture, we assume that the complementary mass-energy fraction $\epsilon_\mathrm{rad} = 1 - \epsilon_\mathrm{isco}$ represents the binding energy lost by the gas particles on their way to the ISCO, and is radiated away. 
\begin{table*}
\caption{Model parameters and description.\label{table_params}}
\centering
\begin{tabular}{cc}
\toprule
\textbf{Parameter} & \textbf{Description}\\
\midrule
$\epsilon_\mathrm{isco}$ & Specific energy of a particle at the ISCO (dimensionless)\\
$r_\mathrm{isco}$ & ISCO radius\\
$a$ & SMBH dimensionless spin magnitude\\
$\phi_\bullet$ & Dimensionless magnetic flux\\
$v_\mathrm{jet}$ & Jet velocity\\
$f_\mathrm{ml}$ & Mass loading factor\\
\bottomrule
\end{tabular}
\end{table*}

By contrast, we define $\eta_\mathrm{BZ}$ as the jet efficiency, namely, the fraction of rest-mass energy effectively reaching the ISCO that is converted into jet power. We then write \citep{tchekhovskoy2011, piana2024}
\begin{equation}
P_\mathrm{BZ} = \eta_\mathrm{BZ} \epsilon_\mathrm{isco}\dot{M}_\mathrm{in}c^2 = 2.8 f(a)\left(\frac{\phi_\bullet}{15}\right)^2\epsilon_\mathrm{isco} \dot{M}_\mathrm{in}c^2,
\label{eq_power_bz}
\end{equation}
where 
\begin{equation}    
f(a)=a^2\left(1+\sqrt{1-a^2}\right)^{-2},
\end{equation}
$\dot{M}_\mathrm{in}$ is the accretion rate onto the sink and $\phi_\bullet$ is the dimensionless magnetic flux parameter in the immediate vicinity of the black hole \citep[see][]{EHT2019}. In this first proof-of-concept implementation, we adopt a fixed magnetic-flux parameter $\phi_\bullet=15$, with the goal of consistently capturing the relative modulation of jet power by spin rather than modeling from first principles the absolute normalization of the Blandford-Znajek process within a fully self-consistent magnetic-flux evolution framework. This simplified formalism allows us to couple black hole feeding, spin evolution, and jet feedback in a controlled and transparent way. A consistent treatment of magnetic-flux evolution will be included in future work.

Within this picture, the SMBH mass evolution needs to take into account both the mass loss during the infall towards the ISCO and the mass carried away by the jet. Therefore, the net mass accreted onto the black hole $\Delta M_{\bullet}$ after accounting for radiative losses is given by
\begin{equation}
\Delta M_{\bullet} = \epsilon_{\text{isco}} \left(1 - f_\mathrm{ml}\right) \dot{M}_{\text{in}}\Delta t,
\label{eq_mlf}
\end{equation}
whereas the mass ejected with the outflow reads 
\begin{equation}
\Delta M_{\text{out}} = \epsilon_{\text{isco}} f_\mathrm{ml} \dot{M}_{\text{in}}\Delta t
\end{equation}
In other words, the sink inflow $\Delta M_{\rm in}$ is first reduced by the ISCO energy-loss factor $\epsilon_{\rm isco}$, and the remaining budget is then divided between the net black-hole growth and the mass loaded into the jet through $f_{\rm ml}$.

The computed jet power is divided into jet kinetic energy and jet internal energy, so we can write
\begin{equation}
P_\mathrm{BZ} = \frac{1}{2}\dot{M}_\mathrm{out}v_\mathrm{jet}^2 + \dot{M}_\mathrm{out}\varepsilon_\mathrm{jet},
\label{eq_power_balance}
\end{equation}
where $v_\mathrm{jet}$ is the velocity of the jet at injection and $\varepsilon_\mathrm{jet}$ its specific internal energy per unit of mass. We set the jet injection temperature on the pc-scale to $T_\mathrm{jet} = 10^8$ K, making it unambiguously hotter than the ambient material, and define the internal jet temperature $\varepsilon_\mathrm{jet} = k_\mathrm{B}T_\mathrm{jet}/[(\gamma -1)\mu \mathrm{m_p}]$, where $k_\mathrm{B}$ is the Boltzmann constant, $\gamma=5/3$ is the adiabatic index for a fully ionized gas, $\mu=0.6$ the mean molecular weight, and $\mathrm{m_p}$ the proton mass. We then solve for the velocity, obtaining
\begin{equation}
v_\mathrm{jet} = \left[2\left(\frac{P_\mathrm{BZ}}{\dot{M}_\mathrm{out}}-\varepsilon_\mathrm{jet}\right)\right]^{1/2}.
\end{equation}
We assume that our jets are characterized by a mass loading factor $f_\mathrm{ml}=0.9$, such that the mass carried by the jet is $\dot{M}_\mathrm{out} = f_\mathrm{ml} \epsilon_\mathrm{isco} \dot{M}_\mathrm{in}$ and that $v_\mathrm{jet}$ remains safely below 0.6c. This choice is consistent with our focus on a low-spin regime characterized by sub-relativistic, baryon-loaded jets. Since $\dot{M}_\mathrm{in}$ is tracked by our simulation, the jet velocity $v_\mathrm{jet}$ is consistently derived at each time step.

We assume that the AGN jet is injected into the simulation along the instantaneous SMBH spin vector, allowing the jet axis to reorient with the evolving spin. 
In order to reduce spurious variations in the injected mass due to grid alignment effects as the AGN jet axis re-orients, we replace the classic binary (on/off) selection of jet injection cells---defined by a Boolean mask for a cylindrical injection region---with a smooth transition scheme, in which cells near the radial and axial boundaries of the injection cylinder receive a fractional contribution. Instead, we assign each cell a weight $w \in [0,1]$ that varies linearly with the signed distance to each boundary over approximately one local cell width. This suppresses artificial orientation-dependent fluctuations in the effective deposition volume and injection rates. The spin evolution model adopted in this work is described in detail in the next section.
This smooth weighting is numerically important because it suppresses purely grid-induced modulation of the effective injection volume as the jet axis precesses. The resulting variability in the injected feedback is therefore more directly tied to the physical spin evolution than to geometric discretization effects.

\begin{figure*}
\centering
\resizebox{\textwidth}{!}{%
\begin{tikzpicture}[
font=\LARGE,
  node distance=0.88cm,
  block/.style={draw, rounded corners, align=center, text width=11.5cm, minimum height=3.5cm, minimum width=12cm, inner sep=2pt},
  side/.style={draw, rounded corners, align=center, text width=11.5cm, minimum height=3.5cm, minimum width=12cm, inner sep=2pt},
  shared/.style={block},
  direct/.style={side, fill=blue!10, fill opacity=0.1, text opacity=1},
  arr/.style={-{Stealth[length=2.5mm,width=1.8mm]}, line width=0.6pt, shorten <=2.2pt, shorten >=2.2pt},
  hybrid/.style={side, fill=orange!10, fill opacity=0.1, text opacity=1}
]

\node[shared] (s1) {Large-scale gas cycle\\ (cooling + turbulence + gravity)\\ $\Rightarrow$ chaotic cold accretion};
\node[shared, below=of s1] (s2) {Resolved accreted angular momentum at sink\\ $\boldsymbol{L}_{\mathrm{acc,sink}}$};

\node[direct, left=2cm of s2] (d1) {\textit{Direct}\\ uses resolved $\boldsymbol{L}_{\mathrm{acc,sink}}$};
\node[direct, below=of d1] (d3) {Accretion torque\\ $\boldsymbol{L}_{\bullet, {\rm acc}} = (1 - f_\mathrm{ml})\boldsymbol{L}_\mathrm{acc, sink}$};

\node[hybrid, right=2cm of s2] (h1) {\textit{Hybrid}\\ uses re-scaled $l_\mathrm{isco}(a)$ but keeps direction $\hat{\boldsymbol{L}}_{\mathrm{acc,sink}}$};
\node[hybrid, below=of h1] (h3) {Accretion torque\\ $\boldsymbol{L}_\mathrm{\bullet, {\rm acc}} = \Delta M_{\bullet} \ell_{\text{isco}} \hat{\boldsymbol{L}}_{\text{acc, sink}}$};

\node[shared, right=2cm of d3] (s3) {Update BH status $M_\bullet$ and $\boldsymbol{a}$\\ 
efficiencies $\epsilon_\mathrm{isco}(a)$ and $\eta_\mathrm{BZ}(a,\phi_\bullet)$};
\node[shared, below=of s3] (s4) {Jet injection and spin-down\\ $P_\mathrm{BZ}$, $\dot M_\mathrm{out}$, $v_\mathrm{jet}$, $\boldsymbol{a}$ };

\draw[arr] (s1.south) -- (s2.north);

\draw[arr] (s2.west) -- (d1.east);
\draw[arr] (s2.east) -- (h1.west);

\draw[arr] (d1.south) -- (d3.north);

\draw[arr] (h1.south) -- (h3.north);

\draw[arr] (d3.east) -- (s3.west);
\draw[arr] (h3.west) -- (s3.east);

\draw[arr] (s3.south) -- (s4.north);
\draw[arr] (s4.east) -- ++(1.0cm,0) |- (s1.east);

\end{tikzpicture}%
}
\caption{Flowchart of the SMBH accretion--spin--feedback coupling used in the simulations. The central branch shows the shared feeding and feedback steps; the two side branches highlight the only difference between the \textit{Direct} and \textit{Hybrid} spin-update prescriptions (an ISCO-based GR closure for the deposited angular-momentum magnitude).}
\label{spin_models}
\end{figure*}
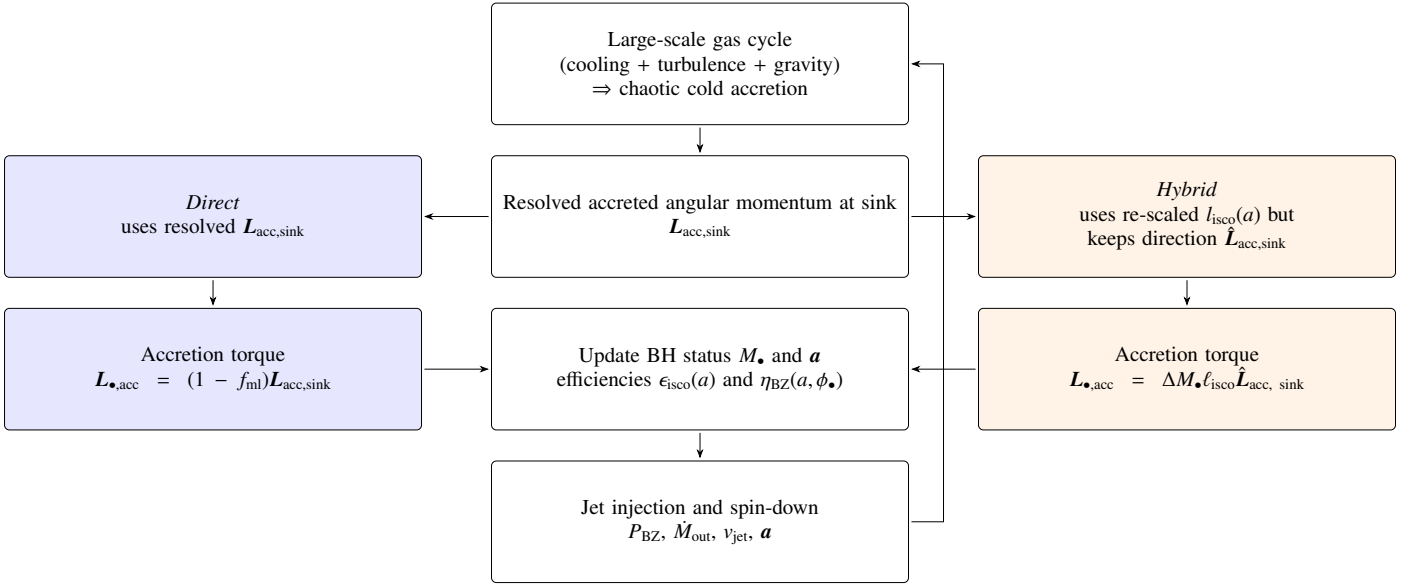

\subsection{SMBH spin model}

The evolution of the supermassive black hole spin is a critical component of our simulation, as it dictates the physical quantities at ISCO and the jet efficiency. To benchmark our model, we implement two different formalisms. A first simpler \textit{Direct} model directly updates the SMBH angular momentum by adding the 3D vector of accreted angular momentum carried by the mass swallowed by the sink particle at the center of the simulation box. This provides a purely resolved sink-scale angular-momentum deposition prescription, without imposing an explicit GR closure on the angular-momentum transfer efficiency. In parallel, our fiducial \textit{Hybrid} model retains the resolved, time-dependent 3D direction of the inflowing angular momentum—thereby capturing the chaotic reorientation of cold accretion—but it determines the magnitude of the angular momentum transferred to the hole using the Kerr thin-disc expressions at the ISCO, accounting for the GR energy and angular momentum budget required for the gas to reach the last stable orbit before capture. These two models are then compared to a \textit{Benchmark} model in which the spin -- set to an initial value of $a_0 = 0.1$ -- and jet evolution are fully decoupled; see Table \ref{table_runs} for a full description of the different simulations. We also show the results from a \textit{Hybrid} run with a higher initial spin $a_0 = 0.5$. What we expect is that when a SMBH grows by consuming material in a stable, coherent accretion, the incoming material adds angular momentum in the same direction, causing the spin to increase, often approaching the maximum possible speed. On the other hand, in a chaotic accretion environment like this one, small, random amounts of matter fall in from different directions and can cause retrograde events that spin the black hole down.

The two prescriptions do not play the same physical role. Our fiducial \textit{Hybrid} model is a closure in which the simulation resolves the instantaneous three-dimensional direction of the inflowing angular momentum down to the sink scale, while the magnitude of the angular momentum transfer to the SMBH is determined through a Kerr-ISCO prescription. By contrast, the \textit{Direct} model is retained only as a control experiment, aimed at highlighting what is overestimated when the sink-scale angular momentum is deposited onto the black hole without such a relativistic closure.

In this sense, the \textit{Direct} model is meant to quantify how sensitive our results are to the assumed ISCO closure by comparison with a purely sink-scale angular-momentum deposition prescription. 
We summarize the coupling between the resolved large-scale gas cycle, the accretion torque, and the spin-dependent jet injection in the flow chart presented in Figure \ref{spin_models}.

In all runs, the SMBH spin magnitude is defined as
\begin{equation}
a = |\boldsymbol{a}| = \frac{c|\boldsymbol{L}_\bullet|}{GM_\bullet^2},
\end{equation}
where $\boldsymbol{L}_\bullet$ is the black hole's angular momentum, $M_\bullet$ is its mass, $c$ is the speed of light, and $G$ is the gravitational constant. Its value is then updated at each time step according to the different prescriptions of each model.

\paragraph{The \textit{Benchmark} model -- decoupled control}
We run a benchmark, control run to isolate the role of spin coupling. In this run, the jet power is fully decoupled from the spin and is computed using a fixed efficiency as 
\begin{equation}
    P_\mathrm{jet} = \epsilon_\mathrm{jet} \dot{M}_\mathrm{in} c^2,
\end{equation}
following the same model used in C26. In this case, the jet is always emitted along the z-axis and the efficiency $\epsilon_\mathrm{jet} = 0.007$ is chosen to match the spin-dependent $\eta_\mathrm{BZ}$ used in our \textit{Hybrid} and \textit{Direct} models evaluated at $t=0$ ($a=0.1$).

\paragraph{The \textit{Direct} model -- resolved control} In this case, spin and jet are fully coupled. To control for the effect of introducing a sub-resolution ISCO closure for SMBH angular momentum evolution, in this model we evolve the SMBH spin by directly depositing the resolved angular momentum of the accreted gas measured at the sink scale. The change in the black hole angular momentum is obtained from the accreted angular momentum vector itself. Then, we simply re-scale the accreted angular momentum according to the mass accreted by the black hole, and we update the SMBH angular momentum as
\begin{equation}
\boldsymbol{L}_{\bullet, \rm{acc}} = \boldsymbol{L}_\mathrm{acc, sink} (1 - f_\mathrm{ml}).
\end{equation}
\begin{figure}
\centering
\includegraphics[width=0.9\linewidth, height=0.81\textheight]{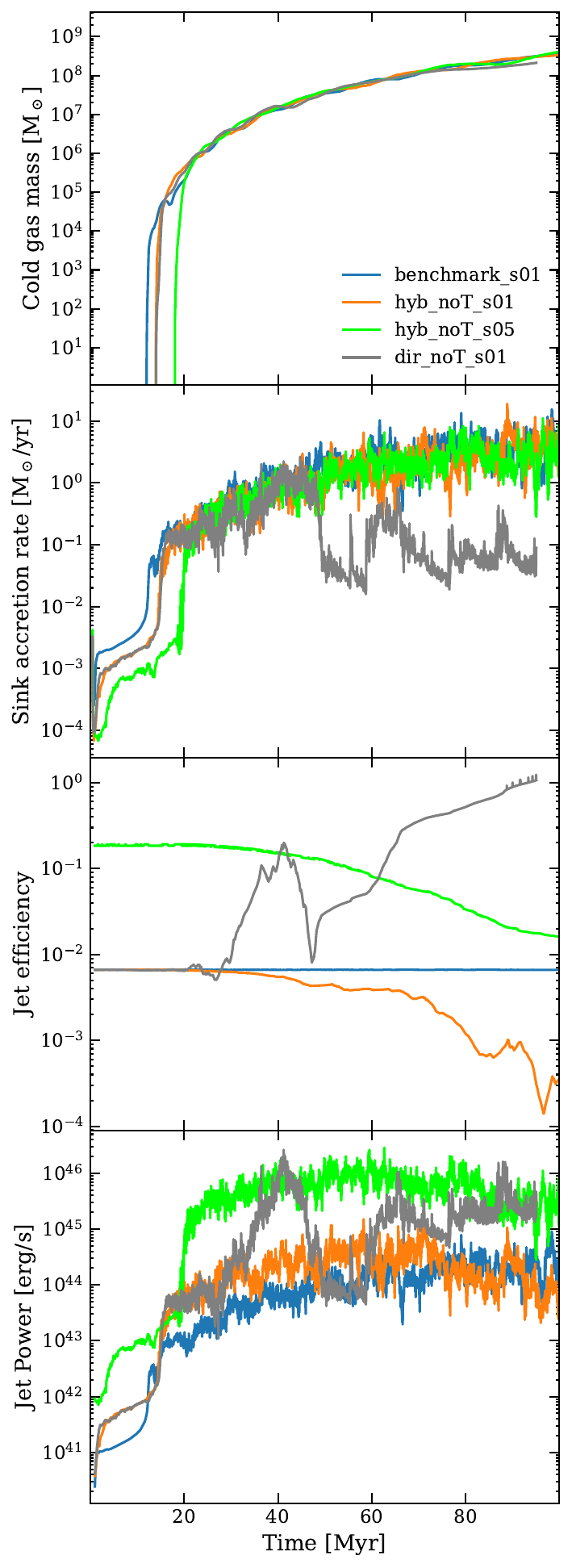}
\vspace{-0.3cm}
\caption{Global accretion and feedback histories for the no-turbulence suite, showing the total cold gas mass (i.e.\ the mass of gas below $T=2000$ K within the central 4 kpc), the sink accretion rate, the jet efficiency and the corresponding instantaneous jet power as a function of time.}
\label{spin}
\end{figure}

We then apply the spin-down effect of the jet torque, assuming that the Blandford-Znajek process extracts rotational energy from the black hole. We model this as a reduction in the magnitude of its angular momentum. The magnitude of the torque is given by:
\begin{equation}
\tau_\mathrm{jet} = \frac{P_\mathrm{BZ}}{\Omega_\mathrm{H}},
\end{equation}
where $\Omega_\mathrm{H}$ is the spin-dependent rotational velocity on the horizon \citep{bardeen1973, tchekhovskoy2011}:
\begin{equation}
\Omega_\mathrm{H} = \frac{ac^3}{2GM_\bullet\left(1+\sqrt{1-a^2}\right)}.
\end{equation}
The loss of angular momentum due to the jet over the time step $\Delta t$ is $\Delta L_\mathrm{jet} = \tau_\mathrm{jet} \Delta t$.
Finally, the new angular momentum of the black hole, $\boldsymbol{L}_{\bullet, \text{new}}$, is computed by adding the accreted component and scaling the resulting magnitude to account for the BZ torque:
\begin{equation}
\boldsymbol{L}_{\bullet , \text{temp}} = \boldsymbol{L}_{\bullet , \text{old}} + \boldsymbol{L}_\mathrm{\bullet , \text{acc}},
\end{equation}
and
\begin{equation}
\boldsymbol{L}_{\bullet, \text{new}} = \boldsymbol{L}_{\bullet, \text{temp}} \left( \frac{|\boldsymbol{L}_{\bullet, \text{temp}}| - \Delta L_{\text{jet}}}{|\boldsymbol{L}_{\bullet, \text{temp}}|} \right).
\end{equation}
The new spin vector $\boldsymbol{a}_{\text{new}}$ is then updated using the new mass and angular momentum
\begin{equation}
\boldsymbol{a}_\mathrm{new} = \frac{c\boldsymbol{L}_{\bullet, \text{new}}}{G(M_\bullet+\Delta M_\bullet)^2}.  
\end{equation}

By construction, the \textit{Direct} model retains the full, time-dependent three-dimensional angular momentum measured at the sink scale, but it does not impose a GR-motivated ISCO closure on the efficiency with which that angular momentum is transferred to the SMBH. It therefore serves mainly as a control case, designed to bracket the impact of unresolved circularization and ISCO physics at a given resolution. For this reason, \textit{Direct} should not be interpreted as a realistic description of horizon-scale accretion, but rather as an intentionally unfiltered sink-scale deposition test.
\begin{table*}
\caption{We list here the setups for the simulations shown in this paper.\label{table_runs}}
\centering
\begin{tabular}{ccccccc}
\toprule
\textbf{Tag} & \textbf{Initial spin} & \textbf{Jet efficiency} & \textbf{Spin model} & \textbf{Turbulence} & \textbf{Box size} & \textbf{Resolution}\\
\midrule
\texttt{benchmark\_s01} & 0.1 & 0.007 - fixed & -- & none & 100 kpc & 1.5 pc\\
\texttt{dir\_noT\_s01} & 0.1 & spin-dependent & \textit{Direct} & none & 100 kpc & 1.5 pc\\
\texttt{hyb\_noT\_s01} & 0.1 & spin-dependent & \textit{Hybrid} & none & 100 kpc & 1.5 pc\\
\texttt{hyb\_noT\_s05} & 0.5 & spin-dependent & \textit{Hybrid} & none & 100 kpc & 1.5 pc\\
\midrule
\texttt{hyb\_lowT\_s01} & 0.1 & spin-dependent & \textit{Hybrid} & low & 100 kpc & 0.7 pc\\
\texttt{hyb\_highT\_s01} & 0.1 & spin-dependent & \textit{Hybrid} & high & 100 kpc & 0.7 pc\\
\bottomrule
\end{tabular}
\end{table*}

\begin{figure}
\includegraphics[width=\columnwidth]{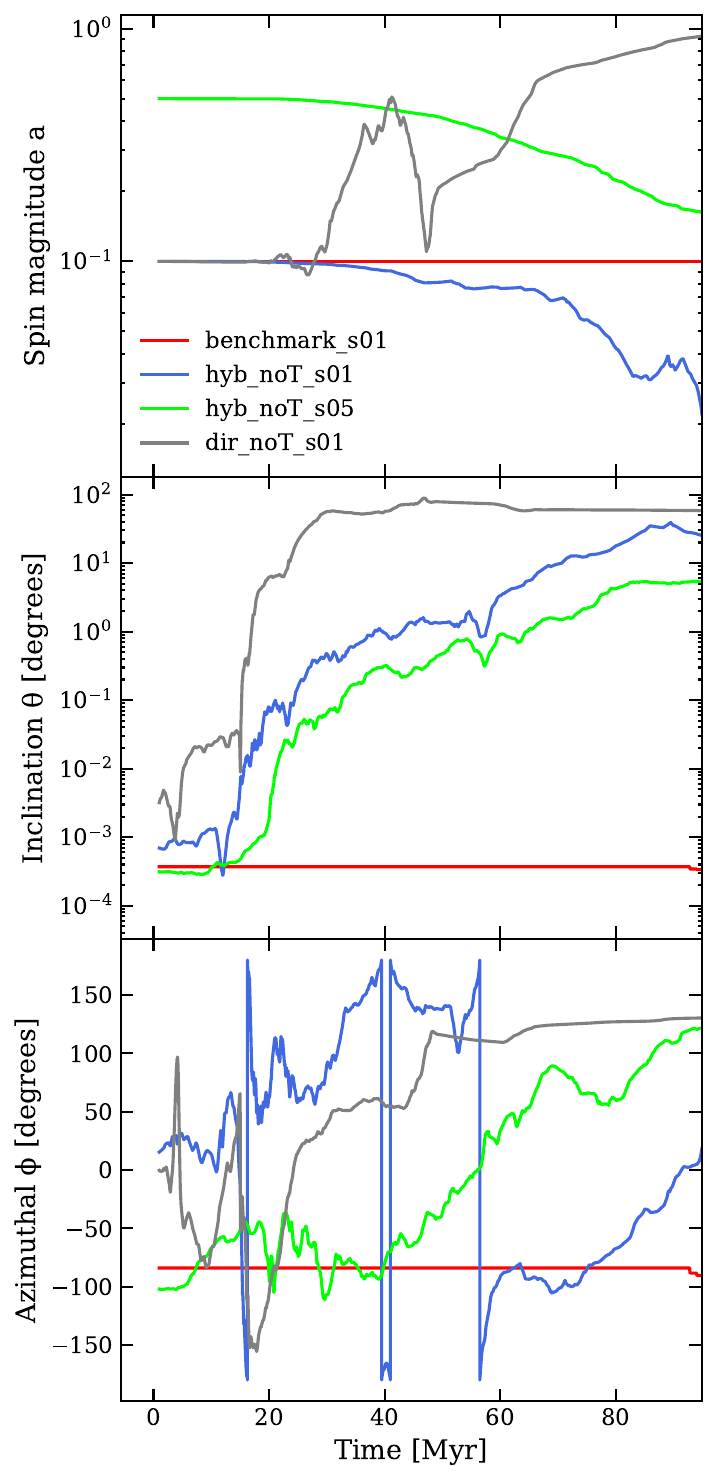}
\caption{Evolution of the dimensionless SMBH spin magnitude $a$ (top panel) and inclination and azimuthal angles $\theta$ and $\phi$ (middle and bottom panel respectively) over the first 100 Myr for the no-turbulence benchmark suite.
}
\label{smbh_diag}%
\end{figure}
\begin{figure*}
\includegraphics[width=\textwidth]{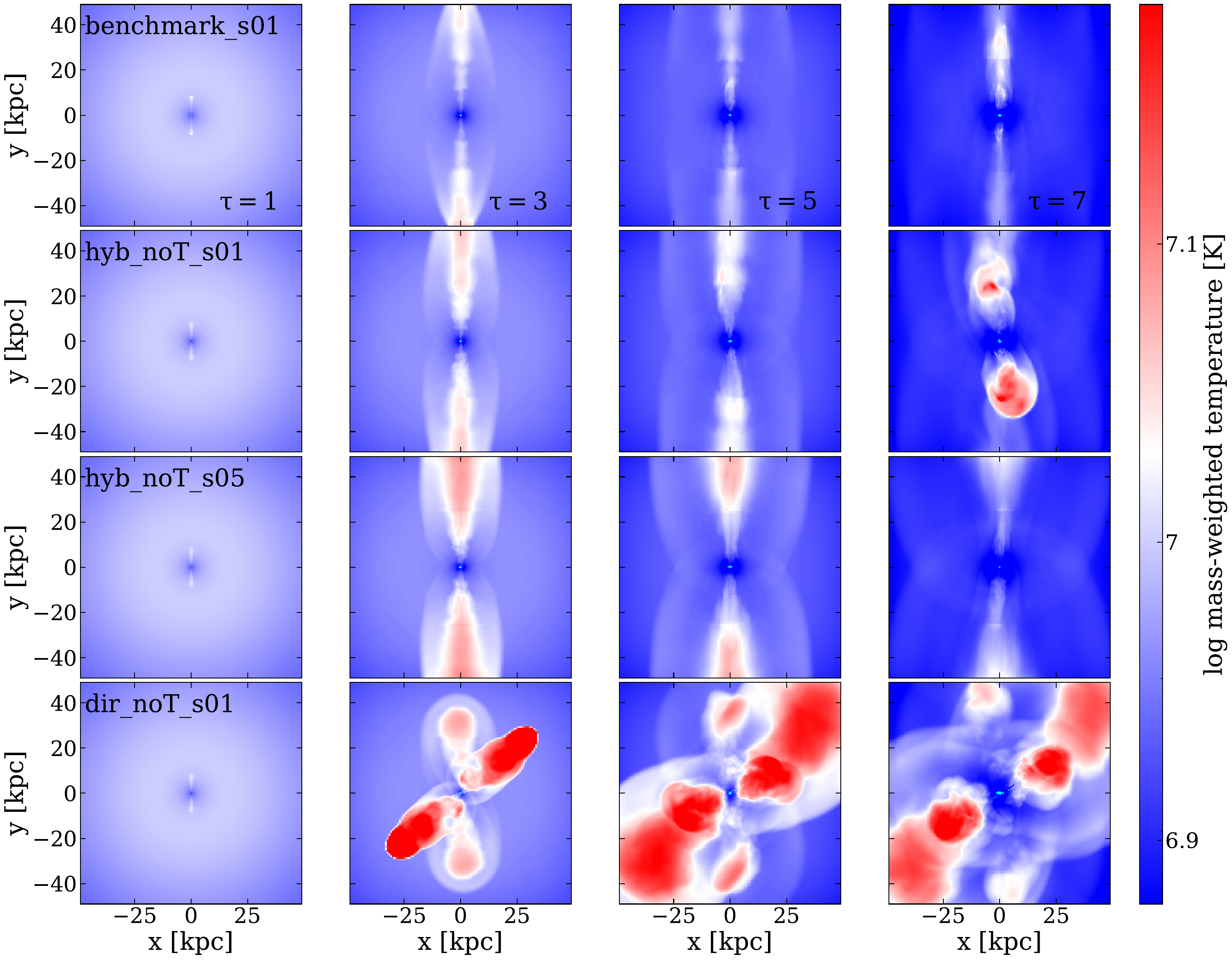}
\caption{Time-evolution of the mass-weighted temperature projection maps of the whole box showing the morphological expansion of the jet structure across the four baseline models at selected epochs ($\tau = t/t_\mathrm{rain}=1,3,5,7$). The tiny cyan contours in the center correspond to regions of cold gas with $T < 200$ K. The top row shows results from the benchmark model in which the spin and jet evolutions are fully decoupled. The second and third rows represent runs using our fiducial, ISCO-closure model but with different initial spin values, respectively 0.1 and 0.5. The bottom row shows the run performed with the control \textit{Direct} model, which overestimates the angular momentum accreted onto the SMBH.}
          \label{temp_proj}%
\end{figure*}

\paragraph{The \textit{Hybrid} method -- fiducial}
In our fiducial model, we assume that the resolved, sink-scale accretion inflow circularizes into a thin accretion disc at the ISCO, whose axis is aligned with the instantaneous sink-measured $\hat{\boldsymbol{L}}_{\rm acc, sink}$. 
In this sense, the model should be interpreted as a hybrid closure between resolved meso-scale feeding and unresolved horizon-scale accretion physics: the simulation provides the torque direction, while general-relativistic disc energetics determine how efficiently that torque is transferred to the black hole.
This closure maximally preserves the torque direction delivered by the resolved CCA flow. In reality, unresolved viscous evolution, Lense-Thirring precession, and Bardeen-Petterson alignment may partially smooth this direction before the gas reaches the ISCO.

First, we determine whether the accretion is prograde or retrograde relative to the black hole spin. This is done by calculating the dot product between the normalized BH spin vector $\hat{\boldsymbol{a}} = \boldsymbol{a}/a$ and the normalized angular momentum vector of the accreted gas $\hat{\boldsymbol{L}}_{\text{acc, sink}}$. If $\hat{\boldsymbol{a}} \cdot \hat{\boldsymbol{L}}_{\text{acc, sink}} \ge 0$, the infalling material at the ISCO is considered co-rotating (prograde); otherwise, it is counter-rotating (retrograde).

Similarly to equation \ref{eq_eps_isco}, we compute the dimensionless ISCO angular-momentum per unit mass-energy as in \citep{bardeen1972}
\begin{equation}
\tilde{\ell}_{\rm isco} =
\frac{r_{\rm isco}^2 \mp 2 a \sqrt{r_{\rm isco}} + a^2}
{\sqrt{r_{\rm isco}}(r_{\rm isco}-2) \pm a},
\end{equation}
where the upper sign corresponds to prograde and the lower sign to retrograde orbits (with the corresponding $r_{\rm isco}$). We then set $\ell_{\rm isco}=\tilde{\ell}_{\rm isco}\,GM/c$.

The change in the black hole's angular momentum vector imparted by accretion, $\boldsymbol{L}_\mathrm{acc, \bullet}$, is then calculated by assuming this net mass carries the ISCO-scaled specific angular momentum per unit mass $\ell_{\text{isco}}$ and is aligned with the direction of the gas angular momentum measured in the simulation, $\hat{\boldsymbol{L}}_{\text{acc, sink}}$:
\begin{equation}
\boldsymbol{L}_{\bullet, \text{acc}} = \Delta M_{\bullet} \ell_{\text{isco}} \hat{\boldsymbol{L}}_{\text{acc, sink}}.
\end{equation}

We then apply the same spin-down effect from the jet torque as in the \textit{Direct} method and compute

\begin{equation}
\boldsymbol{L}_{\bullet, \text{temp}} = \boldsymbol{L}_{\bullet, \text{old}} + \boldsymbol{L}_{\bullet, \text{acc}},
\end{equation}
where now the magnitude of the accreted angular momentum vector $\boldsymbol{L}_{\bullet, \text{acc}}$ is re-scaled according to $\ell_{\text{isco}}$, and 
\begin{equation}
\boldsymbol{L}_{\bullet, \text{new}} = \boldsymbol{L}_{\bullet, \text{temp}} \left( \frac{|\boldsymbol{L}_{\bullet, \text{temp}}| - \Delta L_{\text{jet}}}{|\boldsymbol{L}_{\bullet, \text{temp}}|} \right).
\end{equation}
The new spin vector $\boldsymbol{a}_{\text{new}}$ is then updated using the new mass and angular momentum
\begin{equation}
\boldsymbol{a}_\mathrm{new} = \frac{c\boldsymbol{L}_{\bullet, \text{new}}}{G(M_\bullet+\Delta M_\bullet)^2}.  
\end{equation}

\begin{figure*}
\includegraphics[width=\textwidth]{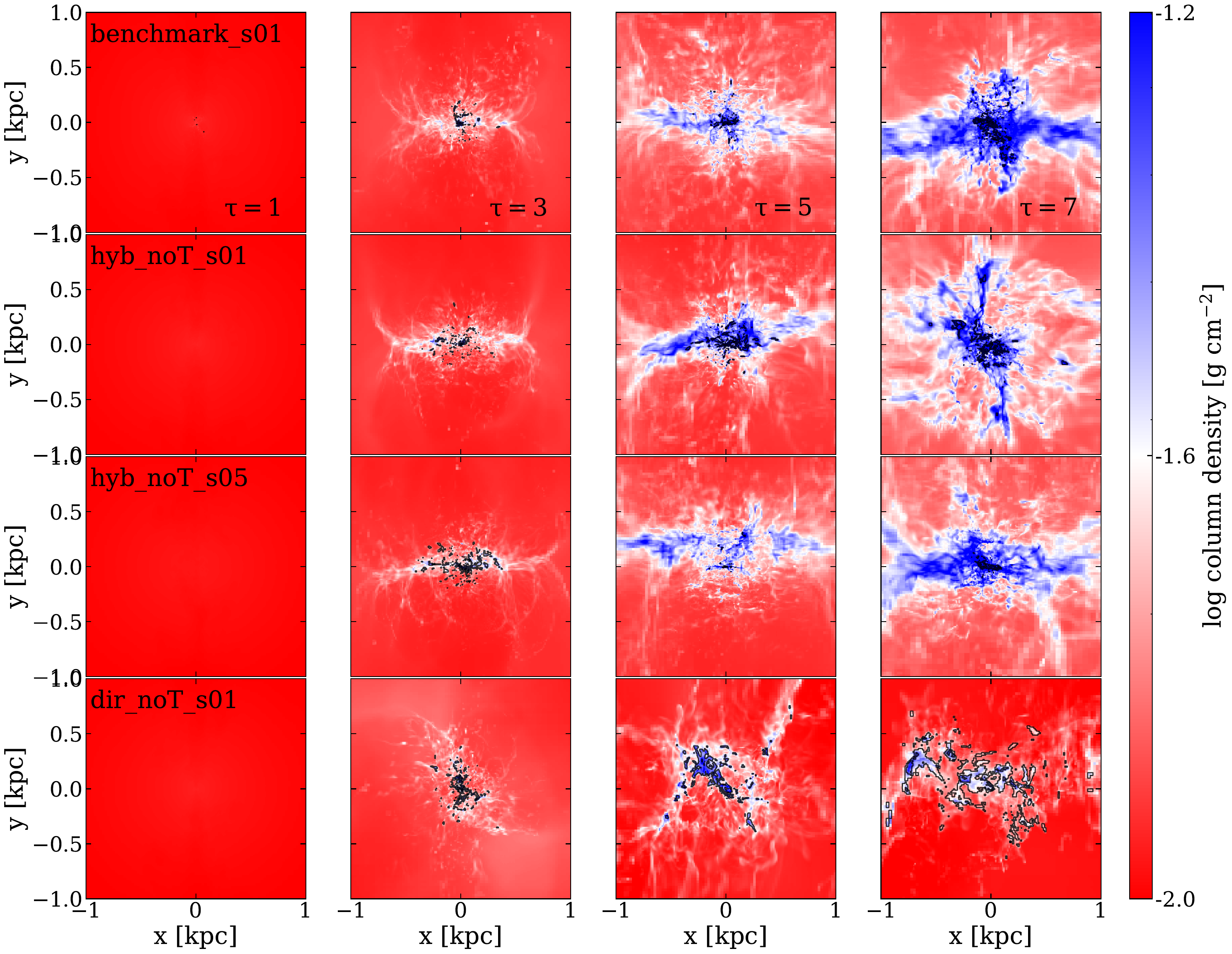}
\caption{Logarithmic column density projections within the central 2 kpc highlighting the structural evolution of the central accretion disc/inflow morphology for the baseline models at selected epochs ($\tau = t/t_\mathrm{rain}=1,3,5,7$). Black contours show molecular gas with $T < 200$ K.}
          \label{dens_proj}%
\end{figure*}
\section{Results: analysis of the spin--jet coupling} \label{s:res1}
\subsection{Self-consistency check and baseline}
To isolate the impact of the SMBH spin evolutionary model on the coupled feeding--feedback cycle, we first analyze a set of no-turbulence benchmark runs (Table~\ref{table_runs}) with a central spatial resolution of 1.5 pc and identical thermodynamic initial conditions. The suite consists of: (i) a \textit{Benchmark} control run with a fixed jet efficiency \texttt{benchmark\_s01}, (ii) two \textit{Hybrid} runs \texttt{hyb\_noT\_s01} and \texttt{hyb\_noT\_s05} with spin-dependent jet power and different initial spin magnitudes ($a_0=0.1$ and $a_0=0.5$), and (iii) a control run with the \textit{Direct} spin model and $a_0=0.1$ \texttt{dir\_noT\_s01}, that directly deposits the resolved sink-scale angular momentum into the SMBH spin without an ISCO closure.

\begin{figure}
\includegraphics[width=\columnwidth]{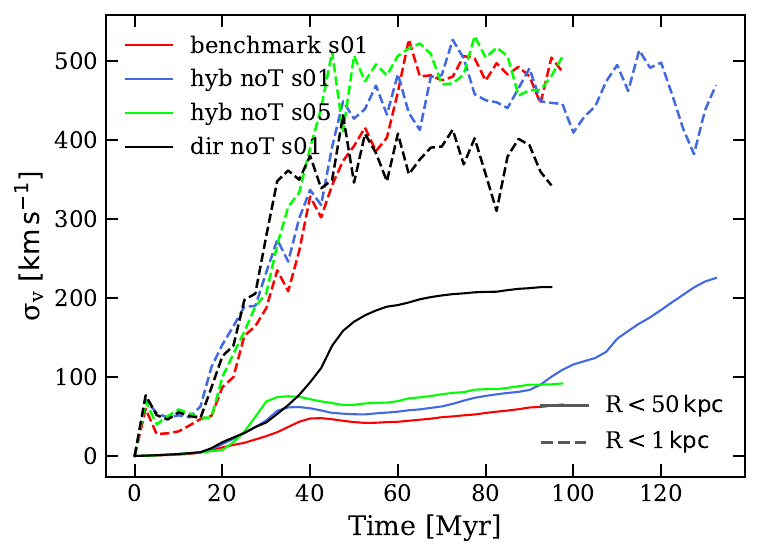}
\caption{Evolution of the volume-weighted velocity dispersion $\sigma_v$ for the no-turbulence suite, averaged across 50-kpc (solid lines) and 1-kpc (dashed lines) regions, for all phases. The late-time rise in the low-spin \textit{Hybrid} model correlates with sustained spin inclination growth. Notice that \texttt{hyb\_noT\_s01} is the only run we evolved for a longer time, being our fiducial run.}
          \label{sigmav}%
\end{figure}
A key advantage of this baseline suite is that the global condensation pathway remains nearly unchanged across the different runs. This allows us to isolate the role of the spin closure on the inner feeding-feedback coupling, rather than on the large-scale thermodynamic build-up of the cold reservoir itself. In other words, the suite is designed to test how the same condensed gas couples differently to the SMBH once the angular-momentum transfer prescription is changed.
\begin{figure*}
\centering
\includegraphics[width=0.95\textwidth]{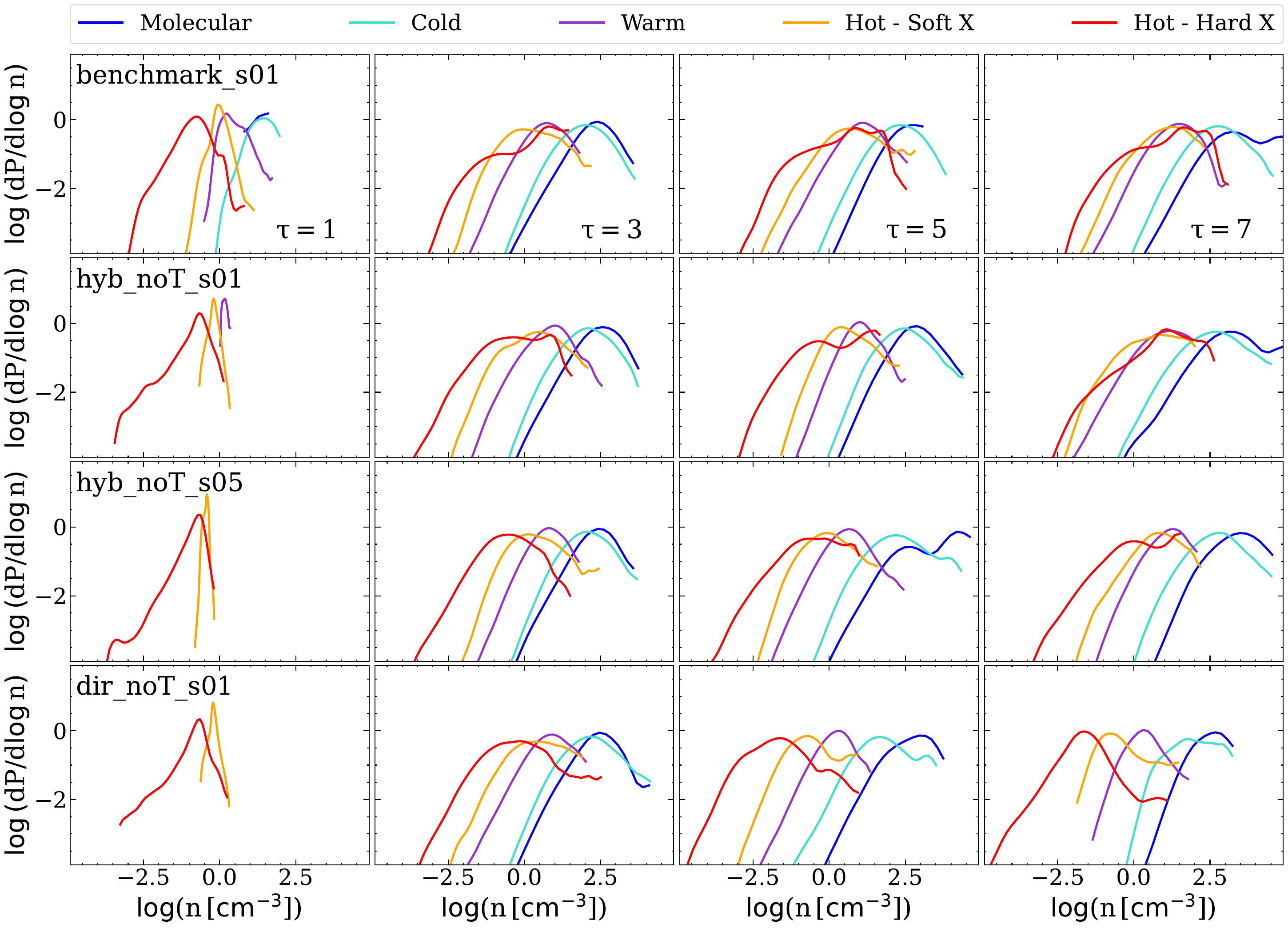}
\caption{Mass-weighted PDFs of gas density for the molecular (blue), cold atomic, warm, hot soft-X, and hot hard-X (red) gas phases (see temperature definitions in \S\ref{s:num}) in the no-turbulence suite, measured within the central 100 pc at four evolutionary times, $\tau\equiv t/t_{\rm rain}=1,3,5,7$.}
          \label{pdf_noturb}
\end{figure*}
The goal of this section comparison is twofold. First, we verify that the \textit{Hybrid} model evolves within physically reasonable spin bounds when coupled to the resolved inflow. We then benchmark our fiducial model against results of the spin evolution derived from analytical models, and use the inferred precession rate as a term of comparison for observational results. Importantly, differences in SMBH growth, jet power, and stirring primarily reflect differences in the central coupling between inflow angular momentum, spin evolution, and feedback, more than in the macroscopic evolution of the total cold gas. 
This is clearly visible in Figure \ref{spin}, which summarizes the mass and energy budgets of the jet-SMBH system. The top panel shows the instantaneous cold gas mass ($T < 2000$ K) within a radius of 4 kpc, and its nearly identical evolution across the suite indicates that the global condensation pathway is only weakly affected by the adopted spin prescription. We can notice, however, that in the high-spin run the onset of cooling is delayed by a few Myr, due to the higher feedback energy injected into the box according to eq.\ \ref{eq_power_bz}. The lower panels, on the other hand, reveal substantial differences in sink accretion, jet efficiency, and jet power, showing that the spin closure mainly controls how the condensed gas couples to the SMBH and how effectively the resulting feedback couples back to the core.
Part of the cold gas falls into the sink, with an accretion rate that is shown in the second panel, and from which the simulation updates the SMBH mass and the jet mass outflow via the mass loading factor parameter (eq.\ \ref{eq_mlf}). In the third panel we plot the jet efficiency as a function of time, while in the bottom we show the evolution of the jet power, computed according to eq.\ \ref{eq_power_bz}. In our initial simplified framework, in which the dimensionless magnetic-flux parameter is kept fixed for all runs, the jet power depends only on the sink accretion rate and on the SMBH spin. Quantitatively, the high-spin run \texttt{hyb\_noT\_s05} maintains jet powers at the $\sim 10^{45}$--$10^{46}$ erg s$^{-1}$ level once cold accretion is established, while the low-spin \textit{Hybrid} run (\texttt{hyb\_noT\_s01}) remains typically in the $\sim 10^{44}$--$10^{45}$ erg s$^{-1}$ range. The benchmark fixed-efficiency run is systematically lower at late times. As expected from the behaviour of the $\eta_\mathrm{BZ}$ efficiency (see eq.\ \ref{eq_power_bz}), the \textit{Direct} run transitions toward high jet powers as the spin is driven upward, reaching values comparable to the high-spin \textit{Hybrid} case.
\begin{figure}
\includegraphics[width=\columnwidth]{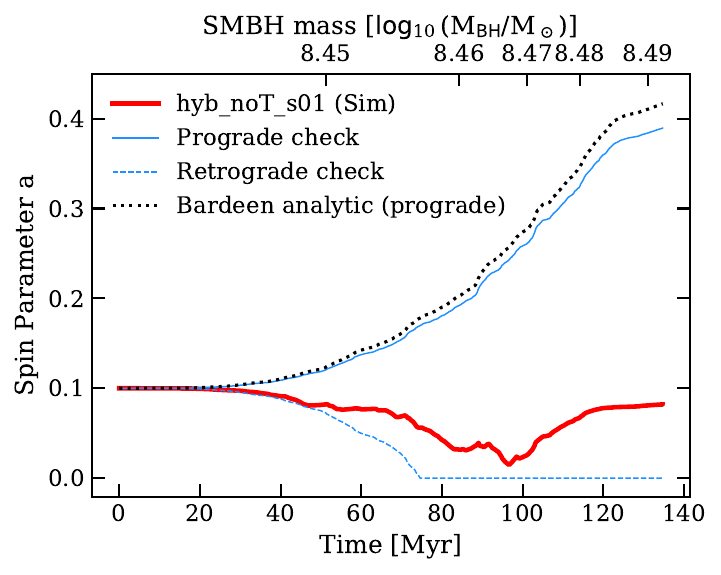}
\caption{Self-consistency checks: spin parameter magnitude plotted as a function of the growing SMBH mass for the fiducial low-spin model, compared against prograde/retrograde analytical checks and the Bardeen analytic prograde limit.}
          \label{validation}%
\end{figure}

Overall, Figure \ref{spin} highlights the key result of this baseline comparison: although the cold-gas reservoir evolves nearly identically across the suite, the central coupling does not. The adopted spin closure leaves the global condensation pathway largely unchanged, but it strongly affects how efficiently the condensed gas feeds the SMBH and how the resulting feedback couples back to the core.

\subsection{Spin evolution and jet reorientation}
Figure \ref{smbh_diag} confirms this picture, showing much higher variability in both the spin magnitude (top panel) and in the inclination and azimuthal angles (bottom panel) for the \texttt{dir\_noT\_s01} run. In fact, by directly adding the resolved sink-scale angular momentum to the SMBH, the \textit{Direct} model overestimates the angular momentum transferred to the black hole, making the accreted gas artificially efficient at reorienting the spin and changing its magnitude. In the fiducial low-spin run \texttt{hyb\_noT\_s01}, instead, fluctuations are smaller, given the ISCO closure, and the spin magnitude decreases from $a\simeq 0.1$ to a few $\times 10^{-2}$ by $\sim 100$ Myr, indicating that (over this interval) mass dilution, intermittent retrograde delivery, and/or BZ spin-down dominate over coherent prograde spin-up. The \textit{Hybrid} run with higher-spin \texttt{hyb\_noT\_s05} shows a similar decline in spin magnitude from $a\simeq 0.5$ to $\sim 0.15-0.2$ over the same period. In general, a tendency towards a low-spin equilibrium is expected in a CCA environment: the SMBH is fed through recurrent, misaligned torque episodes, so spin-up is repeatedly interrupted by stochastic and sometimes retrograde angular-momentum delivery \citep{king2006, king2008, dotti2013}.

We now look at the bottom panel, which describes the deviation from the initial SMBH spin axis. While the \texttt{hyb\_noT\_s01} run reaches inclination angles of the order of $\sim 10-20$ degrees, the \texttt{hyb\_noT\_s05} run remains within a few degrees for most of the evolution. 
The evolution of the direction of the spin axis highlights therefore a second robust trend: low-spin SMBHs are much more susceptible to reorientation, given their lower rotational angular-momentum. 

\subsection{Impact on jet morphology and environment} \label{s:impact}
The differences in spin evolution described above translate directly into different feedback-coupling geometries, and therefore into different morphological and kinematic responses of the surrounding medium.
The morphological imprint of the spin prescription is clearly visible, at different scales, in the temperature and density projection map sequences of Figures \ref{temp_proj} and \ref{dens_proj}. Although, as we have seen, all runs condense a very similar cold-gas reservoir, the way in which the injected jet power couples to the surrounding medium differs substantially. The \texttt{benchmark\_s01} run inflates comparatively regular bipolar cavities aligned with the initial jet direction, whereas the two \textit{Hybrid} models preserve coherent large-scale lobes but with different extents and lateral spreading set by their distinct spin axis evolution. In particular, in Figure \ref{temp_proj}, the higher-spin run \texttt{hyb \_noT \_s05} maintains the most extended and hottest bipolar structure, consistent with its systematically higher Blandford–Znajek efficiency and smaller reorientation angles. By contrast, the \texttt{dir\_noT\_s01} run develops a broader and more disrupted cavity system: because the jet axis wanders rapidly, the deposited energy is distributed over a wider solid angle, weakening the persistence of a single collimated channel and producing a more isotropically stirred core, with turbulence being injected by the jet more diffusely.
This indicates that the spin prescription (through its impact on jet power and direction) modulates the morphology of the cavities excavated by the jet, regulating the subsequent condensation cascade. 

The same trend is also evident in the inner column-density maps shown in Figure \ref{dens_proj}. In the \textit{Hybrid} runs, dense gas remains organized into a relatively compact, flattened central inflow, indicating that the jet continues to couple preferentially to the polar direction without disrupting the equatorial supply of cold material. At late times, the low-spin \textit{Hybrid} case develops stronger warping and asymmetry as the spin axis becomes more mobile, whereas the higher-spin run combines a more stable jet direction with stronger polar clearing. The \textit{Direct} model again stands out as the most disruptive case, even within the 2-kpc radius considered in this figure: repeated changes in jet orientation excavate the central region from different directions, fragmenting the dense gas distribution. This provides a natural morphological explanation for the more suppressed and bursty sink accretion history seen in Figure \ref{spin}. 

\begin{figure}
\includegraphics[width=\columnwidth]{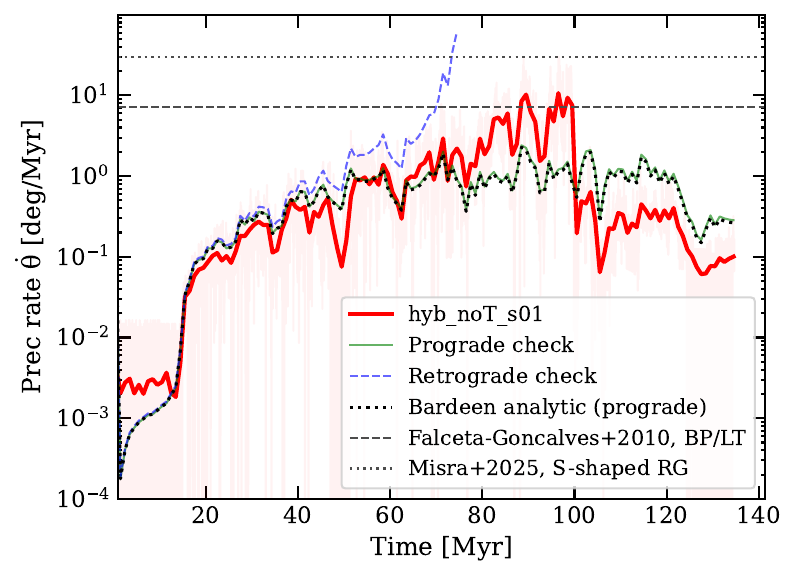}
\caption{Geometric reorientation rate of the SMBH spin axis for the fiducial low-spin \textit{Hybrid} model (\texttt{hyb\_noT\_s01}), compared to the same analytical benchmark models as in Fig.\ \ref{validation}.}
          \label{prec_rate}%
\end{figure}

The near-identical evolution of the total cold-gas mass across the no-turbulence suite suggests that the global thermodynamic pathway to condensation is not strongly altered by the spin prescription at the level of this baseline setup. However, the kinematic state of the full box responds to spin-regulated feedback.
Figure~\ref{sigmav} shows the volume-weighted velocity dispersion $\sigma_v$ of the entire gas within spherical regions of $R=50$ kpc (solid lines) and $R=1$ kpc, calculated after masking high-velocity jet material (cells with $v>1000$ km s$^{-1}$). For each snapshot, we compute the volume-weighted mean velocities $\langle v_i\rangle_w$ and variances $\sigma_i^2=\langle v_i^2\rangle_w-\langle v_i\rangle_w^2$ for $i=x,y,z$, and report the combined dispersion
\begin{equation}
\sigma_v \equiv \left(\sigma_x^2+\sigma_y^2+\sigma_z^2\right)^{1/2}.
\end{equation}
For the region with $R<50$ kpc, the benchmark run shows the slowest growth, remaining below $\sim 100$ km s$^{-1}$ by $\sim 100$ Myr. The high-spin \textit{Hybrid} run (\texttt{hyb\_noT\_s05}) reaches $\sigma_v\sim 100$--130 km s$^{-1}$ after the onset of sustained jet activity, consistent with stronger kinetic injection. The \textit{Direct} run shows the most rapid rise, reaching $\sigma_v\sim 250-300$ km s$^{-1}$ by $\sim 50$--60 Myr and then saturating at that level over the available time span, reflecting the combination of strong jet power and large-amplitude axis reorientation in this model (Fig.~\ref{spin}). Interestingly, the fiducial low-spin \textit{Hybrid} run (\texttt{hyb\_noT\_s01}), which we run up to 140 Myr, shows a delayed but ultimately strong rise in $\sigma_v$ at late times, coincident with sustained growth in the spin inclination (Fig.~\ref{smbh_diag}, bottom panel) and elevated precession rates (see also Figure~\ref{prec_rate}, described later in the text). On the other hand, the velocity dispersion at small scales - for $R < 1$ kpc - saturates at lower values at around $\sigma_v\sim 350-400$ km s$^{-1}$ for the \textit{Direct} run, relatively closer to its full-box $\sigma_v$, indicating that faster spin and jet re-orientation rates correspond to more uniform turbulence. A similar trend, with the two curves partially converging once the jet starts to re-orient at a faster rate, is visible also for our fiducial \texttt{hyb\_noT\_s01}. The other runs, instead, saturate at around $\sigma_v\sim 450-500$ km s$^{-1}$.

This supports the idea that persistent axis variability redistributes kinetic input over a wider solid angle, stirring the gas more evenly. In this sense, the spin prescription affects not only the jet energetics but also the geometry through which the energy is deposited into the inner halo.
In the companion paper P26b, we test this interpretation by implementing different turbulence regimes -- driven and interrupted -- at high resolution.

In Figure \ref{pdf_noturb}, we plot the mass-weighted probability density distributions (PDFs) within the central 100 pc. Throughout the paper, we adopt the same thermodynamic bins used in the companion {\sc BlackHoleWeather} works (see \S\ref{s:num}). 
For each simulation snapshot, radial shell, and thermal phase, we compute a mass-weighted probability density function of the logarithmic gas number density. We define
\begin{equation}
x_i \equiv \log_{10}\left(\frac{n_i}{\mathrm{cm^{-3}}}\right),
\end{equation}
where \(n_i\) is the gas number density in cell \(i\).  For a given thermal phase \(\alpha\) and a radial shell \(R\), the total gas mass entering the PDF is
\begin{equation}
M_{\alpha,R} = \sum_{i\in(\alpha,R)} m_i ,
\end{equation}
where \(m_i\) is the cell mass and the sum is restricted to cells that satisfy both the radial-shell selection and the temperature cut defining phase \(\alpha\). The mass-weighted PDF is then
\begin{equation}
\mathcal{P}_{\alpha,R}(x)=\frac{1}{M_{\alpha,R}}\sum_{i\in(\alpha,R)}m_i\,\delta\!\left(x-x_i\right),
\end{equation}
so that
\begin{equation}
\int \mathcal{P}_{\alpha,R}(x)\,dx = 1 .
\end{equation}
In practice, we estimate this quantity with a finite histogram in \(x=\log_{10}(n/\mathrm{cm^{-3}})\).  For a bin \(j\) of width \(\Delta x_j\), the plotted value is 
\begin{equation}
\mathcal{P}_{\alpha,R,j} = \frac{1}{M_{\alpha,R}\,\Delta x_j} \sum_{i\in(\alpha,R,j)} m_i .
\end{equation}
The figures show
\begin{equation}
\log_{10}\left(\frac{d \mathcal{P}}{d\log_{10} n}\right) \equiv \log_{10}\mathcal{P}_{\alpha,R,j}.
\end{equation}

Figure \ref{pdf_noturb} shows that all no-turbulence models follow nearly the same thermodynamic path toward multiphase condensation. The hot distribution broadens toward higher densities, a warm bridge develops, and cold atomic plus molecular gas build up in the central 100 pc with similar timing and phase occupation across the \textit{Benchmark}, \textit{Direct}, and \textit{Hybrid} runs. This is a useful control result: in the absence of externally driven turbulence, the formation of the central condensed reservoir is regulated mainly by the common cooling-feedback cycle, rather than by the details of the SMBH spin prescription.

The spin closure therefore does not primarily change whether multiphase gas forms, but how the condensed gas couples to the SMBH once it reaches the sink scale. The key model differences emerge in the angular momentum transferred to the hole, the resulting jet power and axis response, and the way feedback stirs the gas. This separation provides the baseline for the turbulent comparison in \S\ref{s:pdfs}: the no-turbulence PDFs isolate the near-degenerate thermodynamic condensation pathway, while the high- and low-turbulence runs test how the weather state changes the radial connectivity, intermittency, and torque coherence of the cold inflow.

\subsection{Fiducial model: self-consistency check}
In Figure~\ref{validation} we plot the spin evolution as a function of time and SMBH mass, with the goal of benchmarking our spin model by comparing our fiducial \texttt{hyb\_noT\_s01} run with three theoretical benchmark curves. These analytical checks are computed in post-processing on the exact, time-dependent black hole mass growth history extracted directly from the simulation, but with a different spin integration model.
The prograde / retrograde checks represent the theoretical spin magnitude and precession rate using our fiducial hybrid ISCO closure and Blandford-Znajek spin-down torque but forcing the orientation of the inflowing gas to be either 100\% aligned (always prograde) or 100\% anti-aligned (always retrograde) with the instantaneous black hole spin, instead of computing the dot product between the accreted material and the spin at each time step. 
The Bardeen analytic prograde curve instead follows the classical spin-up trajectory derived by \citet{bardeen1970}, assuming a steady, thin, prograde accretion disc and neglecting the BZ spin-down. It therefore represents an absolute upper limit for purely accretion-driven spin-up. The purpose of this comparison is not to claim equivalence with idealized thin-disc solutions, but to verify that the simulated spin evolution remains within a physically sensible envelope once the same SMBH mass-growth history is coupled to our \texttt{Hybrid} closure. The Bardeen limit lies less than $10\%$ above our prograde check, which already includes BZ spin-down, showing that the \texttt{Hybrid} model remains well bracketed by the relevant analytic limits despite not resolving the horizon-scale disc.
\begin{figure}
\includegraphics[width=\columnwidth]{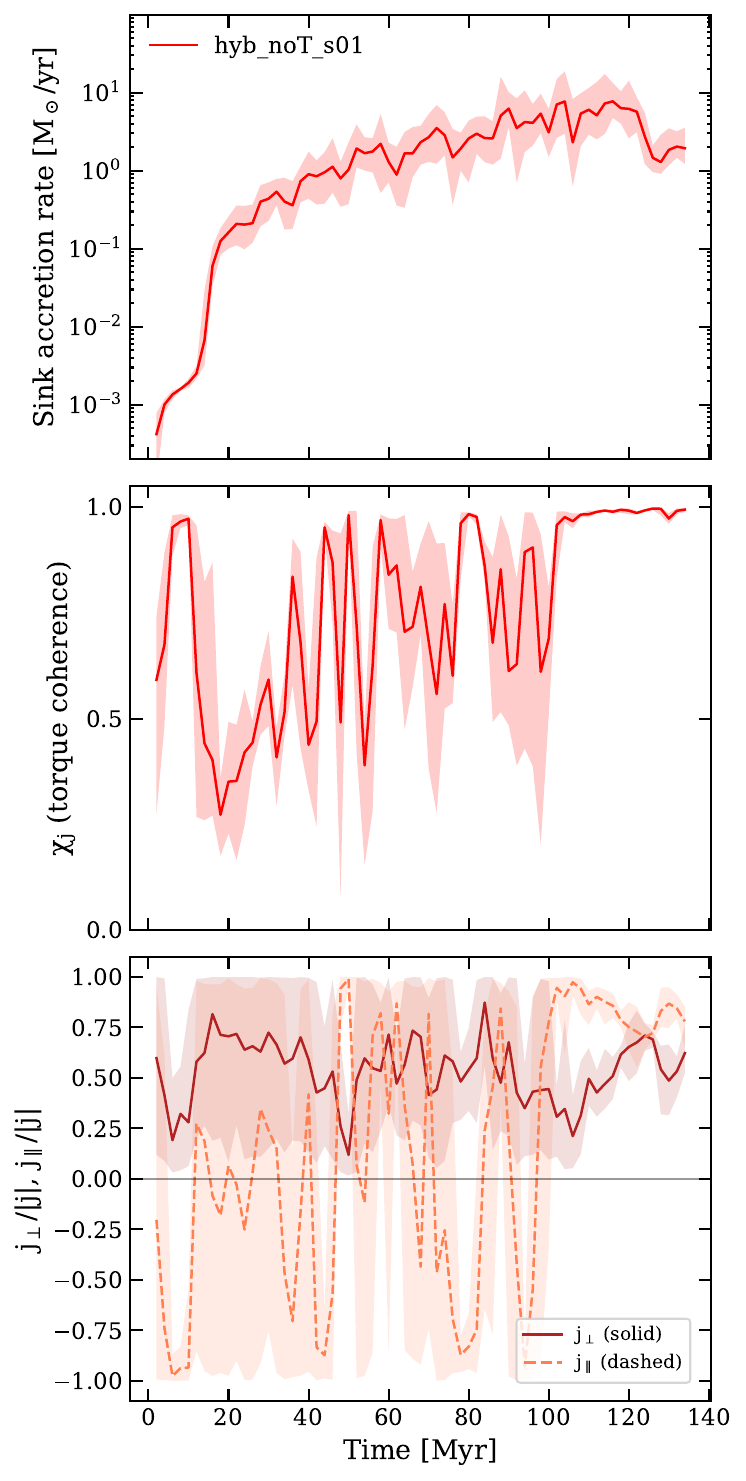}
\caption{Torque-delivery statistics for the fiducial \texttt{Hybrid} run. The top panel shows the sink accretion rate, the middle panel shows the torque-coherence parameter $\chi_\mathrm{j}$ computed over a trailing window $W=2\,\mathrm{Myr}$, and the bottom panel shows the normalized perpendicular and parallel components of the accreted specific angular momentum, $j_\perp/|\mathbf{j}|$ (solid) and $j_\parallel/|\mathbf{j}|$ (dashed). Negative values of $j_\parallel$ mark retrograde torque-delivery episodes, while large $j_\perp$ identifies misaligned accretion capable of reorienting the SMBH spin. Shaded areas represent 1--$\sigma$ deviations from the average.
}
\label{torque_not}%
\end{figure}

\subsection{Torque delivery and jet-axis reorientation}
\begin{figure}
\includegraphics[width=0.95\linewidth]{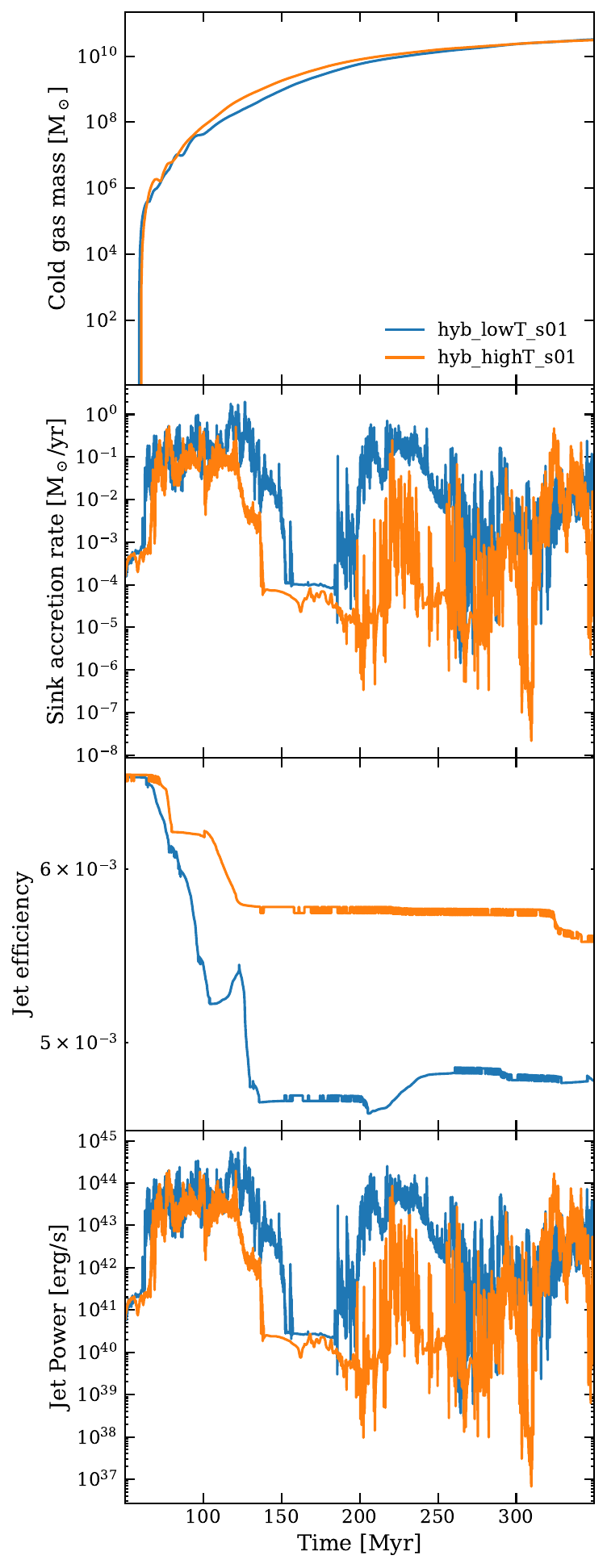}
\caption{Same as Figure \ref{spin} for the two turbulence runs, showing the total cold gas mass, accretion rate onto the sink, jet efficiency and power vs absolute time. Although the total cold-gas reservoirs are comparable, the high-turbulence run shows more intermittent sink accretion and deeper quiescent intervals, which limit the instantaneous jet power despite its slightly higher spin-dependent efficiency.}
\label{spin_turb}%
\end{figure}
In Figure \ref{prec_rate} we compute the geometric spin-axis reorientation rate between successive outputs, which provides a useful diagnostic for comparison with observed systems showing misaligned cavities or bent and S-shaped jet morphologies.
It is calculated as
$\Delta\theta_i=\cos^{-1}(\hat{\boldsymbol{a}}_{i}\cdot\hat{\boldsymbol{a}}_{i+1})$,
where $\dot{\theta}_i=\Delta\theta_i/\Delta t_i$ is reported in deg \ Myr$^{-1}$.
This rate is a cadence-dependent geometric diagnostic, not a directly observed periodic precession frequency. It is used here to connect the simulated jet-axis wandering to systems with misaligned cavities, bent jets, or S-shaped radio morphologies.
The plot illustrates the evolution of the measured geometric reorientation rate for the fiducial \texttt{hyb\_noT\_s01}, compared to the idealized analytical timescales computed with the same analytical models used in Figure \ref{validation}. Crucially, these benchmark frequencies are computed using the corresponding spin curves derived in Figure \ref{validation}, which explains why for the retrograde check we obtain an infinite precession rate (once the spin goes to zero) and for the prograde checks we obtain slower precession rates at later times, when the spin, and hence the gyroscopic inertia, grows higher. 
This divergence of course is only a formal consequence of the analytic estimate as the spin angular momentum tends to zero, and should not be interpreted as a physically meaningful infinite precession.

\begin{figure*}
\includegraphics[width=\textwidth]{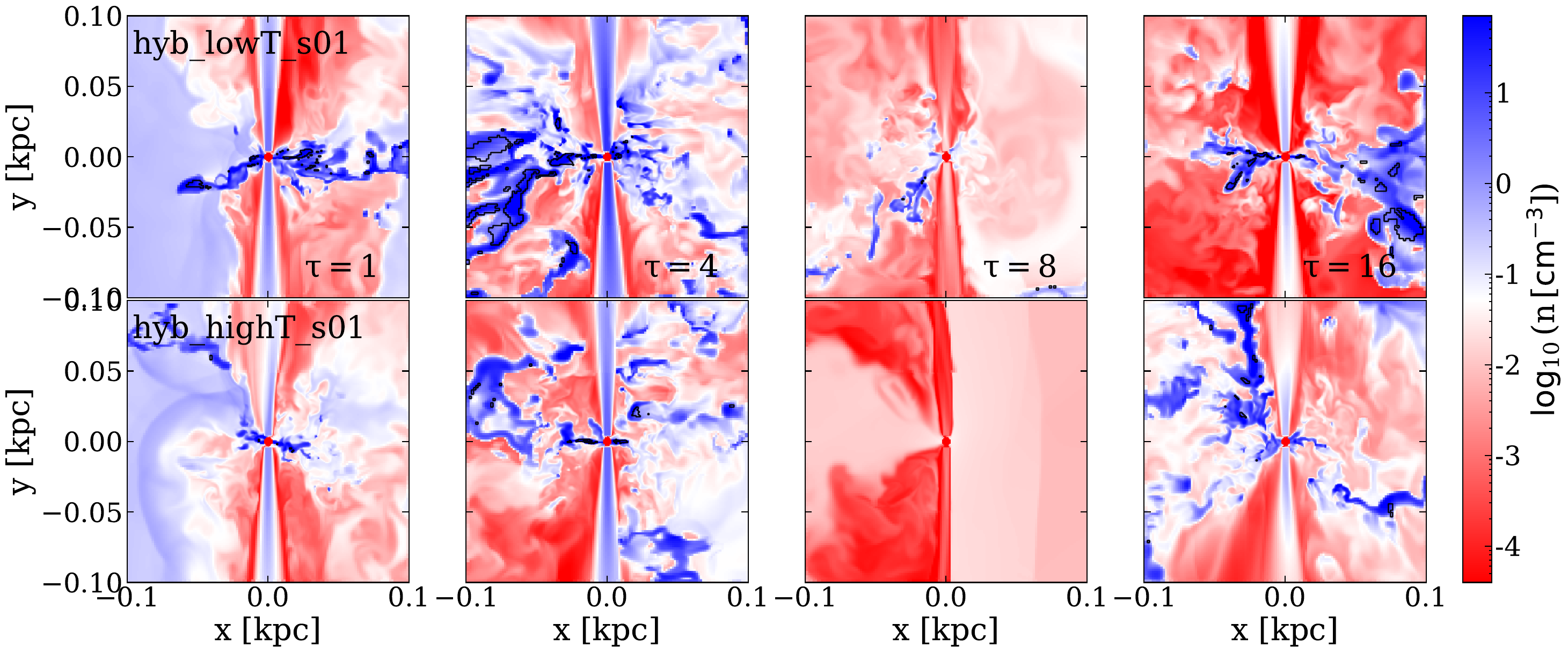}
\caption{Zoom-in into the central 0.1 kpc of the density slices for the high- and low-turbulence Hybrid runs at selected epochs ($\tau = t/t_\mathrm{rain}=1,4,8,16$). Black contours indicate the presence of molecular gas with $T<200$ K.}
\label{slice_turb}%
\end{figure*}

The simulated geometric reorientation rate in the CCA regime naturally remains at $\approx 0.1$ to a few deg Myr$^{-1}$ for most of the evolution, broadly consistent with the range inferred from the observed misaligned cavities in cool-core clusters \citep{falcetagoncalves2010} and S-shaped radio morphologies \citep{misra2025}. 
In particular, between 80 and 100 Myr, the simulated rate spikes by almost an order of magnitude above the analytical prograde solution, approaching $\sim 10$ deg Myr$^{-1}$. This rapid reorientation occurs because chaotic, partially retrograde accretion first spins the SMBH down, reducing its angular-momentum reservoir and therefore its gyroscopic inertia. In this low-spin state, a newly delivered misaligned torque can tilt the spin axis much more efficiently than in the corresponding coherent prograde case. Once the accretion flow becomes coherently prograde again, the SMBH spins back up, rebuilds its angular-momentum reservoir, and the jet axis correspondingly stabilizes. This behaviour illustrates how the time-dependent coherence and sign of the CCA torque regulate jet-axis wandering across the feeding-feedback cycle. It also highlights why spin models that retain stochastic 3D angular-momentum delivery are essential for capturing the short-timescale variability of jet orientation in multiphase AGN environments.

\begin{figure}
\includegraphics[width=\columnwidth]{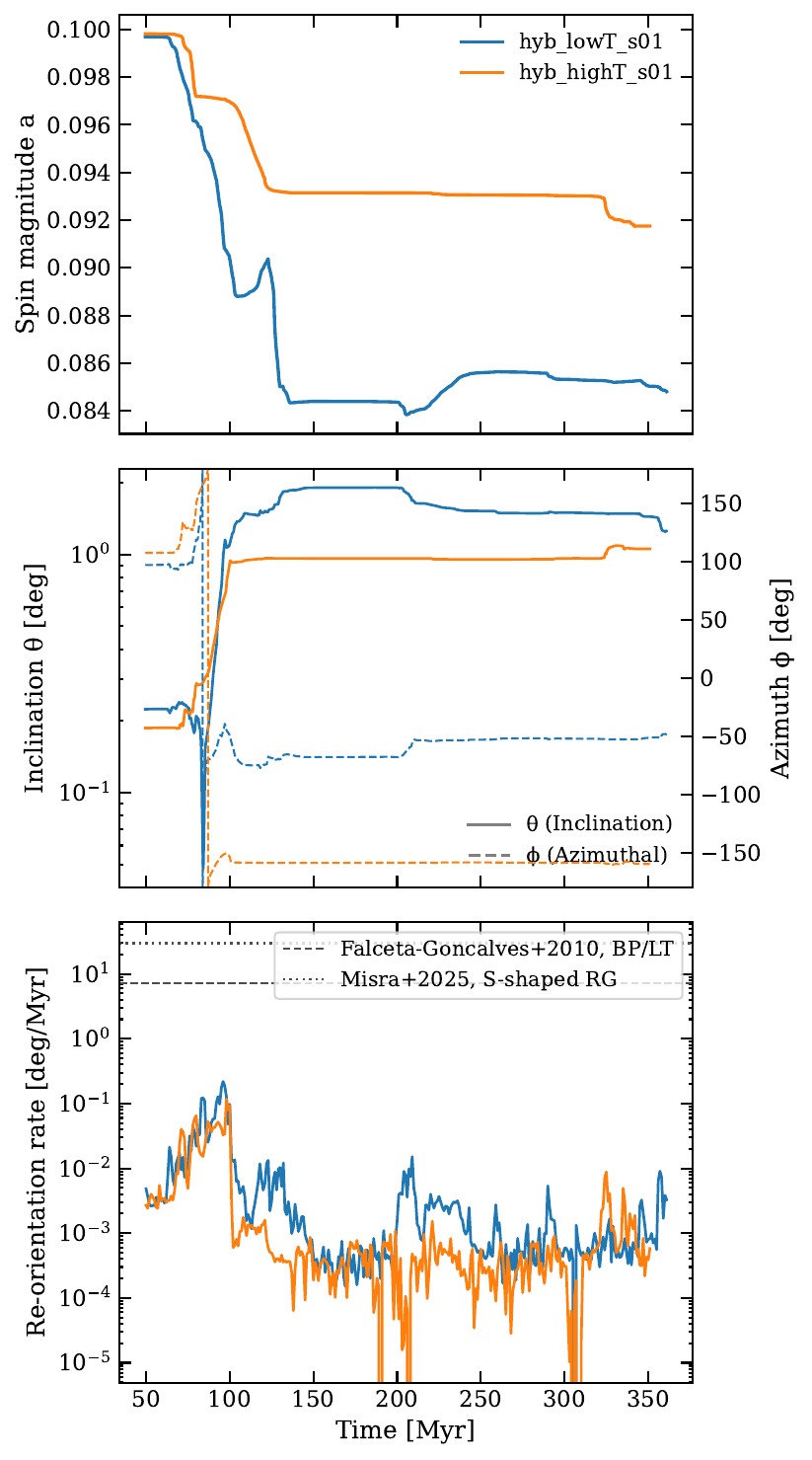}
\caption{Comparative spin evolutions for the low- and high-turbulence environments, showing the evolution of spin magnitude (top), inclination and azimuth angles $\theta$ and $\phi$ (middle), and axis geometric reorientation rate (bottom).}
\label{prec_rate_turb}
\end{figure}

In Figure \ref{torque_not}, we investigate more in depth the statistics of torque delivery in CCA scenarios and how that drives spin evolution. 
This figure provides the clearest physical bridge between multiphase feeding and spin evolution, because it directly quantifies whether the inflow delivers angular momentum coherently or through rapidly cancelling torque episodes.
In the top panel we show the binned accretion rate over a trailing window $W=2$ Myr, in order to show a less noisy trend, with the shaded area representing 1-$\sigma$ deviation from the average. 
In the middle panel, we quantify the coherence of the torque delivered over the same time window by defining the quantity
\begin{equation}
\chi_j(t;W) \equiv 
\frac{\left|\sum\limits_{t_i\in [t-W,t]} \Delta\mathbf{L}(t_i)\right|}
{\sum\limits_{t_i\in [t-W,t]} \left|\Delta\mathbf{L}(t_i)\right|}.
\end{equation}
Values $\chi_j\simeq 1$ indicate a nearly fixed torque direction over $W$ (coherent, disc-like feeding), whereas $\chi_j\ll 1$ indicates strong cancellation from rapidly varying $\Delta\mathbf{L}$ directions (chaotic accretion). A value of $\chi_j\simeq 0$ would indicate a perfectly symmetric or isotropic accretion, so we do not expect to see it even in a fully chaotic scenario. Additionally, we decompose the specific angular momentum of the accreted material, defined as
\begin{equation}
\mathbf{j}_{\rm acc}(t) \equiv \frac{\Delta \mathbf{L}(t)}{\Delta M(t)} \, .
\end{equation}
If we define $\hat{\mathbf{a}}(t)\equiv \mathbf{a}(t)/|\mathbf{a}(t)|$ as the spin unit direction, we can introduce the (signed) component of the angular momentum that is parallel to the spin as
\begin{equation}
j_{\parallel}(t) \equiv \mathbf{j}_{\rm acc}(t)\cdot \hat{\mathbf{a}}(t) \, ,
\end{equation}
and a perpendicular amplitude,
\begin{equation}
j_{\perp}(t) \equiv \left(|\mathbf{j}_{\rm acc}(t)|^2 - j_{\parallel}^2(t)\right)^{1/2} \, .
\end{equation}
Here, $j_{\parallel}>0$ corresponds to prograde delivery (tending to spin up), while $j_{\parallel}<0$ corresponds to retrograde delivery (tending to spin down). Large $j_{\perp}$ indicates a misaligned torque capable of rapid reorientation $\hat{\mathbf{a}}$. In the bottom panel, we can see how the parallel component of the specific angular momentum transitions to strongly negative values around 40, 80 and 100 Myr, indicating pronounced, bursty phases of retrograde accretion. These phases rapidly spin down the SMBH, stripping its gyroscopic inertia and driving the order-of-magnitude spike in the precession rate to nearly 10 deg\,Myr$^{-1}$. Conversely, once the accretion flow becomes persistently prograde and the torque coherence rises toward unity, the SMBH is efficiently spun back up (see Figure \ref{validation}) and the jet axis becomes much more stable. 

\section{Results: spin evolution in turbulent scenarios} \label{s:res2}

We now use the fiducial \textit{Hybrid} prescription to test how explicit halo turbulence modifies the CCA torque-delivery process. The goal is to determine whether turbulence changes (i) the amount of cold gas that condenses, (ii) the efficiency with which this gas reaches the sink, and (iii) the coherence of the angular momentum delivered to the SMBH. This distinction is crucial: a more turbulent halo does not necessarily produce stronger secular spin wandering if the same turbulence fragments the inflow and enhances torque cancellation.

This comparison should be read in the context of the companion {\sc BlackHoleWeather} sequence. B26a isolated the turbulence-only CCA problem, showing that stronger stirring produces a more extended, filament-rich stormy rain, while weaker stirring favours a more compact and centrally retained rainy configuration. C26a then showed that a fixed-axis jet does not erase CCA, but anisotropically reorganizes it through compression, entrainment, shear, and mixing. Here we add the spin-coupled layer: the jet power and direction respond to the angular momentum delivered by the resolved CCA flow. The question is therefore not only how much gas condenses, but how the weather state controls the vector coupling between the cold reservoir, the SMBH spin, and the next feedback episode.

We consider here two higher-resolution simulations, reaching a central cell size of \(0.7\) pc, in which the fiducial Hybrid spin prescription is coupled to explicit turbulent driving in the hot atmosphere. These runs are designed to bracket two distinct CCA environments: a low-turbulence case, \texttt{hyb\_lowT\_s01}, and a high-turbulence case, \texttt{hyb\_highT\_s01}. The difference lies in the prescribed root-mean-square acceleration amplitude of the turbulence driving, which is $a_{\rm rms} = 6.2 \times 10^{-9}$ cm s$^{-2}$ for \texttt{hyb\_lowT\_s01} and $a_{\rm rms} = 1.55 \times 10^{-8}$ cm s$^{-2}$ for \texttt{hyb\_highT\_s01}. Since the spin closure is identical in both simulations, differences in the resulting spin evolution isolate the effect of turbulence on the delivery of mass and angular momentum to the sink, which is 4 cells in radius. In practice, turbulent driving broadens the halo velocity field, enhances mixing, and partly offsets radiative cooling, thereby reducing the mean central accretion rate and altering the coherence of the torque delivered to the SMBH (see B26a).

The central question is therefore whether turbulence acts by increasing stochasticity, by suppressing coherent inflow, or by doing both at different phases of the weather cycle. We quantify this by comparing the cold-gas reservoir, sink accretion history, spin magnitude and reorientation rate, and the coherence and orientation of the angular momentum delivered to the SMBH.
\begin{figure}
\includegraphics[width=\columnwidth]{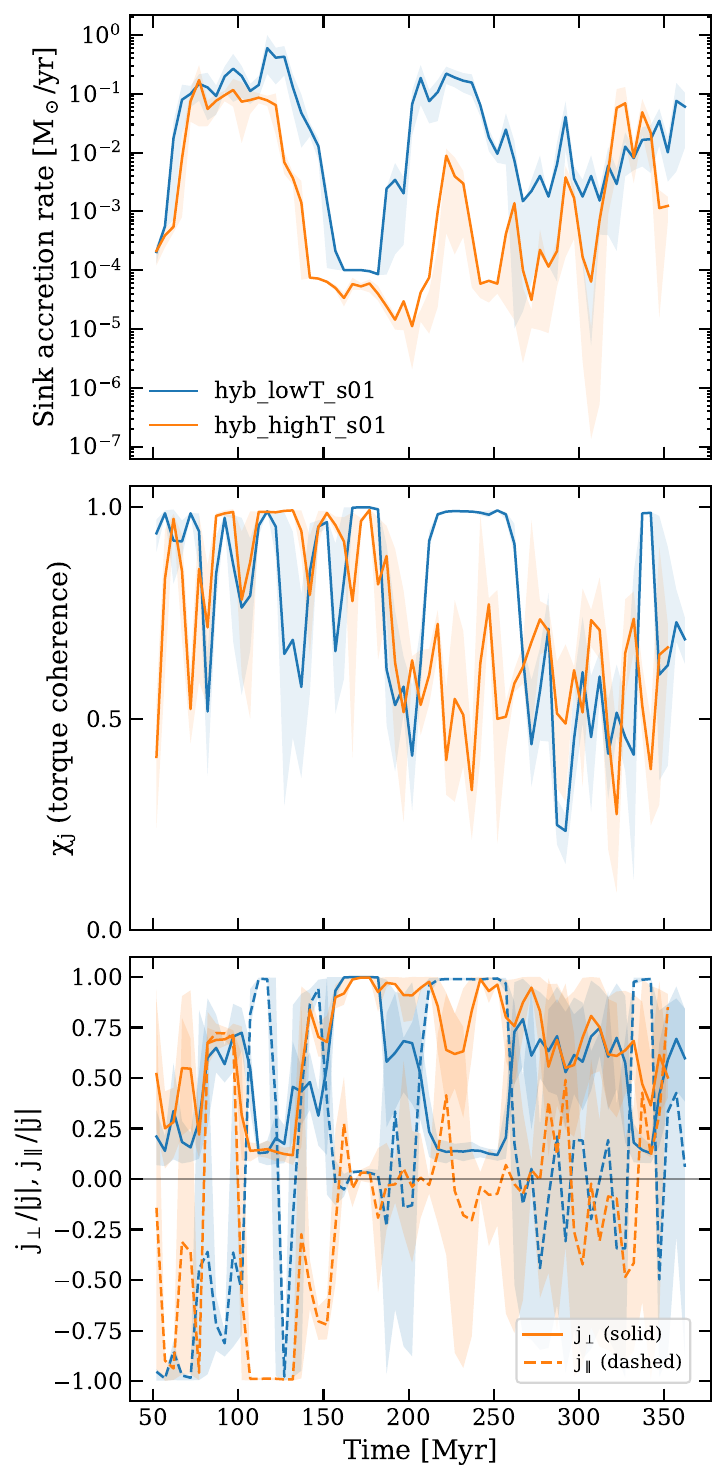}
\caption{Same as Figure~\ref{torque_not}, but for the low- and high-turbulence \textit{Hybrid} runs. This figure provides the key diagnostic link between turbulent CCA feeding and SMBH spin evolution: the low-turbulence case maintains longer coherent torque-delivery episodes, allowing successive accretion events to accumulate more effectively, whereas the high-turbulence case shows more fragmented, cancellation-dominated inflow, leading to a more stable long-term spin evolution.}
\label{torque_turb}%
\end{figure}

For both runs, we first let the simulations run with only turbulence active, keeping both cooling and jets off, to create more realistic initial conditions of velocity dispersion \citep[for direct and indirect observed velocity dispersion values, see also][B26a,b]{gaspari2013, hofmann2016, xrism2025}. The built-up velocity dispersions correspond to \(\sigma_v \simeq 60~{\rm km~s^{-1}}\) for the low-turbulence run and \(\sigma_v \simeq 160~{\rm km~s^{-1}}\) for the high-turbulence run. From that point onward, jets and cooling are activated.
We note that in the turbulent-run figures, the time axis shows the absolute simulation time, including the initial turbulence-only relaxation stage. The dimensionless epochs instead use \(\tau=(t-t_{\rm on})/t_{\rm rain}\), where \(t_{\rm on}\) is the activation time of cooling and jets, and \(t_{\rm rain}=18\) Myr for the turbulent runs.

\subsection{High- vs.~low-turbulence CCA weather}
Figure~\ref{spin_turb} shows that the two turbulent runs undergo broadly similar multiphase condensation cycles, accumulating comparable cold-gas reservoirs over the simulated interval. The first important result is therefore that turbulence affects the delivery of condensed gas to the SMBH more strongly than the total amount of cold gas that forms. The main difference appears in the central feeding history: the low-turbulence run sustains broader, longer-lived, and on average stronger accretion episodes, whereas the high-turbulence run breaks the inflow into shorter, more intermittent bursts separated by deeper troughs in the sink accretion rate $\dot{M_\mathrm{in}}$. 
This difference propagates directly into the feedback channel. Although the high-turbulence run retains a slightly higher spin-dependent jet efficiency because its spin magnitude remains larger, its jet power is often limited by the reduced and more intermittent inflow. The two systems therefore differ primarily in how efficiently and coherently that gas is delivered to the central few parsecs. A more detailed time-series characterization of these duty cycles, including the active accretion fraction, burst durations, and accretion-rate variability amplitude relative to the mean, is presented in the companion paper P26b. Here we use these trends primarily to interpret how turbulence modifies the delivery of mass and angular momentum to the SMBH, and hence the resulting spin evolution.

\begin{figure*}
\includegraphics[width=0.95\textwidth]{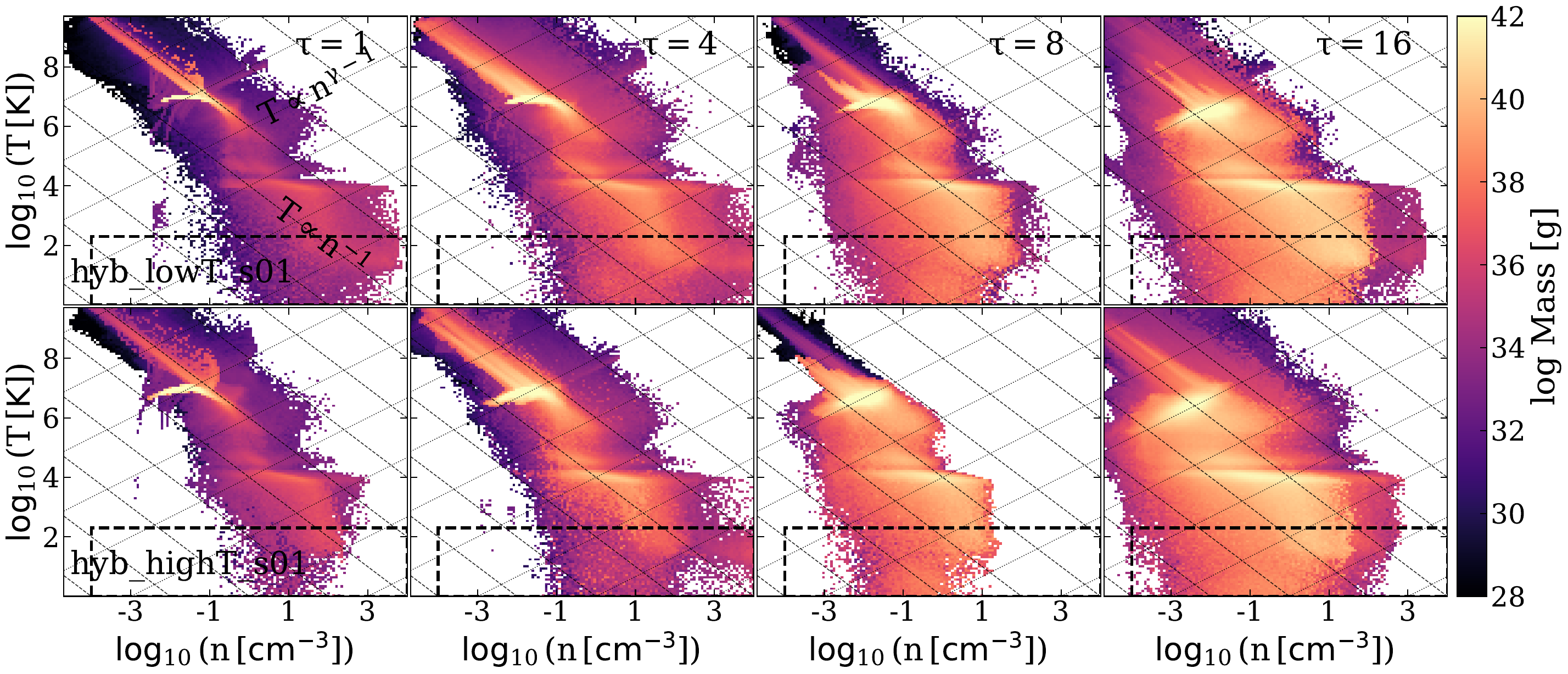}
\caption{Mass-binned density-temperature phase diagrams for the low- and high-turbulence \texttt{Hybrid} runs at selected epochs (\(\tau=t/t_\mathrm{rain}=1,4,8,16\)). Both runs follow a similar broad cooling sequence from the hot phase toward warm, cold, and molecular gas, while turbulence mainly changes the organization and persistence of the condensed high-density material. The dashed box marks the region highlighted with black contours in Figures \ref{dens_proj} and \ref{slice_turb}.}
\label{phase_turb}
\end{figure*}

In the zoom-in density slices shown in Figure~\ref{slice_turb} we can clearly see the cycles that both feeding and feedback go through. The inner region evolves through repeated transitions between rainy, cloudy, and sunny configurations (see \S\ref{sec:weather_cycling}). In rainy phases, dense filaments and compact cold structures reach the central \(\sim 100\) pc, feed the sink, and trigger a stronger jet response. Since we keep the mass loading factor fixed, higher accretion rates correspond to denser jets. Feedback then heats and partially clears the nucleus, lowering the central cold-gas filling factor and pushing the system toward a sunnier state in which the accretion rate drops toward a low, hot-mode baseline, while cold material can persist at larger radii $R>1$ kpc or in fragmented structures (see Appendix \ref{app:large_slices}), but it is less efficiently connected to the sink and therefore delivers mass and angular momentum less coherently. The next rainy episode begins when this gas, or newly condensed material, is again channelled toward the centre. 

This behaviour extends the comparison with both B26a and C26a. In B26a, where no explicit jet is present, the turbulence level mainly controls the radial topology of the cold rain: stronger turbulence spreads condensation to larger radii, while weaker turbulence keeps the cold reservoir more centrally concentrated. In C26a, the fixed-axis jet reshapes this topology by clearing the polar channel and promoting condensation, mixing, and entrainment along the jet--atmosphere interface. In the present spin-coupled runs, the same jet-regulated CCA mechanisms remain active, but the feedback geometry can respond to the evolving SMBH spin. This likely explains why extended sunny intervals with sub-Bondi SMBH accretion rates of \(\sim10^{-5}\)--\(10^{-4}\,M_{\odot}\,{\rm yr^{-1}}\) appear more clearly here than in the fixed-axis C26a,b runs: spin-driven jet-axis variability can redistribute feedback over a broader solid angle, heat or uplift the nuclear gas more efficiently, and temporarily suppress the cold supply to the SMBH. The corresponding duty-cycle differences are quantified in P26b.

In the low-turbulence run, cold structures survive longer and remain more persistently connected to the central few parsecs, producing broader rainy episodes and more sustained accretion. In the high-turbulence run, the same broad condensation cycle is present, but turbulent mixing and shredding break the inflow into shorter, less coherent delivery events separated by deeper quiescent intervals. This recurrent loss and recovery of central connectivity is the physical origin of the different torque-delivery statistics discussed below.

Figure~\ref{prec_rate_turb} shows that these different feeding statistics translate into distinct, although not dramatically divergent, spin histories. The low-turbulence run undergoes the stronger secular spin-down, with the spin magnitude decreasing by \(\simeq 15\%\) from its initial value to \(a \approx 0.085\), whereas the high-turbulence run decreases by \(\simeq 8\%\), remaining closer to \(a \approx 0.092\). The clearer distinction, however, is geometric: the low-turbulence case reaches a larger cumulative inclination excursion, approaching \(\sim 2^\circ\), while the high-turbulence run remains closer to \(\sim 1^\circ\). Both runs display highly variable instantaneous geometric reorientation rates, with short-lived spikes associated with misaligned accretion events, but the net reorientation is larger in the low-turbulence case: in this setup, the high-turbulence run is more intermittent, but it also delivers less persistent net angular momentum to the SMBH. Lower mean accretion rates, shorter feeding episodes, and stronger cancellation between differently oriented inflows reduce both the secular spin-down and the cumulative jet-axis reorientation. Conversely, the low-turbulence run maintains a more coherent connection between the cold reservoir and the sink, allowing misaligned torques to accumulate more efficiently over time.

The origin of this behaviour is clarified by Figure~\ref{torque_turb}, where we compare the torque-delivery diagnostics for \texttt{hyb\_lowT\_s01} and \texttt{hyb\_highT\_s01}. The middle panel shows that the angular momentum delivered to the sink can remain coherent over Myr windows even in a turbulent CCA flow. In both runs, \(\chi_j\) intermittently reaches values \(\sim 0.6\)--1, indicating that the accreted material is not drawn from a fully isotropic distribution of clumps, but often arrives through transiently preferred directions. The key difference is the persistence of these coherent windows. The low-turbulence run maintains longer intervals of high coherence, particularly during the phases of sustained accretion, whereas the high-turbulence run is more fragmented: multiple clouds, filaments, or streams with different angular-momentum orientations contribute within the same time window, producing sharper drops in \(\chi_j\) and stronger cancellation of the net torque.

\begin{table*}
\begin{tabular}{lrrrrrrrrrr}
\toprule
Run & $\langle \dot M_{\rm sink}\rangle$ & $(a_\mathrm{f}- a_0)/a_0$ & $\theta_{\rm max}$ & median $\chi_j$ & $f(\chi_j>0.7)$ & $f(j_\parallel<0)$ & $\langle \lambda_\mathrm{Edd} \rangle$ & $f(\lambda_\mathrm{Edd} > 0.01)$ \\
\midrule
\texttt{hyb\_lowT\_s01} & 0.0669 & -0.149 & 2.13 & 0.92 & 0.645 & 0.444 & 0.0107 & 0.701 \\
\texttt{hyb\_highT\_s01} & 0.0177 & -0.0808 & 1.12 & 0.692 & 0.495 & 0.599 & 0.00284 & 0.383 \\
\bottomrule
\end{tabular}
\caption{Statistical comparison between the low- and high-turbulence runs: average sink accretion rate, fractional spin change, maximum inclination angle, median torque coherence, coherence duty cycle, retrograde duty cycle, mean Eddington ratio, and Eddington-ratio duty cycle.}\label{table_turb_comp}
\end{table*}

The bottom panel confirms that the delivered angular momentum is strongly three-dimensional in both simulations. The perpendicular component, \(j_\perp/|j|\), is often large, showing that a substantial fraction of the torque budget is in principle available to tilt the spin axis. At the same time, the signed parallel component, \(j_\parallel/|j|\), repeatedly changes sign, marking alternating prograde and retrograde delivery. Retrograde intervals reduce the spin magnitude and make the SMBH easier to reorient, whereas prograde intervals rebuild the spin reservoir and tend to stabilize the jet axis. 
The different balance between these components, together with the different coherence time encoded in \(\chi_j\), provides the direct link between turbulent CCA feeding and the spin histories shown in Figure~\ref{prec_rate_turb}. 
We provide a more quantitative comparison between the high- and low-turbulence runs in Table~\ref{table_turb_comp}. The higher average accretion rate of the low-turbulence run is associated with a larger cumulative inclination excursion, higher median torque coherence, and a higher Eddington-ratio duty cycle.

\begin{figure*}
\includegraphics[width=0.9\textwidth]{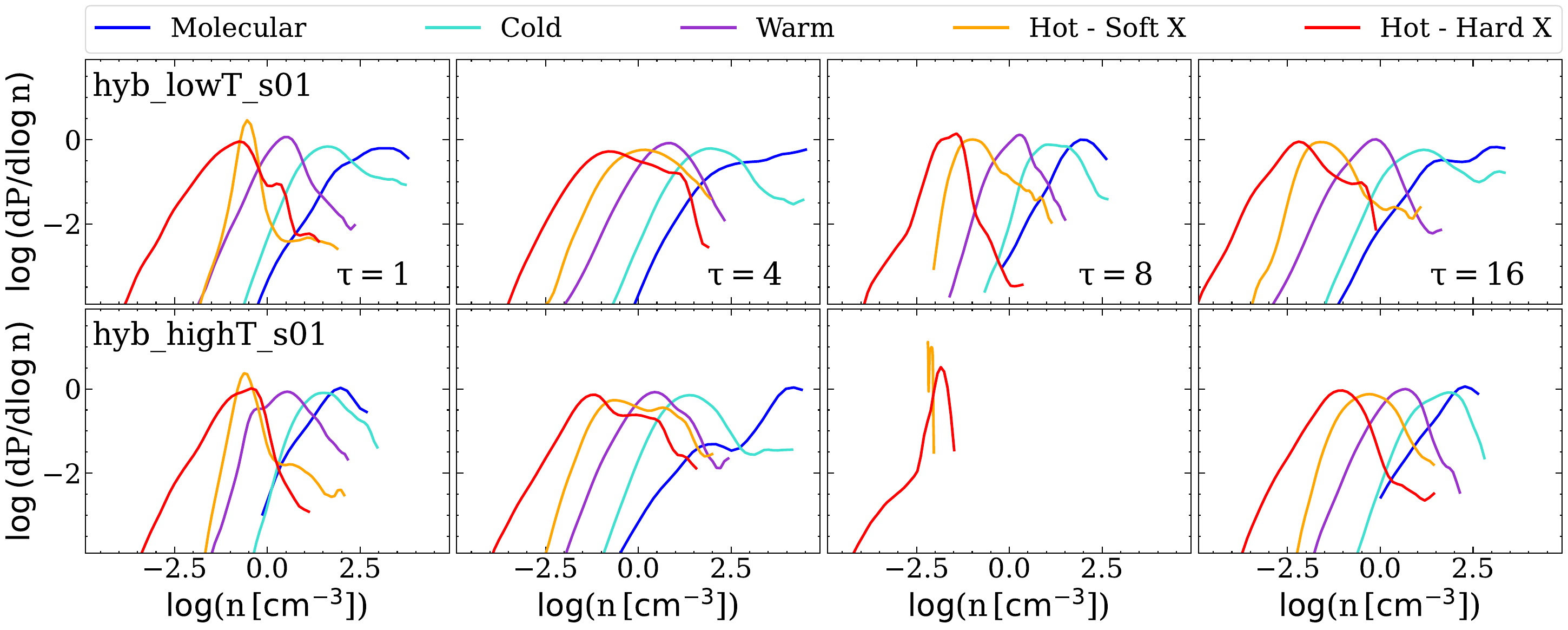}
\caption{Mass-Weighted PDFs of gas density for the 5 phases within the central \(100\) pc for the low- and high-turbulence \texttt{Hybrid} runs at selected epochs (\(\tau=t/t_\mathrm{rain}=1,4,8,16\)). The PDFs trace the recurrent repopulation and clearing of the nuclear region across the weather cycle: the low-turbulence run shows more persistent multiphase central occupancy, whereas the high-turbulence run more frequently returns to a hot-dominated state. This indicates that stronger turbulence does not suppress condensation globally, but reduces the residence time and volume filling of cold gas in the central accretion region, consistent with the reduced torque coherence shown in Figure~\ref{torque_turb}.}
\label{pdf_turb}
\end{figure*}
\subsection{Thermodynamic structure of the turbulent runs} \label{s:pdfs}

Figures~\ref{phase_turb} and \ref{pdf_turb} show the thermodynamic structure of the turbulent runs during the recurrent weather cycle discussed above. Figure~\ref{phase_turb} presents the mass-binned distribution in the \((n,T)\) plane, while Figure~\ref{pdf_turb} shows the corresponding mass-weighted density probability density functions (PDFs) within the central \(100\) pc. The distinction between these diagnostics is important: the phase diagrams emphasize where most of the gas mass resides in thermodynamic space, whereas the PDFs trace the volume filling and phase occupancy of the nuclear region. Both simulations populate a similar broad cooling sequence, extending from the hot X-ray phase toward a warm, cold, and molecular gas. This confirms that explicit turbulent driving does not prevent multiphase condensation. Instead, turbulence primarily regulates how the condensed gas is organized, how long it remains connected to the centre, and how efficiently it repopulates the central accretion region.

The reference lines in Figure~\ref{phase_turb} help clarify the thermodynamic nature of this cooling path. Isobaric evolution follows \(T\propto n^{-1}\), while adiabatic compression or expansion follows \(T\propto n^{\gamma-1}\), with \(\gamma=5/3\). Much of the transition from the hot phase toward dense cold gas proceeds broadly toward higher density and lower temperature, closer to an isobaric condensation path than to pure adiabatic compression. This is expected for gas cooling while remaining approximately pressure-coupled to the surrounding halo. Purely adiabatic compression would instead heat the gas as its density rises, and cannot by itself generate the cold and molecular branch. Deviations from the isobaric direction, together with intermediate-temperature material between the hot and cold phases, trace the combined effects of turbulent mixing, jet-driven compression or expansion, shocks, and reheating.

The difference between the two runs is clearest in the central \(100\) pc PDFs. 
In the low-turbulence run (top row), the inner region shows a broad multiphase distribution at \(\tau=1\) and \(\tau=4\), becomes hot-dominated at \(\tau=8\) as the cold gas is reduced and shifts to lower densities, and is repopulated by molecular gas by \(\tau=16\).
This sequence traces the feeding and feedback cycle: cold gas first reaches the centre, feedback then clears or heats the nucleus, and the inner region is later refilled by newly delivered condensed material. In the high-turbulence run, by contrast, the central \(100\) pc is strongly multiphase at around \(\tau=4\), while the PDFs at \(\tau=1\) and \(\tau=16\) are more strongly dominated by the hot phases. At \(\tau=8\), instead, we see an example of what we mean by sunny weather, in which the central region is completely devoid of cold gas. In general, in the high-turbulence case, the cold-gas residence time and volume filling factor in the nuclear region are reduced.
This lowers the mean sink accretion rate and limits the ability of the jet to sustain a long-lived multiphase nuclear weather cycle. By contrast, the low-turbulence run allows cold structures to remain more coherently connected to the sink, producing stronger accretion episodes and a more effective jet response. This behaviour is consistent with the torque diagnostics in Figure~\ref{torque_turb}. 
The hot-dominated PDFs within the central \(100\,{\rm pc}\) (micro-meso scale) during sunny phases indicate a temporary interruption of central delivery, not a global shutdown of condensation. The zoom-out density slices and outer-shell PDFs in Appendix~\ref{app:B} show that cold gas can persist at larger radii, remain fragmented or weakly connected to the sink, and later re-enter the inner accretion cycle.

This behaviour provides a useful thermodynamic link with both B26a and C26a. Relative to the no-jet B26a runs, where turbulence alone produces a sharper contrast between extended stormy rain and compact rainy condensation, the jet partially washes out the difference by redistributing low-entropy gas through uplift, compression, and interface mixing. This is especially visible at meso and inner-macro scales, where jet-regulated condensation can maintain warm/cold material even when the central \(100\,{\rm pc}\) is temporarily depleted. Relative to the fixed-axis C26a runs, the spin-coupled models retain the same basic jet-regulated CCA pathway, but the nuclear phase occupancy becomes more intermittent: the centre can enter genuinely hot-dominated \sunny\ intervals, while cold gas persists at larger radii and later reconnects to the sink. Thus, spin coupling does not appear to change the existence of the CCA condensation cascade; it mainly modifies how feedback geometry regulates the residence time, clearing, and reconnection of cold gas in the nuclear region.

\section{Discussion} \label{s:disc}
\subsection{Comparison with other work} \label{s:comparison}
SMBH spin evolution has been studied with semi-analytic models, cosmological simulations, sub-grid accretion-disc prescriptions, and horizon-scale GRMHD calculations. These approaches have established a broad physical picture: coherent accretion tends to spin black holes up, while mergers and randomly oriented accretion episodes can reduce the mean spin and increase the scatter \citep[e.g.][]{volonteri2005,king2006,king2008,perego2009,dotti2013}. Recent cosmological implementations have extended this framework by including spin-dependent feedback and disc-mediated angular-momentum transfer. For example, \citet{bustamante2019} implemented SMBH spin evolution in \texttt{AREPO}, while \citet{sala2024} studied spin evolution with \texttt{OPENGADGET3}. Related cosmological studies based on \texttt{Horizon-AGN}, \texttt{NewHorizon}, and \texttt{Galactica} further connected the spin magnitude and orientation to the assembly history and the angular momentum of the host galaxy \citep{beckmann2024,beckmann2025,peirani2024}. These works provide the large-scale demographic context for our study, but their cosmological resolution necessarily requires sub-grid prescriptions for the final delivery of mass and angular momentum to the SMBH.

A second set of studies has focused more directly on the unresolved accretion flow around the black hole. \citet{dubois2014spin} showed that turbulent ISM angular momentum can randomize SMBH spin evolution, while also suppressing accretion when cold dense gas is inefficiently supplied. \citet{fiacconi2018} introduced a moving-mesh \texttt{AREPO} model in which the SMBH is coupled to a sub-grid thin $\alpha$-disc and evolves through accretion and Bardeen-Petterson torques. \citet{talbot2021,talbot2022} coupled such disc-mediated spin evolution with spin-driven Blandford-Znajek jets, allowing the jet power and direction to respond to the evolving BH-disc system. More recently, \citet{koudmani2024} and \citet{husko2026} emphasized that the predicted spin evolution depends sensitively on the assumed accretion-disc state, including transitions between radiatively efficient discs, truncated discs, winds, and jet-dominated modes.

Our \texttt{Hybrid} model shares with these works the idea that the angular momentum delivered by the hydro solver should not be deposited directly onto the SMBH without an inner closure. This is precisely what our \texttt{Direct} control run demonstrates: using the sink-scale angular momentum without an ISCO filter strongly overestimates spin variability and jet-axis wandering. The difference is that our model is designed specifically for the chaotic cold accretion regime. In this regime, the simulation resolves the multiphase clouds and filaments that deliver rapidly varying torque directions down to the sink scale. We therefore use the resolved inflow to set the instantaneous torque direction, while the Kerr ISCO prescription sets the angular-momentum magnitude transferred to the SMBH. This does not replace a full viscous-disc model, but it preserves the time-dependent three-dimensional torque information that is central to CCA.

Our results also connect to simulations of jet reorientation in group and cluster atmospheres. Prescribed reorienting or precessing jets can distribute energy over a wider solid angle, inflate multiple detached cavities, and produce curved or misaligned radio structures \citep[e.g.][]{cielo2018reorienting,horton2020}. In those works, however, the jet direction is imposed through a chosen precession law or reorientation pattern. In our simulations, the jet axis follows the evolving SMBH spin vector, which is itself driven by the stochastic angular-momentum delivery of the CCA flow. The feedback geometry therefore emerges from the coupled feeding-spin-feedback cycle rather than from an externally prescribed jet motion.

At smaller scales, GRMHD simulations provide the relativistic basis for the spin-dependent feedback channel. Magnetically arrested discs can launch highly efficient Blandford-Znajek outflows when magnetic flux accumulates near rapidly spinning black holes, with total efficiencies that can exceed the rest-mass energy flux \citep[e.g.][]{tchekhovskoy2011}. Tilted-disc simulations further show that misaligned accretion can produce precessing jets and complex disc-spin-jet coupling, with the magnetic flux regulating how strongly the inner disc and the jet align with the BH spin \citep[e.g.][]{liska2018}. These calculations resolve the horizon-scale plasma physics that our group-scale \texttt{AthenaPK} simulations cannot capture. In contrast, they cannot yet follow the condensation, fragmentation, turbulent stirring, and feedback-regulated recycling of gas from tens of picoseconds down to the central parsec. Our approach is therefore best viewed as a meso-to-micro closure: the large-scale simulation supplies the CCA torque direction, and the ISCO closure supplies the unresolved relativistic angular-momentum transfer.

The distinctive contribution of the present work is to place the evolution of the SMBH spin inside the CCA weather cycle. Previous CCA studies showed that cold clouds and filaments condensing out of a turbulent hot halo collide, mix, and cancel the angular momentum, driving strong AGN variability. Here we extend that picture from mass accretion to spin evolution: the coherence, sign, and three-dimensional orientation of the delivered angular momentum determine whether the SMBH is spun down and reoriented, or spun up and stabilized. 

Finally, we point out here that the exact normalization of the spin-down rate and the fastest reorientation episodes remain subject to the fixed magnetic-flux prescription and inner-disc alignment physics, to be addressed in future work.\\

\subsection{Spin-regulated BH weather states}
\label{sec:weather_cycling}

The above results place the validated \textit{Hybrid} spin model within the broader \textsc{BlackHoleWeather} cycle framework (\citealt{gaspari2020}, B26a,b, C26a,b). Here we use the weather language to interpret how the spin-coupled turbulent runs move between different feeding states; the in-depth analysis of these weather states, including inflow continuity, torque coherence, \(j_\parallel\), \(j_\perp\), accretion variability, projected CCA diagnostics, and the schematic weather diagram, is developed in the companion paper P26b (which also includes more extended turbulent/CCA runs). 

A weather state is a scale-dependent configuration of the hot atmosphere and its multiphase condensate, defined by where cold/warm gas forms, how efficiently it remains connected to the sink, and, in the spin-coupled problem, how coherently it delivers angular momentum to the SMBH.
We find four phenomenological weather states. A \emph{rainy} state is tied to centrally connected precipitation, in which cold/molecular gas reaches the micro scale, sustains sink feeding, and can deliver coherent angular momentum. A \emph{stormy} state is extended, filamentary, and burst dominated, with a broad hot--warm--cold bridge over meso-to-macro radii and rapidly varying torque delivery. A \emph{cloudy} state contains fragmented or jet-processed multiphase gas, but with inefficient or intermittent nuclear delivery. A \emph{sunny} state is a hot-dominated or feedback-cleared central configuration, where the cold-gas filling factor is low and the sink accretion rate approaches a weak hot-mode baseline. These states are scale dependent and can overlap: the central \(\sim100\,{\rm pc}\) may be sunny while the inner kpc remains cloudy or stormy.

The turbulent \textit{Hybrid} runs show that the spin response is controlled by where the system sits in this weather cycle. Both low- and high-turbulence models form multiphase gas, but the low-turbulence run preserves longer rainy connections between the cold reservoir and the sink. These connected intervals sustain stronger sink feeding, stronger secular spin-down, and a larger cumulative reorientation. The high-turbulence run is more disturbed, but its inflow is also more fragmented and cancellation-prone: cold gas survives at larger radii, yet its delivery to the SMBH is less coherent, so the long-term spin response is weaker. Stronger turbulence therefore does not automatically imply stronger secular spin wandering; it can instead stabilize the SMBH spin by shortening the coherence time of the feeding channel.

The companion papers isolate successive layers of the same {\sc BlackHoleWeather} cycle. B26a establish the turbulence-controlled feeding baseline, showing how cold rain becomes either compact and centrally retained or extended and stormy. C26a add fixed-axis mechanical feedback, showing that jets reshape the multiphase bridge through compression, entrainment, and mixing, without globally suppressing condensation. The present work adds the spin-coupled layer: the same weather state also regulates the direction and coherence of the torque delivered to the SMBH, while the evolving spin returns this information to the halo through the orientation and power of subsequent jet episodes.

The weather cycle therefore becomes a vector feedback loop. Rainy states favour coherent reservoir-fed torque delivery; stormy states can generate bursty misaligned torques but also enhanced cancellation; cloudy states weaken central delivery; and sunny states correspond to torque-starved hot-mode intervals. The spin vector thus stores a finite memory of recent CCA feeding and feeds it back into the next jet direction. P26b quantifies this vector cycle in time and phase space, connecting it to inflow continuity, torque coherence, variability, and observable multiphase kinematics.

\section{Conclusions}
\label{s:conc}

In this work, we introduced and tested a time-dependent SMBH spin framework for chaotic cold accretion (CCA) environments in a GPU-accelerated AMR code. The simulations resolve the multiphase inflow and its three-dimensional angular momentum down to the sub-Bondi, parsec-scale sink region, while the unresolved final transfer to the SMBH is handled through a relativistic ISCO closure. We first used a no-turbulence benchmark suite to isolate the role of the spin prescription, and then applied our fiducial \texttt{Hybrid} model to low- and high-turbulence CCA runs to determine how weather-regulated feeding controls spin evolution and jet-axis reorientation. Our main conclusions are as follows.
\begin{itemize}
\item The large-scale reservoir of cold gas is nearly independent of whether the jet is fixed, spin-coupled, or rapidly reorienting over the simulated time span. In the no-turbulence suite, the \texttt{Benchmark}, \texttt{Direct}, and \texttt{Hybrid} runs develop almost identical cold gas masses and similar nuclear phase-PDF trajectories. The spin prescription instead controls how the condensed gas couples to the SMBH, how the jet responds, and how feedback is geometrically redistributed.\\

\vspace{-0.1cm}
\item A relativistic ISCO closure is required for physically plausible spin evolution. The \texttt{Hybrid} model remains bracketed by analytic prograde and retrograde limits for the same SMBH mass-growth history. By contrast, the \texttt{Direct} model strongly overestimates spin variability and jet-axis wandering by depositing sink-scale angular momentum without filtering its magnitude through unresolved circularization and ISCO physics. The resolved torque direction is valuable, but the transferred angular-momentum magnitude must be regularized by inner relativistic accretion physics.\\

\vspace{-0.1cm}
\item Spin modulates the inner feeding-feedback loop. Although the global condensation pathway is nearly unchanged across the no-turbulence models, the central accretion rate, jet efficiency, jet power, and feedback geometry differ substantially. Spin-dependent jets therefore change where and how energy is deposited into the inner atmosphere, even when the total condensed gas reservoir is almost unchanged.\\

\vspace{-0.1cm}
\item Low-spin SMBHs are easier to reorient. In the no-turbulence suite, the low-spin \texttt{Hybrid} run reaches inclination changes of order $10$--$20^\circ$, whereas the higher-spin run remains much more stable. This follows from angular-momentum inertia: at lower spin, a given misaligned CCA torque tilts the SMBH spin vector more efficiently. The resulting geometric reorientation rates, typically $\sim0.1$ to a few deg Myr$^{-1}$ with short spikes during rapid spin-down episodes, should be interpreted as cadence-dependent jet-axis wandering diagnostics rather than strictly periodic precession frequencies.\\

\vspace{-0.1cm}
\item In turbulent CCA, torque coherence and central connectivity matter more than turbulence amplitude alone. The low-turbulence run preserves longer rainy connections between the cold reservoir and the sink, producing stronger secular spin-down and a larger cumulative inclination excursion. The high-turbulence run is more disturbed, but its inflow is more fragmented and cancellation-dominated, shortening the coherence time of the delivered angular momentum. Stronger turbulence can therefore stabilize long-term spin evolution by disrupting coherent angular-momentum channels, even while multiphase gas survives at larger radii.

\end{itemize}

Taken together, these results show that the \textsc{BlackHoleWeather} cycle becomes a vector feedback loop once SMBH spin is evolved self-consistently. In the fixed-jet BHW runs, the weather state mainly regulates when feedback is triggered and how much power is injected. In the spin-coupled runs, the same weather state also regulates the direction and coherence of the torque delivered to the SMBH, and therefore where the next jet points.
Rainy states favour coherent reservoir-fed torque delivery; stormy states can generate strong instantaneous misaligned torques but also enhanced cancellation; cloudy states retain multiphase gas while weakening central delivery; and sunny states correspond to torque-starved hot-mode intervals. The spin vector therefore stores a finite memory of past CCA feeding episodes and returns that memory to the halo through the orientation and power of subsequent jet activity.
The main implication is that CCA is not only a stochastic fuel supply, but also a chaotic torque engine. The same multiphase weather cycle that regulates SMBH feeding also controls how the hole is spun, how the jet is redirected, and how feedback anisotropy is imprinted onto the surrounding atmosphere. The companion paper P26b quantifies this vector weather cycle in time and phase space, connecting it to inflow continuity, torque coherence, accretion variability, and observable multiphase kinematics.\\

\begin{acknowledgements}
    The {\sc BlackHoleWeather} authors acknowledge key funding support from the European Research Council (ERC) under the European Union's Horizon Europe research and innovation programme (Consolidator Grant BlackHoleWeather, No.~101086804; PI: Gaspari). Views and opinions expressed are, however, those of the author(s) only and do not necessarily reflect those of the European Union or the European Research Council Executive Agency; neither the European Union nor the granting authority can be held responsible for them.
The authors acknowledge ISCRA for awarding this project access to the LEONARDO supercomputer, owned by the EuroHPC Joint Undertaking, hosted by CINECA (Italy).
We acknowledge EuroHPC Joint Undertaking for awarding us access to Jupiter at FZJ, Germany.
VO acknowledges support from the DICYT ESO-Chile Comite Mixto PS 1757, Fondecyt Regular 1251702, and CIRAS-AI Project, code FIUF137139-USACH.
FMM carried out part of the research activities described in this paper with contribution of the Next Generation EU funds within the National Recovery and Resilience Plan (PNRR), Mission 4 - Education and Research, Component 2 - From Research to Business (M4C2), Investment Line 3.1 - Strengthening and creation of Research Infrastructures, Project IR0000034 – “STILES - Strengthening the Italian Leadership in ELT and SKA.
PT acknowledges support from NASA NNH22ZDA001N Astrophysics Data and Analysis Program under award 24-ADAP24-0011.
The authors thank Hung-Yi Pu, Roberto Serafinelli, Silvano Molendi, Amirnezam Amiri, and Fabrizio Fiore for comments and discussions. The authors thank Philipp Grete for support with the \texttt{AthenaPK} code.
We thank the organizers and participants of the following conferences for the stimulating discussions that helped improve this work: `BlackHoleWeather I' (Sexten, ITA) and `Massive Black Hole Spin' (Edinburgh, UK).

\end{acknowledgements}
%
%
\bibliography{lib}{}

@string{jan="January"}

@string{feb="February"}

@string{mar="March"}

@string{apr="April"}

@string{may="May"}

@string{jun="June"}

@string{jul="July"}

@string{aug="August"}

@string{sep="September"}

@string{oct="October"}

@string{nov="November"}

@string{dec="December"}

@ARTICLE{schmidt2009,
  author  = {{Schmidt}, W. and {Federrath}, C. and {Hupp}, M. and {Kern}, S. and {Niemeyer}, J.~C.},
  title   = {{Numerical simulations of compressively driven interstellar turbulence. I. Isothermal gas}},
  journal = {Astronomy \& Astrophysics},
  year    = {2009},
  volume  = {494},
  number  = {1},
  pages   = {127--145},
  doi     = {10.1051/0004-6361:200809967},
  eprint  = {0809.1321},
  archivePrefix = {arXiv},
  primaryClass  = {astro-ph}
}

@ARTICLE{gaspari2013_turb,
       author = {{Gaspari}, M. and {Churazov}, E.},
        title = "{Constraining turbulence and conduction in the hot ICM through density perturbations}",
      journal = {\aap},
         year = 2013,
        month = nov,
       volume = {559},
          eid = {A78},
        pages = {A78},
          doi = {10.1051/0004-6361/201322295},
archivePrefix = {arXiv},
       eprint = {1307.4397},
 primaryClass = {astro-ph.CO},
       adsurl = {https://ui.adsabs.harvard.edu/abs/2013A&A...559A..78G}
}

@ARTICLE{grete2018,
  author  = {{Grete}, P. and {O'Shea}, B.~W. and {Beckwith}, K.},
  title   = {{As a Matter of Force---Systematic Biases in Idealized Turbulence Simulations}},
  journal = {The Astrophysical Journal Letters},
  year    = {2018},
  volume  = {858},
  number  = {2},
  pages   = {L19},
  doi     = {10.3847/2041-8213/aac0f5},
  eprint  = {1803.05481},
  archivePrefix = {arXiv},
  primaryClass  = {physics.flu-dyn}
}

@ARTICLE{cielo2018reorienting,
  author  = {{Cielo}, S. and {Babul}, A. and {Antonuccio-Delogu}, V. and {Silk}, J. and {Volonteri}, M.},
  title   = {Feedback from reorienting AGN jets. I. Jet-ICM coupling, cavity properties and global energetics},
  journal = {Astronomy \& Astrophysics},
  year    = {2018},
  volume  = {617},
  pages   = {A58},
  doi     = {10.1051/0004-6361/201832582}
}

@ARTICLE{fiacconi2018,
  author  = {{Fiacconi}, D. and {Sijacki}, D. and {Pringle}, J.~E.},
  title   = {Galactic nuclei evolution with spinning black holes: method and implementation},
  journal = {Monthly Notices of the Royal Astronomical Society},
  year    = {2018},
  volume  = {477},
  pages   = {3807--3835},
  doi     = {10.1093/mnras/sty893}
}

@ARTICLE{dubois2014spin,
  author  = {{Dubois}, Y. and {Volonteri}, M. and {Silk}, J. and {Devriendt}, J. and {Slyz}, A.},
  title   = {Black hole evolution -- II. Spinning black holes in a supernova-driven turbulent interstellar medium},
  journal = {Monthly Notices of the Royal Astronomical Society},
  year    = {2014},
  volume  = {440},
  pages   = {2333--2346},
  doi     = {10.1093/mnras/stu425}
}

@ARTICLE{sala2024,
  author  = {{Sala}, L. and {Valentini}, M. and {Biffi}, V. and {Dolag}, K.},
  title   = {Supermassive black hole spin evolution in cosmological simulations with OPENGADGET3},
  journal = {Astronomy \& Astrophysics},
  year    = {2024},
  volume  = {685},
  pages   = {A92},
  doi     = {10.1051/0004-6361/202348925}
}

@ARTICLE{beckmann2025,
  author  = {{Beckmann}, R.~S. and {Dubois}, Y. and {Volonteri}, M. and {Dong-Paez}, C.~A. and {Peirani}, S. and {Piotrowska}, J.~M. and {Martin}, G. and {Kraljic}, K. and {Devriendt}, J. and {Pichon}, C. and {Yi}, S.~K.},
  title   = {Black hole spin evolution across cosmic time from the NewHorizon simulation},
  journal = {Monthly Notices of the Royal Astronomical Society},
  year    = {2025},
  volume  = {536},
  pages   = {1838--1856},
  doi     = {10.1093/mnras/stae2595}
}

@ARTICLE{peirani2024,
  author  = {{Peirani}, S. and {Suto}, Y. and {Beckmann}, R.~S. and {Volonteri}, M. and {Lin}, Y.-T. and {Dubois}, Y. and {Yi}, S.~K. and {Pichon}, C. and {Kraljic}, K. and {Park}, M. and {Devriendt}, J. and {Han}, S. and {Chen}, W.-H.},
  title   = {Cosmic evolution of black hole spin and galaxy orientations: Clues from the NewHorizon and Galactica simulations},
  journal = {Astronomy \& Astrophysics},
  year    = {2024},
  volume  = {686},
  pages   = {A233},
  doi     = {10.1051/0004-6361/202349101}
}

@ARTICLE{koudmani2024,
  author  = {{Koudmani}, S. and {Somerville}, R.~S. and {Sijacki}, D. and {Bourne}, M.~A. and {Jiang}, Y.-F. and {Profit}, K.},
  title   = {A unified accretion disc model for supermassive black holes in galaxy formation simulations: method and implementation},
  journal = {Monthly Notices of the Royal Astronomical Society},
  year    = {2024},
  volume  = {532},
  pages   = {60--91},
  doi     = {10.1093/mnras/stae1422}
}

@ARTICLE{husko2026,
  author  = {{Hu{\v{s}}ko}, F. and {Lacey}, C.~G. and {Schaye}, J. and {Schaller}, M. and {Chaikin}, E. and {Ploeckinger}, S. and {Ben{\'i}tez-Llambay}, A. and {Richings}, A.~J. and {Trayford}, J.~W.},
  title   = {A hybrid active galactic nucleus feedback model with spinning black holes, winds and jets},
  journal = {Monthly Notices of the Royal Astronomical Society},
  year    = {2026},
  volume  = {547},
  pages   = {stag324},
  doi     = {10.1093/mnras/stag324}
}

@ARTICLE{liska2018,
  author  = {{Liska}, M. and {Tchekhovskoy}, A. and {Ingram}, A. and {van der Klis}, M.},
  title   = {Formation of precessing jets by tilted black hole discs in 3D general relativistic MHD simulations},
  journal = {Monthly Notices of the Royal Astronomical Society: Letters},
  year    = {2018},
  volume  = {474},
  pages   = {L81--L85},
  doi     = {10.1093/mnrasl/slx174}
}

@article{Bardeen1970,
  author  = {Bardeen, J. M.},
  title   = {Kerr Metric Black Holes},
  journal = {Nature},
  year    = {1970},
  volume  = {226},
  number  = {5240},
  pages   = {64--65},
  doi     = {10.1038/226064a0}
}

@article{perego2009,
  author  = {Perego, A. and Dotti, M. and Colpi, M. and Volonteri, M.},
  year    = {2009},
  title   = {Mass and spin co-evolution during the alignment of a black hole in a warped accretion disc},
  journal = {MNRAS},
  volume  = {399},
  number  = {4},
  pages   = {2249--2263},
  doi     = {10.1111/j.1365-2966.2009.15427.x}
}

@article{dotti2013,
  author  = {Dotti, M. and Colpi, M. and Pallini, S. and Perego, A. and Volonteri, M.},
  year    = {2013},
  title   = {On the Orientation and Magnitude of the Black Hole Spin in Galactic Nuclei},
  journal = {ApJ},
  volume  = {762},
  number  = {2},
  pages   = {68},
  doi     = {10.1088/0004-637X/762/2/68}
}

@article{reynolds2021,
  author  = {Reynolds, Christopher S.},
  year    = {2021},
  title   = {Observational Constraints on Black Hole Spin},
  journal = {ARA\&A},
  volume  = {59},
  pages   = {117}
}

@article{volonteri2005,
  author  = {Volonteri, Marta and Madau, Piero and Quataert, Eliot and Rees, Martin J.},
  year    = {2005},
  title   = {The Distribution and Cosmic Evolution of Massive Black Hole Spins},
  journal = {ApJ},
  volume  = {620},
  pages   = {69}
}

@article{McDonald2018,
  author  = {McDonald, M. and Gaspari, M. and McNamara, B. R. and Tremblay, G. R.},
  year    = {2018},
  title   = {Revisiting the Cooling Flow Problem in Galaxies, Groups, and Clusters of Galaxies},
  journal = {ApJ},
  volume  = {858},
  number  = {1},
  pages   = {45},
  doi     = {10.3847/1538-4357/aabace}
}

@article{Tremblay2018,
  author  = {Tremblay, G. R. and Combes, F. and Oonk, J. B. R. and Russell, H. R. and McDonald, M. A. and Gaspari, M. and Salome, P. and Hamer, S. and Mathews, W. G. and Fabian, A. C. and others},
  year    = {2018},
  title   = {A Galaxy-scale Fountain of Cold Molecular Gas Pumped by a Black Hole},
  journal = {ApJ},
  volume  = {865},
  number  = {1},
  pages   = {13},
  doi     = {10.3847/1538-4357/aad6dd}
}

@article{Maccagni2021,
  author  = {Maccagni, F. M. and Serra, P. and Gaspari, M. and Kleiner, D. and Morokuma-Matsui, K. and Oosterloo, T. A. and Onodera, M. and Kamphuis, P. and Loi, F. and Thorat, K. and Ramatsoku, M. and Smirnov, O. M. and White, S. V.},
  year    = {2021},
  title   = {AGN feeding and feedback in Fornax A. Kinematical analysis of the multi-phase ISM},
  journal = {A\&A},
  volume  = {656},
  pages   = {A45},
  doi     = {10.1051/0004-6361/202141143}
}

@article{Temi2022,
  author  = {Temi, P. and Gaspari, M. and Brighenti, F. and Werner, N. and Grossov{\'a}, R. and Gitti, M. and Sun, M. and Amblard, A. and Simionescu, A.},
  year    = {2022},
  title   = {Probing Multiphase Gas in Local Massive Elliptical Galaxies via Multiwavelength Observations},
  journal = {ApJ},
  volume  = {928},
  number  = {2},
  pages   = {150},
  doi     = {10.3847/1538-4357/ac5036}
}

@article{Olivares2022,
  author  = {Olivares, V. and Salom{\'e}, P. and Hamer, S. L. and Combes, F. and Gaspari, M. and Kolokythas, K. and O'Sullivan, E. and Beckmann, R. S. and Babul, A. and Polles, F. L. and Lehnert, M. and Loubser, S. I. and Donahue, M. and Gendron-Marsolais, M.-L. and Lagos, P. and Pineau des For{\^e}ts, G. and Godard, B. and Rose, T. and Tremblay, G. R. and Ferland, G. and Guillard, P.},
  year    = {2022},
  title   = {Gas condensation in brightest group galaxies unveiled with MUSE. Morphology and kinematics of the ionized gas},
  journal = {A\&A},
  volume  = {666},
  pages   = {A94},
  doi     = {10.1051/0004-6361/202142475}
}

@article{Reefe2025,
  author  = {Reefe, M. and McDonald, M. and Chatzikos, M. and Seebeck, J. and Mushotzky, R. and Veilleux, S. and Allen, S. W. and Bayliss, M. and Calzadilla, M. and Canning, R. and Floyd, B. and Gaspari, M. and Hlavacek-Larrondo, J. and McNamara, B. and Russell, H. and Sharon, K. and Somboonpanyakul, T.},
  year    = {2025},
  title   = {Directly imaging the cooling flow in the Phoenix cluster},
  journal = {Nature},
  volume  = {638},
  pages   = {360--364},
  doi     = {10.1038/s41586-024-08369-x}
}

@article{Olivares2025,
  author  = {Olivares, V. and Picquenot, A. and Su, Y. and Gaspari, M. and Gendron-Marsolais, M.-L. and Polles, F. L. and Nulsen, P.},
  year    = {2025},
  title   = {An H$\alpha$--X-ray surface-brightness correlation for filaments in cooling-flow clusters},
  journal = {Nature Astronomy},
  volume  = {9},
  number  = {3},
  pages   = {449--457},
  doi     = {10.1038/s41550-024-02473-8}
}

@article{Romero2025,
  author  = {Romero, C. E. and Gaspari, M. and Schellenberger, G. and Benson, B. A. and Bleem, L. E. and Bulbul, E. and Forman, W. R. and Kraft, R. P. and Nulsen, P. and Reichardt, C. L. and Sarkar, A. and Somboonpanyakul, T. and Su, Y.},
  year    = {2025},
  title   = {SZ-X-Ray Surface Brightness Fluctuations in the SPT-XMM Clusters},
  journal = {ApJ},
  volume  = {985},
  number  = {2},
  pages   = {248},
  doi     = {10.3847/1538-4357/adcd74}
}

@article{Omoruyi2026,
  author  = {Omoruyi, O. and Tremblay, G. and Baum, S. A. and Clarke, T. E. and Dabhade, P. and Fabian, A. C. and Gaspari, M. and Gulati, S. and Kharb, P. and Markevitch, M. and Nulsen, P. E. J. and O'Dea, C. P. and Randall, S. and Raychaudhury, S. and Vaddi, S. and Vikhlinin, A. and ZuHone, J.},
  year    = {2026},
  title   = {A Deep Chandra View of A2597: Bubbles, Shocks, Cold Fueling, and a Plasma Depletion Layer},
  journal = {ApJ},
  volume  = {997},
  number  = {1},
  pages   = {114},
  doi     = {10.3847/1538-4357/ae2006}
}

@article{misra2025,
  author  = {Misra, Arpita and Jamrozy, Marek and We{\.z}gowiec, Marek and Kozie{\l}-Wierzbowska, Dorota},
  year    = {2025},
  title   = {Multiwavelength investigations of PKS 2300-18: S-shaped radio quasar with precessing jets and double-peaked broad emission-line spectrum},
  journal = {Monthly Notices of the Royal Astronomical Society},
  volume  = {536},
  number  = {3},
  pages   = {2025},
}

@INPROCEEDINGS{aalto2015,
       author = {{Aalto}, S.},
        title = "{Galaxies and Galaxy Nuclei: From Hot Cores to Cold Outflows}",
    booktitle = {Revolution in Astronomy with ALMA: The Third Year},
         year = 2015,
       editor = {{Iono}, D. and {Tatematsu}, K. and {Wootten}, A. and {Testi}, L.},
       series = {Astronomical Society of the Pacific Conference Series},
       volume = {499},
        month = dec,
        pages = {85},
       adsurl = {https://ui.adsabs.harvard.edu/abs/2015ASPC..499...85A}
}

@ARTICLE{barbani2023,
       author = {{Barbani}, Filippo and {Pascale}, Raffaele and {Marinacci}, Federico and {Sales}, Laura V. and {Vogelsberger}, Mark and {Torrey}, Paul and {Li}, Hui},
        title = "{Galactic coronae in Milky Way-like galaxies: the role of stellar feedback in gas accretion}",
      journal = {\mnras},
         year = 2023,
        month = sep,
       volume = {524},
       number = {3},
        pages = {4091-4108},
          doi = {10.1093/mnras/stad2152},
archivePrefix = {arXiv},
       eprint = {2306.11791},
 primaryClass = {astro-ph.GA},
       adsurl = {https://ui.adsabs.harvard.edu/abs/2023MNRAS.524.4091B}
}

@ARTICLE{barbani2025,
       author = {{Barbani}, Filippo and {Pascale}, Raffaele and {Marinacci}, Federico and {Torrey}, Paul and {Sales}, Laura V. and {Li}, Hui and {Vogelsberger}, Mark},
        title = "{Understanding the baryon cycle: Fueling star formation via inflows in Milky Way-like galaxies}",
      journal = {\aap},
         year = 2025,
        month = may,
       volume = {697},
          eid = {A121},
        pages = {A121},
          doi = {10.1051/0004-6361/202452608},
archivePrefix = {arXiv},
       eprint = {2504.01075},
 primaryClass = {astro-ph.GA},
       adsurl = {https://ui.adsabs.harvard.edu/abs/2025A&A...697A.121B}
}

@ARTICLE{barbani2026a,
author = {{Barbani}, Filippo and {Gaspari}, Massimo and {Cammelli}, Vieri and et al.},
title = "{BlackHoleWeather – Chaotic cold accretion across the meso scale: Morphology and thermodynamics}",
journal = {\aap},
year = 2026,
volume = {Submitted},
number = {},
eid = {},
pages = {},
doi = {},
adsurl = {}
}

@ARTICLE{barbani2026b,
author = {{Barbani}, Filippo and {Gaspari}, Massimo and {Piana}, Olmo and et al.},
title = "{BlackHoleWeather – Chaotic cold accretion across the meso scale: Variability and kinematics}",
journal = {\aap},
year = 2026,
volume = {Submitted},
number = {},
eid = {},
pages = {},
doi = {},
adsurl = {}
}

@ARTICLE{bardeen1972,
       author = {{Bardeen}, James M. and {Press}, William H. and {Teukolsky}, Saul A.},
        title = "{Rotating Black Holes: Locally Nonrotating Frames, Energy Extraction, and Scalar Synchrotron Radiation}",
      journal = {\apj},
         year = 1972,
        month = dec,
       volume = {178},
        pages = {347-370},
          doi = {10.1086/151796},
       adsurl = {https://ui.adsabs.harvard.edu/abs/1972ApJ...178..347B}
}

@ARTICLE{bardeen1973,
       author = {{Bardeen}, J.~M. and {Carter}, B. and {Hawking}, S.~W.},
        title = "{The four laws of black hole mechanics}",
      journal = {Communications in Mathematical Physics},
         year = 1973,
        month = jun,
       volume = {31},
       number = {2},
        pages = {161-170},
          doi = {10.1007/BF01645742},
       adsurl = {https://ui.adsabs.harvard.edu/abs/1973CMaPh..31..161B}
}

@ARTICLE{bardeen1975,
       author = {{Bardeen}, James M. and {Petterson}, Jacobus A.},
        title = "{The Lense-Thirring Effect and Accretion Disks around Kerr Black Holes}",
      journal = {\apjl},
         year = 1975,
        month = jan,
       volume = {195},
        pages = {L65},
          doi = {10.1086/181711},
       adsurl = {https://ui.adsabs.harvard.edu/abs/1975ApJ...195L..65B}
}

@ARTICLE{beckmann2019,
       author = {{Beckmann}, R.~S. and {Dubois}, Y. and {Guillard}, P. and {Salome}, P. and {Olivares}, V. and {Polles}, F. and {Cadiou}, C. and {Combes}, F. and {Hamer}, S. and {Lehnert}, M.~D. and {Pineau des Forets}, G.},
        title = "{Dense gas formation and destruction in a simulated Perseus-like galaxy cluster with spin-driven black hole feedback}",
      journal = {\aap},
         year = 2019,
        month = nov,
       volume = {631},
          eid = {A60},
        pages = {A60},
          doi = {10.1051/0004-6361/201936188},
archivePrefix = {arXiv},
       eprint = {1909.01329},
 primaryClass = {astro-ph.GA},
       adsurl = {https://ui.adsabs.harvard.edu/abs/2019A&A...631A..60B}
}

@ARTICLE{beckmann2024,
       author = {{Beckmann}, R.~S. and {Smethurst}, R.~J. and {Simmons}, B.~D. and {Coil}, A. and {Dubois}, Y. and {Garland}, I.~L. and {Lintott}, C.~J. and {Martin}, G. and {Peirani}, S. and {Pichon}, C.},
        title = "{Supermassive black holes in merger-free galaxies have higher spins which are preferentially aligned with their host galaxy}",
      journal = {\mnras},
         year = 2024,
        month = feb,
       volume = {527},
       number = {4},
        pages = {10867-10877},
          doi = {10.1093/mnras/stad1795},
       adsurl = {https://ui.adsabs.harvard.edu/abs/2024MNRAS.52710867B}
}

@ARTICLE{blandford1977,
       author = {{Blandford}, R.~D. and {Znajek}, R.~L.},
        title = "{Electromagnetic extraction of energy from Kerr black holes.}",
      journal = {\mnras},
         year = 1977,
        month = may,
       volume = {179},
        pages = {433-456},
          doi = {10.1093/mnras/179.3.433},
       adsurl = {https://ui.adsabs.harvard.edu/abs/1977MNRAS.179..433B}
}

@BOOK{brenneman2013,
       author = {{Brenneman}, Laura},
        title = "{Measuring the Angular Momentum of Supermassive Black Holes}",
         year = 2013,
          doi = {10.1007/978-1-4614-7771-6},
       adsurl = {https://ui.adsabs.harvard.edu/abs/2013mams.book.....B}
}

@ARTICLE{bruni2021,
       author = {{Bruni}, G. and {Brienza}, M. and {Panessa}, F. and {Bassani}, L. and {Dallacasa}, D. and {Venturi}, T. and {Baldi}, R.~D. and {Botteon}, A. and {Drabent}, A. and {Malizia}, A. and {Massaro}, F. and {R{\"o}ttgering}, H.~J.~A. and {Ubertini}, P. and {Ursini}, F. and {van Weeren}, R.~J.},
        title = "{Hard X-ray selected giant radio galaxies - III. The LOFAR view}",
      journal = {\mnras},
         year = 2021,
        month = jun,
       volume = {503},
       number = {4},
        pages = {4681-4699},
          doi = {10.1093/mnras/stab623},
archivePrefix = {arXiv},
       eprint = {2103.00999},
 primaryClass = {astro-ph.HE},
       adsurl = {https://ui.adsabs.harvard.edu/abs/2021MNRAS.503.4681B}
}

@ARTICLE{bustamante2019,
       author = {{Bustamante}, Sebastian and {Springel}, Volker},
        title = "{Spin evolution and feedback of supermassive black holes in cosmological simulations}",
      journal = {\mnras},
         year = 2019,
        month = dec,
       volume = {490},
       number = {3},
        pages = {4133-4153},
          doi = {10.1093/mnras/stz2836},
archivePrefix = {arXiv},
       eprint = {1902.04651},
 primaryClass = {astro-ph.GA},
       adsurl = {https://ui.adsabs.harvard.edu/abs/2019MNRAS.490.4133B}
}

@ARTICLE{cammelli2025,
       author = {{Cammelli}, Vieri and {Monaco}, Pierluigi and {Tan}, Jonathan C. and {Singh}, Jasbir and {Fontanot}, Fabio and {De Lucia}, Gabriella and {Hirschmann}, Michaela and {Xie}, Lizhi},
        title = "{The formation of supermassive black holes from Population III.1 seeds. III. Galaxy evolution and black hole growth from semi-analytic modelling}",
      journal = {\mnras},
         year = 2025,
        month = jan,
       volume = {536},
       number = {1},
        pages = {851-870},
          doi = {10.1093/mnras/stae2663},
archivePrefix = {arXiv},
       eprint = {2407.09949},
 primaryClass = {astro-ph.GA},
       adsurl = {https://ui.adsabs.harvard.edu/abs/2025MNRAS.536..851C}
}

@ARTICLE{cammelli2026a,
author = {{Cammelli}, Vieri and {Gaspari}, Massimo and {Piana}, Olmo and et al.},
title = "{BlackHoleWeather – Chaotic cold accretion across the meso scale: Morphology and thermodynamics}",
journal = {\aap},
year = 2026,
volume = {Submitted},
number = {},
eid = {},
pages = {},
doi = {},
adsurl = {}
}

@ARTICLE{cammelli2026b,
author = {{Cammelli}, Vieri and {Gaspari}, Massimo and {Barbani}, Filippo and et al.},
title = "{BlackHoleWeather – Chaotic cold accretion across the meso scale: Variability and kinematics}",
journal = {\aap},
year = 2026,
volume = {Submitted},
number = {},
eid = {},
pages = {},
doi = {},
adsurl = {}
}

@ARTICLE{cavagnolo2009,
       author = {{Cavagnolo}, Kenneth W. and {Donahue}, Megan and {Voit}, G. Mark and {Sun}, Ming},
        title = "{Intracluster Medium Entropy Profiles for a Chandra Archival Sample of Galaxy Clusters}",
      journal = {\apjs},
         year = 2009,
        month = may,
       volume = {182},
       number = {1},
        pages = {12-32},
          doi = {10.1088/0067-0049/182/1/12},
archivePrefix = {arXiv},
       eprint = {0902.1802},
 primaryClass = {astro-ph.CO},
       adsurl = {https://ui.adsabs.harvard.edu/abs/2009ApJS..182...12C}
}

@ARTICLE{cielo2018,
       author = {{Cielo}, Salvatore and {Bieri}, Rebekka and {Volonteri}, Marta and {Wagner}, Alexander Y. and {Dubois}, Yohan},
        title = "{AGN feedback compared: jets versus radiation}",
      journal = {\mnras},
         year = 2018,
        month = jun,
       volume = {477},
       number = {1},
        pages = {1336-1355},
          doi = {10.1093/mnras/sty708},
archivePrefix = {arXiv},
       eprint = {1712.03955},
 primaryClass = {astro-ph.GA},
       adsurl = {https://ui.adsabs.harvard.edu/abs/2018MNRAS.477.1336C}
}

@ARTICLE{danekhar2024,
       author = {{Danehkar}, A.},
        title = "{Relativistic reflection modeling in AGN and related variability from PCA: a brief review}",
      journal = {Frontiers in Astronomy and Space Sciences},
         year = 2024,
        month = oct,
       volume = {11},
          eid = {1479301},
        pages = {1479301},
          doi = {10.3389/fspas.2024.1479301},
archivePrefix = {arXiv},
       eprint = {2410.01852},
 primaryClass = {astro-ph.HE},
       adsurl = {https://ui.adsabs.harvard.edu/abs/2024FrASS..1179301D}
}

@ARTICLE{dimatteo2005,
   author = {{Di Matteo}, T. and {Springel}, V. and {Hernquist}, L.},
    title = "{Energy input from quasars regulates the growth and activity of black holes and their host galaxies}",
  journal = {\nat},
   eprint = {astro-ph/0502199},
     year = 2005,
    month = feb,
   volume = 433,
    pages = {604-607},
      doi = {10.1038/nature03335},
   adsurl = {http://adsabs.harvard.edu/abs/2005Natur.433..604D}
}

@ARTICLE{dubois2014,
   author = {{Dubois}, Y. and {Pichon}, C. and {Welker}, C. and {Le Borgne}, D. and 
	{Devriendt}, J. and {Laigle}, C. and {Codis}, S. and {Pogosyan}, D. and 
	{Arnouts}, S. and {Benabed}, K. and {Bertin}, E. and {Blaizot}, J. and 
	{Bouchet}, F. and {Cardoso}, J.-F. and {Colombi}, S. and {de Lapparent}, V. and 
	{Desjacques}, V. and {Gavazzi}, R. and {Kassin}, S. and {Kimm}, T. and 
	{McCracken}, H. and {Milliard}, B. and {Peirani}, S. and {Prunet}, S. and 
	{Rouberol}, S. and {Silk}, J. and {Slyz}, A. and {Sousbie}, T. and 
	{Teyssier}, R. and {Tresse}, L. and {Treyer}, M. and {Vibert}, D. and 
	{Volonteri}, M.},
    title = "{Dancing in the dark: galactic properties trace spin swings along the cosmic web}",
  journal = {\mnras},
archivePrefix = "arXiv",
   eprint = {1402.1165},
     year = 2014,
    month = oct,
   volume = 444,
    pages = {1453-1468},
      doi = {10.1093/mnras/stu1227},
   adsurl = {http://adsabs.harvard.edu/abs/2014MNRAS.444.1453D}
}

@article{edwards2014,
title = {Kokkos: Enabling manycore performance portability through polymorphic memory access patterns},
journal = {Journal of Parallel and Distributed Computing},
volume = {74},
number = {12},
pages = {3202-3216},
year = {2014},
note = {Domain-Specific Languages and High-Level Frameworks for High-Performance Computing},
issn = {0743-7315},
doi = {https://doi.org/10.1016/j.jpdc.2014.07.003},
url = {https://www.sciencedirect.com/science/article/pii/S0743731514001257},
author = {H. {Carter Edwards} and Christian R. Trott and Daniel Sunderland}
}

@ARTICLE{EHT2019,
       author = {{Event Horizon Telescope Collaboration} and {Akiyama}, Kazunori and {Alberdi}, Antxon and {Alef}, Walter and {Asada}, Keiichi and {Azulay}, Rebecca and {Baczko}, Anne-Kathrin and {Ball}, David and {Balokovi{\'c}}, Mislav and {Barrett}, John and {Bintley}, Dan and {Blackburn}, Lindy and {Boland}, Wilfred and {Bouman}, Katherine L. and {Bower}, Geoffrey C. and {Bremer}, Michael and {Brinkerink}, Christiaan D. and {Brissenden}, Roger and {Britzen}, Silke and {Broderick}, Avery E. and {Broguiere}, Dominique and {Bronzwaer}, Thomas and {Byun}, Do-Young and {Carlstrom}, John E. and {Chael}, Andrew and {Chan}, Chi-kwan and {Chatterjee}, Shami and {Chatterjee}, Koushik and {Chen}, Ming-Tang and {Chen}, Yongjun and {Cho}, Ilje and {Christian}, Pierre and {Conway}, John E. and {Cordes}, James M. and {Crew}, Geoffrey B. and {Cui}, Yuzhu and {Davelaar}, Jordy and {De Laurentis}, Mariafelicia and {Deane}, Roger and {Dempsey}, Jessica and {Desvignes}, Gregory and {Dexter}, Jason and {Doeleman}, Sheperd S. and {Eatough}, Ralph P. and {Falcke}, Heino and {Fish}, Vincent L. and {Fomalont}, Ed and {Fraga-Encinas}, Raquel and {Friberg}, Per and {Fromm}, Christian M. and {G{\'o}mez}, Jos{\'e} L. and {Galison}, Peter and {Gammie}, Charles F. and {Garc{\'\i}a}, Roberto and {Gentaz}, Olivier and {Georgiev}, Boris and {Goddi}, Ciriaco and {Gold}, Roman and {Gu}, Minfeng and {Gurwell}, Mark and {Hada}, Kazuhiro and {Hecht}, Michael H. and {Hesper}, Ronald and {Ho}, Luis C. and {Ho}, Paul and {Honma}, Mareki and {Huang}, Chih-Wei L. and {Huang}, Lei and {Hughes}, David H. and {Ikeda}, Shiro and {Inoue}, Makoto and {Issaoun}, Sara and {James}, David J. and {Jannuzi}, Buell T. and {Janssen}, Michael and {Jeter}, Britton and {Jiang}, Wu and {Johnson}, Michael D. and {Jorstad}, Svetlana and {Jung}, Taehyun and {Karami}, Mansour and {Karuppusamy}, Ramesh and {Kawashima}, Tomohisa and {Keating}, Garrett K. and {Kettenis}, Mark and {Kim}, Jae-Young and {Kim}, Junhan and {Kim}, Jongsoo and {Kino}, Motoki and {Koay}, Jun Yi and {Koch}, Patrick M. and {Koyama}, Shoko and {Kramer}, Michael and {Kramer}, Carsten and {Krichbaum}, Thomas P. and {Kuo}, Cheng-Yu and {Lauer}, Tod R. and {Lee}, Sang-Sung and {Li}, Yan-Rong and {Li}, Zhiyuan and {Lindqvist}, Michael and {Liu}, Kuo and {Liuzzo}, Elisabetta and {Lo}, Wen-Ping and {Lobanov}, Andrei P. and {Loinard}, Laurent and {Lonsdale}, Colin and {Lu}, Ru-Sen and {MacDonald}, Nicholas R. and {Mao}, Jirong and {Markoff}, Sera and {Marrone}, Daniel P. and {Marscher}, Alan P. and {Mart{\'\i}-Vidal}, Iv{\'a}n and {Matsushita}, Satoki and {Matthews}, Lynn D. and {Medeiros}, Lia and {Menten}, Karl M. and {Mizuno}, Yosuke and {Mizuno}, Izumi and {Moran}, James M. and {Moriyama}, Kotaro and {Moscibrodzka}, Monika and {Mul{\ensuremath{\ddot{}}}ler}, Cornelia and {Nagai}, Hiroshi and {Nagar}, Neil M. and {Nakamura}, Masanori and {Narayan}, Ramesh and {Narayanan}, Gopal and {Natarajan}, Iniyan and {Neri}, Roberto and {Ni}, Chunchong and {Noutsos}, Aristeidis and {Okino}, Hiroki and {Olivares}, H{\'e}ctor and {Oyama}, Tomoaki and {{\"O}zel}, Feryal and {Palumbo}, Daniel C.~M. and {Patel}, Nimesh and {Pen}, Ue-Li and {Pesce}, Dominic W. and {Pi{\'e}tu}, Vincent and {Plambeck}, Richard and {PopStefanija}, Aleksandar and {Porth}, Oliver and {Prather}, Ben and {Preciado-L{\'o}pez}, Jorge A. and {Psaltis}, Dimitrios and {Pu}, Hung-Yi and {Ramakrishnan}, Venkatessh and {Rao}, Ramprasad and {Rawlings}, Mark G. and {Raymond}, Alexander W. and {Rezzolla}, Luciano and {Ripperda}, Bart and {Roelofs}, Freek and {Rogers}, Alan and {Ros}, Eduardo and {Rose}, Mel and {Roshanineshat}, Arash and {Rottmann}, Helge and {Roy}, Alan L. and {Ruszczyk}, Chet and {Ryan}, Benjamin R. and {Rygl}, Kazi L.~J. and {S{\'a}nchez}, Salvador and {S{\'a}nchez-Arguelles}, David and {Sasada}, Mahito and {Savolainen}, Tuomas and {Schloerb}, F. Peter and {Schuster}, Karl-Friedrich and {Shao}, Lijing and {Shen}, Zhiqiang and {Small}, Des and {Sohn}, Bong Won and {SooHoo}, Jason and {Tazaki}, Fumie and {Tiede}, Paul and {Tilanus}, Remo P.~J. and {Titus}, Michael and {Toma}, Kenji and {Torne}, Pablo and {Trent}, Tyler and {Trippe}, Sascha and {Tsuda}, Shuichiro and {van Bemmel}, Ilse and {van Langevelde}, Huib Jan and {van Rossum}, Daniel R. and {Wagner}, Jan and {Wardle}, John and {Weintroub}, Jonathan and {Wex}, Norbert and {Wharton}, Robert and {Wielgus}, Maciek and {Wong}, George N. and {Wu}, Qingwen and {Young}, Andr{\'e} and {Young}, Ken and {Younsi}, Ziri and {Yuan}, Feng},
        title = "{First M87 Event Horizon Telescope Results. V. Physical Origin of the Asymmetric Ring}",
      journal = {\apjl},
         year = 2019,
        month = apr,
       volume = {875},
       number = {1},
          eid = {L5},
        pages = {L5},
          doi = {10.3847/2041-8213/ab0f43},
archivePrefix = {arXiv},
       eprint = {1906.11242},
 primaryClass = {astro-ph.GA},
       adsurl = {https://ui.adsabs.harvard.edu/abs/2019ApJ...875L...5E}
}

@ARTICLE{falcetagoncalves2010,
       author = {{Falceta-Gon{\c{c}}alves}, D. and {Caproni}, A. and {Abraham}, Z. and {Teixeira}, D.~M. and {de Gouveia Dal Pino}, E.~M.},
        title = "{Precessing Jets and X-ray Bubbles from NGC 1275 (3C 84) in the Perseus Galaxy Cluster: A View from Three-dimensional Numerical Simulations}",
      journal = {\apjl},
         year = 2010,
        month = apr,
       volume = {713},
       number = {1},
        pages = {L74-L78},
          doi = {10.1088/2041-8205/713/1/L74},
archivePrefix = {arXiv},
       eprint = {1003.2406},
 primaryClass = {astro-ph.CO},
       adsurl = {https://ui.adsabs.harvard.edu/abs/2010ApJ...713L..74F}
}

@ARTICLE{fournier2024,
       author = {{Fournier}, M. and {Grete}, P. and {Br{\"u}ggen}, M. and {Glines}, F.~W. and {O'Shea}, B.~W.},
        title = "{The properties of magnetised cold filaments in a cool-core galaxy cluster}",
      journal = {\aap},
         year = 2024,
        month = nov,
       volume = {691},
          eid = {A239},
        pages = {A239},
          doi = {10.1051/0004-6361/202451031},
archivePrefix = {arXiv},
       eprint = {2406.05044},
 primaryClass = {astro-ph.GA},
       adsurl = {https://ui.adsabs.harvard.edu/abs/2024A&A...691A.239F}
}

@ARTICLE{garofalo2010,
       author = {{Garofalo}, D. and {Evans}, D.~A. and {Sambruna}, R.~M.},
        title = "{The evolution of radio-loud active galactic nuclei as a function of black hole spin}",
      journal = {\mnras},
         year = 2010,
        month = aug,
       volume = {406},
       number = {2},
        pages = {975-986},
          doi = {10.1111/j.1365-2966.2010.16797.x},
archivePrefix = {arXiv},
       eprint = {1004.1166},
 primaryClass = {astro-ph.CO},
       adsurl = {https://ui.adsabs.harvard.edu/abs/2010MNRAS.406..975G}
}

@ARTICLE{gaspari2012,
       author = {{Gaspari}, M. and {Ruszkowski}, M. and {Sharma}, P.},
        title = "{Cause and Effect of Feedback: Multiphase Gas in Cluster Cores Heated by AGN Jets}",
      journal = {\apj},
         year = 2012,
        month = feb,
       volume = {746},
       number = {1},
          eid = {94},
        pages = {94},
          doi = {10.1088/0004-637X/746/1/94},
archivePrefix = {arXiv},
       eprint = {1110.6063},
 primaryClass = {astro-ph.CO},
       adsurl = {https://ui.adsabs.harvard.edu/abs/2012ApJ...746...94G}
}

@ARTICLE{gaspari2013,
       author = {{Gaspari}, M. and {Ruszkowski}, M. and {Oh}, S. Peng},
        title = "{Chaotic cold accretion on to black holes}",
      journal = {\mnras},
         year = 2013,
        month = jul,
       volume = {432},
       number = {4},
        pages = {3401-3422},
          doi = {10.1093/mnras/stt692},
archivePrefix = {arXiv},
       eprint = {1301.3130},
 primaryClass = {astro-ph.CO},
       adsurl = {https://ui.adsabs.harvard.edu/abs/2013MNRAS.432.3401G}
}

@ARTICLE{gaspari2015,
       author = {{Gaspari}, M. and {Brighenti}, F. and {Temi}, P.},
        title = "{Chaotic cold accretion on to black holes in rotating atmospheres}",
      journal = {\aap},
         year = 2015,
        month = jul,
       volume = {579},
          eid = {A62},
        pages = {A62},
          doi = {10.1051/0004-6361/201526151},
archivePrefix = {arXiv},
       eprint = {1407.7531},
 primaryClass = {astro-ph.GA},
       adsurl = {https://ui.adsabs.harvard.edu/abs/2015A&A...579A..62G}
}

@ARTICLE{gaspari2017,
       author = {{Gaspari}, M. and {Temi}, P. and {Brighenti}, F.},
        title = "{Raining on black holes and massive galaxies: the top-down multiphase condensation model}",
      journal = {\mnras},
         year = 2017,
        month = apr,
       volume = {466},
       number = {1},
        pages = {677-704},
          doi = {10.1093/mnras/stw3108},
archivePrefix = {arXiv},
       eprint = {1608.08216},
 primaryClass = {astro-ph.GA},
       adsurl = {https://ui.adsabs.harvard.edu/abs/2017MNRAS.466..677G}
}

@ARTICLE{gaspari2017b,
       author = {{Gaspari}, Massimo and {S{\k{a}}dowski}, Aleksander},
        title = "{Unifying the Micro and Macro Properties of AGN Feeding and Feedback}",
      journal = {\apj},
         year = 2017,
        month = mar,
       volume = {837},
       number = {2},
          eid = {149},
        pages = {149},
          doi = {10.3847/1538-4357/aa61a3},
archivePrefix = {arXiv},
       eprint = {1701.07030},
 primaryClass = {astro-ph.HE},
       adsurl = {https://ui.adsabs.harvard.edu/abs/2017ApJ...837..149G}
}

@ARTICLE{gaspari2020,
       author = {{Gaspari}, Massimo and {Tombesi}, Francesco and {Cappi}, Massimo},
        title = "{Linking macro-, meso- and microscales in multiphase AGN feeding and feedback}",
      journal = {Nature Astronomy},
         year = 2020,
        month = jan,
       volume = {4},
        pages = {10-13},
          doi = {10.1038/s41550-019-0970-1},
archivePrefix = {arXiv},
       eprint = {2001.04985},
 primaryClass = {astro-ph.GA},
       adsurl = {https://ui.adsabs.harvard.edu/abs/2020NatAs...4...10G}
}

@ARTICLE{gerosa2019,
       author = {{Gerosa}, Davide and {Lima}, Alicia and {Berti}, Emanuele and {Sperhake}, Ulrich and {Kesden}, Michael and {O'Shaughnessy}, Richard},
        title = "{Wide nutation: binary black-hole spins repeatedly oscillating from full alignment to full anti-alignment}",
      journal = {Classical and Quantum Gravity},
         year = 2019,
        month = may,
       volume = {36},
       number = {10},
          eid = {105003},
        pages = {105003},
          doi = {10.1088/1361-6382/ab14ae},
archivePrefix = {arXiv},
       eprint = {1811.05979},
 primaryClass = {gr-qc},
       adsurl = {https://ui.adsabs.harvard.edu/abs/2019CQGra..36j5003G}
}

@ARTICLE{greenhill2003,
       author = {{Greenhill}, L.~J. and {Booth}, R.~S. and {Ellingsen}, S.~P. and {Herrnstein}, J.~R. and {Jauncey}, D.~L. and {McCulloch}, P.~M. and {Moran}, J.~M. and {Norris}, R.~P. and {Reynolds}, J.~E. and {Tzioumis}, A.~K.},
        title = "{A Warped Accretion Disk and Wide-Angle Outflow in the Inner Parsec of the Circinus Galaxy}",
      journal = {\apj},
         year = 2003,
        month = jun,
       volume = {590},
       number = {1},
        pages = {162-173},
          doi = {10.1086/374862},
archivePrefix = {arXiv},
       eprint = {astro-ph/0302533},
 primaryClass = {astro-ph},
       adsurl = {https://ui.adsabs.harvard.edu/abs/2003ApJ...590..162G}
}

@ARTICLE{grete2023,
       author = {{Grete}, Philipp and {Dolence}, Joshua C. and {Miller}, Jonah M. and {Brown}, Joshua and {Ryan}, Ben and {Gaspar}, Andrew and {Glines}, Forrest and {Swaminarayan}, Sriram and {Lippuner}, Jonas and {Solomon}, Clell J. and {Shipman}, Galen and {Junghans}, Christoph and {Holladay}, Daniel and {Stone}, James M. and {Roberts}, Luke F.},
        title = "{Parthenon -- a performance portable block-structured adaptive mesh refinement framework}",
      journal = {arXiv e-prints},
         year = 2022,
        month = feb,
          eid = {arXiv:2202.12309},
        pages = {arXiv:2202.12309},
          doi = {10.48550/arXiv.2202.12309},
archivePrefix = {arXiv},
       eprint = {2202.12309},
 primaryClass = {cs.DC},
       adsurl = {https://ui.adsabs.harvard.edu/abs/2022arXiv220212309G}
}

@ARTICLE{grete2025,
       author = {{Grete}, Philipp and {O'Shea}, Brian W. and {Glines}, Forrest W. and {Prasad}, Deovrat and {Wibking}, Benjamin D. and {Fournier}, Martin and {Br{\"u}ggen}, Marcus and {Voit}, G. Mark},
        title = "{The XMAGNET Exascale MHD Simulations of SMBH Feedback in Galaxy Groups and Clusters: Overview and Preliminary Cluster Results}",
      journal = {\apj},
         year = 2025,
        month = aug,
       volume = {988},
       number = {2},
          eid = {155},
        pages = {155},
          doi = {10.3847/1538-4357/adde45},
archivePrefix = {arXiv},
       eprint = {2502.13213},
 primaryClass = {astro-ph.GA},
       adsurl = {https://ui.adsabs.harvard.edu/abs/2025ApJ...988..155G}
}

@ARTICLE{guo2023,
       author = {{Guo}, Minghao and {Stone}, James M. and {Kim}, Chang-Goo and {Quataert}, Eliot},
        title = "{Toward Horizon-scale Accretion onto Supermassive Black Holes in Elliptical Galaxies}",
      journal = {\apj},
         year = 2023,
        month = mar,
       volume = {946},
       number = {1},
          eid = {26},
        pages = {26},
          doi = {10.3847/1538-4357/acb81e},
archivePrefix = {arXiv},
       eprint = {2211.05131},
 primaryClass = {astro-ph.HE},
       adsurl = {https://ui.adsabs.harvard.edu/abs/2023ApJ...946...26G}
}

@ARTICLE{hernquist1990,
       author = {{Hernquist}, Lars},
        title = "{An Analytical Model for Spherical Galaxies and Bulges}",
      journal = {\apj},
         year = 1990,
        month = jun,
       volume = {356},
        pages = {359},
          doi = {10.1086/168845},
       adsurl = {https://ui.adsabs.harvard.edu/abs/1990ApJ...356..359H}
}

@ARTICLE{hofmann2016,
       author = {{Hofmann}, F. and {Sanders}, J.~S. and {Nandra}, K. and {Clerc}, N. and {Gaspari}, M.},
        title = "{Thermodynamic perturbations in the X-ray halo of 33 clusters of galaxies observed with Chandra ACIS}",
      journal = {\aap},
         year = 2016,
        month = jan,
       volume = {585},
          eid = {A130},
        pages = {A130},
          doi = {10.1051/0004-6361/201526925},
archivePrefix = {arXiv},
       eprint = {1510.08445},
 primaryClass = {astro-ph.CO},
       adsurl = {https://ui.adsabs.harvard.edu/abs/2016A&A...585A.130H}
}

@ARTICLE{horton2020,
       author = {{Horton}, Maya A. and {Krause}, Martin G.~H. and {Hardcastle}, Martin J.},
        title = "{3D hydrodynamic simulations of large-scale precessing jets: radio morphology}",
      journal = {\mnras},
         year = 2020,
        month = dec,
       volume = {499},
       number = {4},
        pages = {5765-5781},
          doi = {10.1093/mnras/staa3020},
archivePrefix = {arXiv},
       eprint = {2010.00480},
 primaryClass = {astro-ph.GA},
       adsurl = {https://ui.adsabs.harvard.edu/abs/2020MNRAS.499.5765H}
}

@ARTICLE{king2006,
       author = {{King}, A.~R. and {Pringle}, J.~E.},
        title = "{Growing supermassive black holes by chaotic accretion}",
      journal = {\mnras},
         year = 2006,
        month = nov,
       volume = {373},
       number = {1},
        pages = {L90-L92},
          doi = {10.1111/j.1745-3933.2006.00249.x},
archivePrefix = {arXiv},
       eprint = {astro-ph/0609598},
 primaryClass = {astro-ph},
       adsurl = {https://ui.adsabs.harvard.edu/abs/2006MNRAS.373L..90K}
}

@ARTICLE{king2008,
       author = {{King}, A.~R. and {Pringle}, J.~E. and {Hofmann}, J.~A.},
        title = "{The evolution of black hole mass and spin in active galactic nuclei}",
      journal = {\mnras},
         year = 2008,
        month = apr,
       volume = {385},
       number = {3},
        pages = {1621-1627},
          doi = {10.1111/j.1365-2966.2008.12943.x},
archivePrefix = {arXiv},
       eprint = {0801.1564},
 primaryClass = {astro-ph},
       adsurl = {https://ui.adsabs.harvard.edu/abs/2008MNRAS.385.1621K}
}

@ARTICLE{kormendy2013,
   author = {{Kormendy}, J. and {Ho}, L.~C.},
    title = "{Coevolution (Or Not) of Supermassive Black Holes and Host Galaxies}",
  journal = {\araa},
archivePrefix = "arXiv",
   eprint = {1304.7762},
     year = 2013,
    month = aug,
   volume = 51,
    pages = {511-653},
      doi = {10.1146/annurev-astro-082708-101811},
   adsurl = {http://adsabs.harvard.edu/abs/2013ARA%26A..51..511K}
}

@ARTICLE{krause2019,
       author = {{Krause}, Martin G.~H. and {Hardcastle}, Martin J. and {Shabala}, Stanislav S.},
        title = "{Probing gaseous halos of galaxies with radio jets}",
      journal = {\aap},
         year = 2019,
        month = jul,
       volume = {627},
          eid = {A113},
        pages = {A113},
          doi = {10.1051/0004-6361/201935762},
archivePrefix = {arXiv},
       eprint = {1905.13506},
 primaryClass = {astro-ph.GA},
       adsurl = {https://ui.adsabs.harvard.edu/abs/2019A&A...627A.113K}
}

@ARTICLE{krolik2015,
       author = {{Krolik}, Julian H. and {Hawley}, John F.},
        title = "{A Steady-state Alignment Front in an Accretion Disk Subjected to Lense-thirring Torques}",
      journal = {\apj},
         year = 2015,
        month = jun,
       volume = {806},
       number = {1},
          eid = {141},
        pages = {141},
          doi = {10.1088/0004-637X/806/1/141},
archivePrefix = {arXiv},
       eprint = {1505.01050},
 primaryClass = {astro-ph.HE},
       adsurl = {https://ui.adsabs.harvard.edu/abs/2015ApJ...806..141K}
}

@ARTICLE{lowell2024,
       author = {{Lowell}, Beverly and {Jacquemin-Ide}, Jonatan and {Tchekhovskoy}, Alexander and {Duncan}, Alex},
        title = "{Rapid Black Hole Spin-down by Thick Magnetically Arrested Disks}",
      journal = {\apj},
         year = 2024,
        month = jan,
       volume = {960},
       number = {1},
          eid = {82},
        pages = {82},
          doi = {10.3847/1538-4357/ad09af},
archivePrefix = {arXiv},
       eprint = {2302.01351},
 primaryClass = {astro-ph.HE},
       adsurl = {https://ui.adsabs.harvard.edu/abs/2024ApJ...960...82L}
}

@ARTICLE{lu2005,
       author = {{Lu}, Ju-Fu and {Zhou}, Bo-Yan},
        title = "{Observational Evidence of Jet Precession in Galactic Nuclei Caused by Accretion Disks}",
      journal = {\apjl},
         year = 2005,
        month = dec,
       volume = {635},
       number = {1},
        pages = {L17-L20},
          doi = {10.1086/499333},
archivePrefix = {arXiv},
       eprint = {astro-ph/0511212},
 primaryClass = {astro-ph},
       adsurl = {https://ui.adsabs.harvard.edu/abs/2005ApJ...635L..17L}
}

@ARTICLE{martinez2011,
       author = {{Mart{\'\i}nez-Sansigre}, Alejo and {Rawlings}, Steve},
        title = "{Observational constraints on the spin of the most massive black holes from radio observations}",
      journal = {\mnras},
         year = 2011,
        month = jul,
       volume = {414},
       number = {3},
        pages = {1937-1964},
          doi = {10.1111/j.1365-2966.2011.18512.x},
archivePrefix = {arXiv},
       eprint = {1102.2228},
 primaryClass = {astro-ph.HE},
       adsurl = {https://ui.adsabs.harvard.edu/abs/2011MNRAS.414.1937M}
}

@ARTICLE{mukherjee2025,
       author = {{Mukherjee}, Dipanjan},
        title = "{Jet Feedback on kpc Scales: A Review}",
      journal = {Galaxies},
         year = 2025,
        month = sep,
       volume = {13},
       number = {5},
          eid = {102},
        pages = {102},
          doi = {10.3390/galaxies13050102},
archivePrefix = {arXiv},
       eprint = {2506.03888},
 primaryClass = {astro-ph.GA},
       adsurl = {https://ui.adsabs.harvard.edu/abs/2025Galax..13..102M}
}

@ARTICLE{narayan2022,
       author = {{Narayan}, Ramesh and {Chael}, Andrew and {Chatterjee}, Koushik and {Ricarte}, Angelo and {Curd}, Brandon},
        title = "{Jets in magnetically arrested hot accretion flows: geometry, power, and black hole spin-down}",
      journal = {\mnras},
         year = 2022,
        month = apr,
       volume = {511},
       number = {3},
        pages = {3795-3813},
          doi = {10.1093/mnras/stac285},
archivePrefix = {arXiv},
       eprint = {2108.12380},
 primaryClass = {astro-ph.HE},
       adsurl = {https://ui.adsabs.harvard.edu/abs/2022MNRAS.511.3795N}
}

@ARTICLE{navarro1996,
   author = {{Navarro}, J.~F. and {Frenk}, C.~S. and {White}, S.~D.~M.},
    title = "{The Structure of Cold Dark Matter Halos}",
  journal = {\apj},
   eprint = {astro-ph/9508025},
     year = 1996,
    month = may,
   volume = 462,
    pages = {563},
      doi = {10.1086/177173},
   adsurl = {http://adsabs.harvard.edu/abs/1996ApJ...462..563N}
}

@INCOLLECTION{nixon2016,
       author = {{Nixon}, Chris and {King}, Andrew},
        title = "{Warp Propagation in Astrophysical Discs}",
    booktitle = {Lecture Notes in Physics, Berlin Springer Verlag},
         year = 2016,
       editor = {{Haardt}, Francesco and {Gorini}, Vittorio and {Moschella}, Ugo and {Treves}, Aldo and {Colpi}, Monica},
       volume = {905},
        pages = {45},
          doi = {10.1007/978-3-319-19416-5_2},
       adsurl = {https://ui.adsabs.harvard.edu/abs/2016LNP...905...45N}
}

@INPROCEEDINGS{novikov1973,
       author = {{Novikov}, I.~D. and {Thorne}, K.~S.},
        title = "{Astrophysics of black holes.}",
    booktitle = {Black Holes (Les Astres Occlus)},
         year = 1973,
       editor = {{Dewitt}, C. and {Dewitt}, B.~S.},
        month = jan,
        pages = {343-450},
       adsurl = {https://ui.adsabs.harvard.edu/abs/1973blho.conf..343N}
}

@ARTICLE{osullivan2017,
       author = {{O'Sullivan}, Ewan and {Ponman}, Trevor J. and {Kolokythas}, Konstantinos and {Raychaudhury}, Somak and {Babul}, Arif and {Vrtilek}, Jan M. and {David}, Laurence P. and {Giacintucci}, Simona and {Gitti}, Myriam and {Haines}, Chris P.},
        title = "{The Complete Local Volume Groups Sample - I. Sample selection and X-ray properties of the high-richness subsample}",
      journal = {\mnras},
         year = 2017,
        month = dec,
       volume = {472},
       number = {2},
        pages = {1482-1505},
          doi = {10.1093/mnras/stx2078},
archivePrefix = {arXiv},
       eprint = {1708.03555},
 primaryClass = {astro-ph.GA},
       adsurl = {https://ui.adsabs.harvard.edu/abs/2017MNRAS.472.1482O}
}

@ARTICLE{piana2021,
       author = {{Piana}, Olmo and {Dayal}, Pratika and {Volonteri}, Marta and {Choudhury}, Tirthankar Roy},
        title = "{The mass assembly of high-redshift black holes}",
      journal = {\mnras},
         year = 2021,
        month = jan,
       volume = {500},
       number = {2},
        pages = {2146-2158},
          doi = {10.1093/mnras/staa3363},
archivePrefix = {arXiv},
       eprint = {2009.13505},
 primaryClass = {astro-ph.GA},
       adsurl = {https://ui.adsabs.harvard.edu/abs/2021MNRAS.500.2146P}
}

@ARTICLE{piana2024,
       author = {{Piana}, Olmo and {Pu}, Hung-Yi and {Wu}, Kinwah},
        title = "{Super-Eddington accretion in high-redshift black holes and the emergence of jetted AGN}",
      journal = {\mnras},
         year = 2024,
        month = may,
       volume = {530},
       number = {2},
        pages = {1732-1748},
          doi = {10.1093/mnras/stae851},
archivePrefix = {arXiv},
       eprint = {2403.15106},
 primaryClass = {astro-ph.GA},
       adsurl = {https://ui.adsabs.harvard.edu/abs/2024MNRAS.530.1732P}
}

@ARTICLE{piana2026b,
author = {{Piana}, Olmo and {Gaspari}, Massimo and {Cammelli}, Vieri and et al.},
title = "{BlackHoleWeather – Chaotic cold accretion across the meso scale: Variability and kinematics}",
journal = {\aap},
year = 2026,
volume = {Submitted},
number = {},
eid = {},
pages = {},
doi = {},
adsurl = {}
}

@ARTICLE{prasad2026,
       author = {{Prasad}, Deovrat and {Grete}, Philipp and {O'Shea}, Brian W. and {Glines}, Forrest W. and {Voit}, G. Mark and {van de Voort}, Freeke and {Fournier}, Martin and {Wibking}, Benjamin D.},
        title = "{XMAGNET: kinetic, thermal, and magnetic AGN feedback in massive galaxies at halo masses {\ensuremath{\sim}}10$^{13.5}$ M$_{{\ensuremath{\odot}}}$}",
      journal = {\mnras},
         year = 2026,
        month = jan,
       volume = {545},
       number = {3},
          eid = {staf2155},
        pages = {staf2155},
          doi = {10.1093/mnras/staf2155},
archivePrefix = {arXiv},
       eprint = {2508.17508},
 primaryClass = {astro-ph.GA},
       adsurl = {https://ui.adsabs.harvard.edu/abs/2026MNRAS.545f2155P}
}

@ARTICLE{pu2020,
       author = {{Pu}, Hung-Yi and {Takahashi}, Masaaki},
        title = "{Properties of Trans-fast Magnetosonic Jets in Black Hole Magnetospheres}",
      journal = {\apj},
         year = 2020,
        month = mar,
       volume = {892},
       number = {1},
          eid = {37},
        pages = {37},
          doi = {10.3847/1538-4357/ab77ab},
archivePrefix = {arXiv},
       eprint = {2002.08185},
 primaryClass = {astro-ph.IM},
       adsurl = {https://ui.adsabs.harvard.edu/abs/2020ApJ...892...37P}
}

@INCOLLECTION{rezzolla2016,
       author = {{Rezzolla}, Luciano},
        title = "{An Introduction to Astrophysical Black Holes and Their Dynamical Production}",
    booktitle = {Lecture Notes in Physics, Berlin Springer Verlag},
         year = 2016,
       editor = {{Haardt}, Francesco and {Gorini}, Vittorio and {Moschella}, Ugo and {Treves}, Aldo and {Colpi}, Monica},
       volume = {905},
        pages = {1},
          doi = {10.1007/978-3-319-19416-5_1},
       adsurl = {https://ui.adsabs.harvard.edu/abs/2016LNP...905....1R}
}

@ARTICLE{ricarte2025,
       author = {{Ricarte}, Angelo and {Natarajan}, Priyamvada and {Narayan}, Ramesh and {Palumbo}, Daniel C.~M.},
        title = "{Multimessenger Probes of Supermassive Black Hole Spin Evolution}",
      journal = {\apj},
         year = 2025,
        month = feb,
       volume = {980},
       number = {1},
          eid = {136},
        pages = {136},
          doi = {10.3847/1538-4357/ad9ea9},
archivePrefix = {arXiv},
       eprint = {2410.07477},
 primaryClass = {astro-ph.HE},
       adsurl = {https://ui.adsabs.harvard.edu/abs/2025ApJ...980..136R}
}

@ARTICLE{schure2009,
       author = {{Schure}, K.~M. and {Kosenko}, D. and {Kaastra}, J.~S. and {Keppens}, R. and {Vink}, J.},
        title = "{A new radiative cooling curve based on an up-to-date plasma emission code}",
      journal = {\aap},
         year = 2009,
        month = dec,
       volume = {508},
       number = {2},
        pages = {751-757},
          doi = {10.1051/0004-6361/200912495},
archivePrefix = {arXiv},
       eprint = {0909.5204},
 primaryClass = {astro-ph.GA},
       adsurl = {https://ui.adsabs.harvard.edu/abs/2009A&A...508..751S}
}

@ARTICLE{sesana2014,
       author = {{Sesana}, A. and {Barausse}, E. and {Dotti}, M. and {Rossi}, E.~M.},
        title = "{Linking the Spin Evolution of Massive Black Holes to Galaxy Kinematics}",
      journal = {\apj},
         year = 2014,
        month = oct,
       volume = {794},
       number = {2},
          eid = {104},
        pages = {104},
          doi = {10.1088/0004-637X/794/2/104},
archivePrefix = {arXiv},
       eprint = {1402.7088},
 primaryClass = {astro-ph.CO},
       adsurl = {https://ui.adsabs.harvard.edu/abs/2014ApJ...794..104S}
}

@ARTICLE{sikora2007,
       author = {{Sikora}, Marek and {Stawarz}, {\L}ukasz and {Lasota}, Jean-Pierre},
        title = "{Radio Loudness of Active Galactic Nuclei: Observational Facts and Theoretical Implications}",
      journal = {\apj},
         year = 2007,
        month = apr,
       volume = {658},
       number = {2},
        pages = {815-828},
          doi = {10.1086/511972},
archivePrefix = {arXiv},
       eprint = {astro-ph/0604095},
 primaryClass = {astro-ph},
       adsurl = {https://ui.adsabs.harvard.edu/abs/2007ApJ...658..815S}
}

@ARTICLE{stone2020,
       author = {{Stone}, James M. and {Tomida}, Kengo and {White}, Christopher J. and {Felker}, Kyle G.},
        title = "{The Athena++ Adaptive Mesh Refinement Framework: Design and Magnetohydrodynamic Solvers}",
      journal = {\apjs},
         year = 2020,
        month = jul,
       volume = {249},
       number = {1},
          eid = {4},
        pages = {4},
          doi = {10.3847/1538-4365/ab929b},
archivePrefix = {arXiv},
       eprint = {2005.06651},
 primaryClass = {astro-ph.IM},
       adsurl = {https://ui.adsabs.harvard.edu/abs/2020ApJS..249....4S}
}

@ARTICLE{talbot2021,
       author = {{Talbot}, Rosie Y. and {Bourne}, Martin A. and {Sijacki}, Debora},
        title = "{Blandford-Znajek jets in galaxy formation simulations: method and implementation}",
      journal = {\mnras},
         year = 2021,
        month = jul,
       volume = {504},
       number = {3},
        pages = {3619-3650},
          doi = {10.1093/mnras/stab804},
archivePrefix = {arXiv},
       eprint = {2011.10580},
 primaryClass = {astro-ph.GA},
       adsurl = {https://ui.adsabs.harvard.edu/abs/2021MNRAS.504.3619T}
}

@ARTICLE{talbot2022,
       author = {{Talbot}, Rosie Y. and {Sijacki}, Debora and {Bourne}, Martin A.},
        title = "{Blandford-Znajek jets in galaxy formation simulations: exploring the diversity of outflows produced by spin-driven AGN jets in Seyfert galaxies}",
      journal = {\mnras},
         year = 2022,
        month = aug,
       volume = {514},
       number = {3},
        pages = {4535-4559},
          doi = {10.1093/mnras/stac1566},
archivePrefix = {arXiv},
       eprint = {2111.01801},
 primaryClass = {astro-ph.GA},
       adsurl = {https://ui.adsabs.harvard.edu/abs/2022MNRAS.514.4535T}
}

@ARTICLE{tchekhovskoy2010,
       author = {{Tchekhovskoy}, Alexander and {Narayan}, Ramesh and {McKinney}, Jonathan C.},
        title = "{Black Hole Spin and The Radio Loud/Quiet Dichotomy of Active Galactic Nuclei}",
      journal = {\apj},
         year = 2010,
        month = mar,
       volume = {711},
       number = {1},
        pages = {50-63},
          doi = {10.1088/0004-637X/711/1/50},
archivePrefix = {arXiv},
       eprint = {0911.2228},
 primaryClass = {astro-ph.HE},
       adsurl = {https://ui.adsabs.harvard.edu/abs/2010ApJ...711...50T}
}

@ARTICLE{tchekhovskoy2011,
       author = {{Tchekhovskoy}, Alexander and {Narayan}, Ramesh and {McKinney}, Jonathan C.},
        title = "{Efficient generation of jets from magnetically arrested accretion on a rapidly spinning black hole}",
      journal = {\mnras},
         year = 2011,
        month = nov,
       volume = {418},
       number = {1},
        pages = {L79-L83},
          doi = {10.1111/j.1745-3933.2011.01147.x},
archivePrefix = {arXiv},
       eprint = {1108.0412},
 primaryClass = {astro-ph.HE},
       adsurl = {https://ui.adsabs.harvard.edu/abs/2011MNRAS.418L..79T}
}

@ARTICLE{ubertosi2023,
       author = {{Ubertosi}, F. and {Gitti}, M. and {Brighenti}, F. and {Olivares}, V. and {O'Sullivan}, E. and {Schellenberger}, G.},
        title = "{Waking the monster: The onset of AGN feedback in galaxy clusters hosting young central radio galaxies}",
      journal = {\aap},
         year = 2023,
        month = may,
       volume = {673},
          eid = {A52},
        pages = {A52},
          doi = {10.1051/0004-6361/202345894},
archivePrefix = {arXiv},
       eprint = {2303.04821},
 primaryClass = {astro-ph.GA},
       adsurl = {https://ui.adsabs.harvard.edu/abs/2023A&A...673A..52U}
}

@ARTICLE{volonteri2013,
       author = {{Volonteri}, M. and {Sikora}, M. and {Lasota}, J.-P. and {Merloni}, A.},
        title = "{The Evolution of Active Galactic Nuclei and their Spins}",
      journal = {\apj},
         year = 2013,
        month = oct,
       volume = {775},
       number = {2},
          eid = {94},
        pages = {94},
          doi = {10.1088/0004-637X/775/2/94},
archivePrefix = {arXiv},
       eprint = {1210.1025},
 primaryClass = {astro-ph.HE},
       adsurl = {https://ui.adsabs.harvard.edu/abs/2013ApJ...775...94V}
}

@ARTICLE{Xrism2025,
       author = {{XRISM Collaboration} and {Audard}, Marc and {Awaki}, Hisamitsu and {Ballhausen}, Ralf and {Bamba}, Aya and {Behar}, Ehud and {Boissay-Malaquin}, Rozenn and {Brenneman}, Laura and {Brown}, Gregory V. and {Corrales}, Lia and {Costantini}, Elisa and {Cumbee}, Renata and {Diaz Trigo}, Maria and {Done}, Chris and {Dotani}, Tadayasu and {Ebisawa}, Ken and {Eckart}, Megan E. and {Eckert}, Dominique and {Eguchi}, Satoshi and {Enoto}, Teruaki and {Ezoe}, Yuichiro and {Foster}, Adam and {Fujimoto}, Ryuichi and {Fujita}, Yutaka and {Fukazawa}, Yasushi and {Fukushima}, Kotaro and {Furuzawa}, Akihiro and {Gallo}, Luigi and {Garc{\'\i}a}, Javier A. and {Gu}, Liyi and {Guainazzi}, Matteo and {Hagino}, Kouichi and {Hamaguchi}, Kenji and {Hatsukade}, Isamu and {Hayashi}, Katsuhiro and {Hayashi}, Takayuki and {Hell}, Natalie and {Hodges-Kluck}, Edmund and {Hornschemeier}, Ann and {Ichinohe}, Yuto and {Ishida}, Manabu and {Ishikawa}, Kumi and {Ishisaki}, Yoshitaka and {Kaastra}, Jelle and {Kallman}, Timothy and {Kara}, Erin and {Katsuda}, Satoru and {Kanemaru}, Yoshiaki and {Kelley}, Richard and {Kilbourne}, Caroline and {Kitamoto}, Shunji and {Kobayashi}, Shogo and {Kohmura}, Takayoshi and {Kubota}, Aya and {Leutenegger}, Maurice and {Loewenstein}, Michael and {Maeda}, Yoshitomo and {Markevitch}, Maxim and {Matsumoto}, Hironori and {Matsushita}, Kyoko and {McCammon}, Dan and {McNamara}, Brian and {Mernier}, Fran{\c{c}}ois and {Miller}, Eric D. and {Miller}, Jon M. and {Mitsuishi}, Ikuyuki and {Mizumoto}, Misaki and {Mizuno}, Tsunefumi and {Mori}, Koji and {Mukai}, Koji and {Murakami}, Hiroshi and {Mushotzky}, Richard and {Nakajima}, Hiroshi and {Nakazawa}, Kazuhiro and {Ness}, Jan-Uwe and {Nobukawa}, Kumiko and {Nobukawa}, Masayoshi and {Noda}, Hirofumi and {Odaka}, Hirokazu and {Ogawa}, Shoji and {Ogorzalek}, Anna and {Okajima}, Takashi and {Ota}, Naomi and {Paltani}, Stephane and {Petre}, Robert and {Plucinsky}, Paul and {Porter}, Frederick S. and {Pottschmidt}, Katja and {Sato}, Kosuke and {Sato}, Toshiki and {Sawada}, Makoto and {Seta}, Hiromi and {Shidatsu}, Megumi and {Simionescu}, Aurora and {Smith}, Randall and {Suzuki}, Hiromasa and {Szymkowiak}, Andrew and {Takahashi}, Hiromitsu and {Takeo}, Mai and {Tamagawa}, Toru and {Tamura}, Keisuke and {Tanaka}, Takaaki and {Tanimoto}, Atsushi and {Tashiro}, Makoto and {Terada}, Yukikatsu and {Terashima}, Yuichi and {Tsuboi}, Yohko and {Tsujimoto}, Masahiro and {Tsunemi}, Hiroshi and {Tsuru}, Takeshi and {Uchida}, Hiroyuki and {Uchida}, Nagomi and {Uchida}, Yuusuke and {Uchiyama}, Hideki and {Ueda}, Yoshihiro and {Uno}, Shinichiro and {Vink}, Jacco and {Watanabe}, Shin and {Williams}, Brian J. and {Yamada}, Satoshi and {Yamada}, Shinya and {Yamaguchi}, Hiroya and {Yamaoka}, Kazutaka and {Yamasaki}, Noriko and {Yamauchi}, Makoto and {Yamauchi}, Shigeo and {Yaqoob}, Tahir and {Yoneyama}, Tomokage and {Yoshida}, Tessei and {Yukita}, Mihoko and {Zhuravleva}, Irina and {Bartalesi}, Tommaso and {Ettori}, Stefano and {Kosarzycki}, Roman and {Lovisari}, Lorenzo and {Rose}, Tom and {Sarkar}, Arnab and {Sun}, Ming and {Tamhane}, Prathamesh},
        title = "{XRISM Reveals Low Nonthermal Pressure in the Core of the Hot, Relaxed Galaxy Cluster A2029}",
      journal = {\apjl},
         year = 2025,
        month = mar,
       volume = {982},
       number = {1},
          eid = {L5},
        pages = {L5},
          doi = {10.3847/2041-8213/ada7cd},
archivePrefix = {arXiv},
       eprint = {2501.05514},
 primaryClass = {astro-ph.HE},
       adsurl = {https://ui.adsabs.harvard.edu/abs/2025ApJ...982L...5X}
}
\bibliographystyle{aa}
\clearpage
\onecolumn

\appendix
\section{Thermodynamic radial profiles of the no-turbulence suite}\label{app:A}
Figure~\ref{profiles_not} shows the mass-weighted radial profiles of density, temperature, pressure, and velocity magnitude for the four no-turbulence simulations. Gas is separated into the five temperature phases used throughout the paper: hot hard-X, hot soft-X, warm, cold, and molecular gas. Each profile represents a mass-weighted average within 3D spherical shells centred on the SMBH, extending out to \(\simeq50\,{\rm kpc}\). The line intensity encodes temporal evolution, from early (faint, $t/t_{\text{rain}}=0$) to late (bright, $t/t_{\text{rain}}=7$) stages.

\begin{figure}[h]
\includegraphics[width=\textwidth]{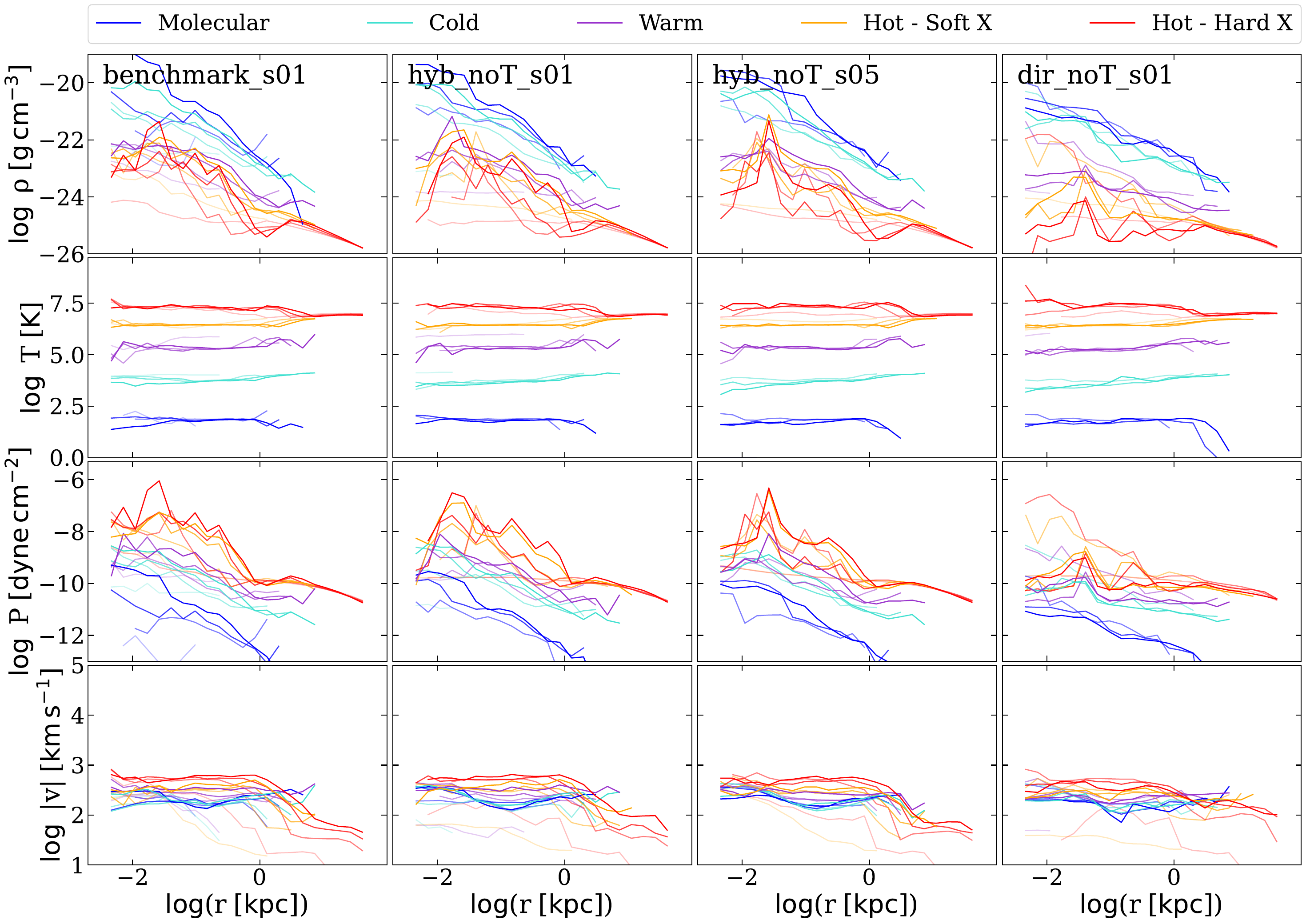}
\caption{Time evolution of the mass-weighted radial profiles of density, temperature, pressure, and velocity magnitude for the no-turbulence suite at selected epochs \((\tau=t/t_{\rm rain}=1,3,5,7)\).}
\label{profiles_not}
\end{figure}

\section{Macro-scale multiphase structure in the turbulent runs}\label{app:B}

We show complementary density slices and phase distributions on larger spatial scales. These diagnostics demonstrate that, during sunny phases, cold gas can be depleted from the immediate nuclear region while still surviving in the inner kpc or halo, either in fragmented structures or in partially coherent filaments that may later re-enter the central feeding cycle.

\subsection{Density structure}
\label{app:large_slices}

\begin{figure}[h]
\includegraphics[width=\textwidth]{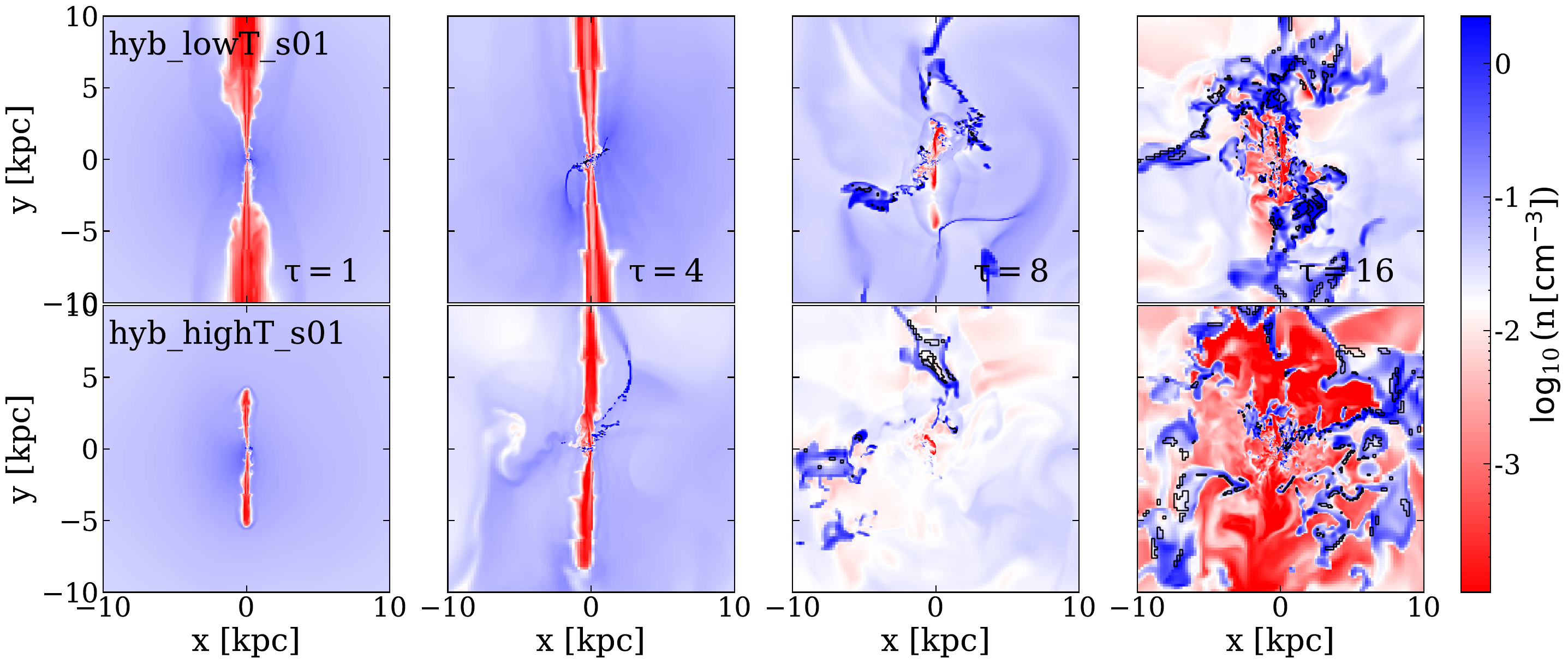}
\includegraphics[width=\textwidth]{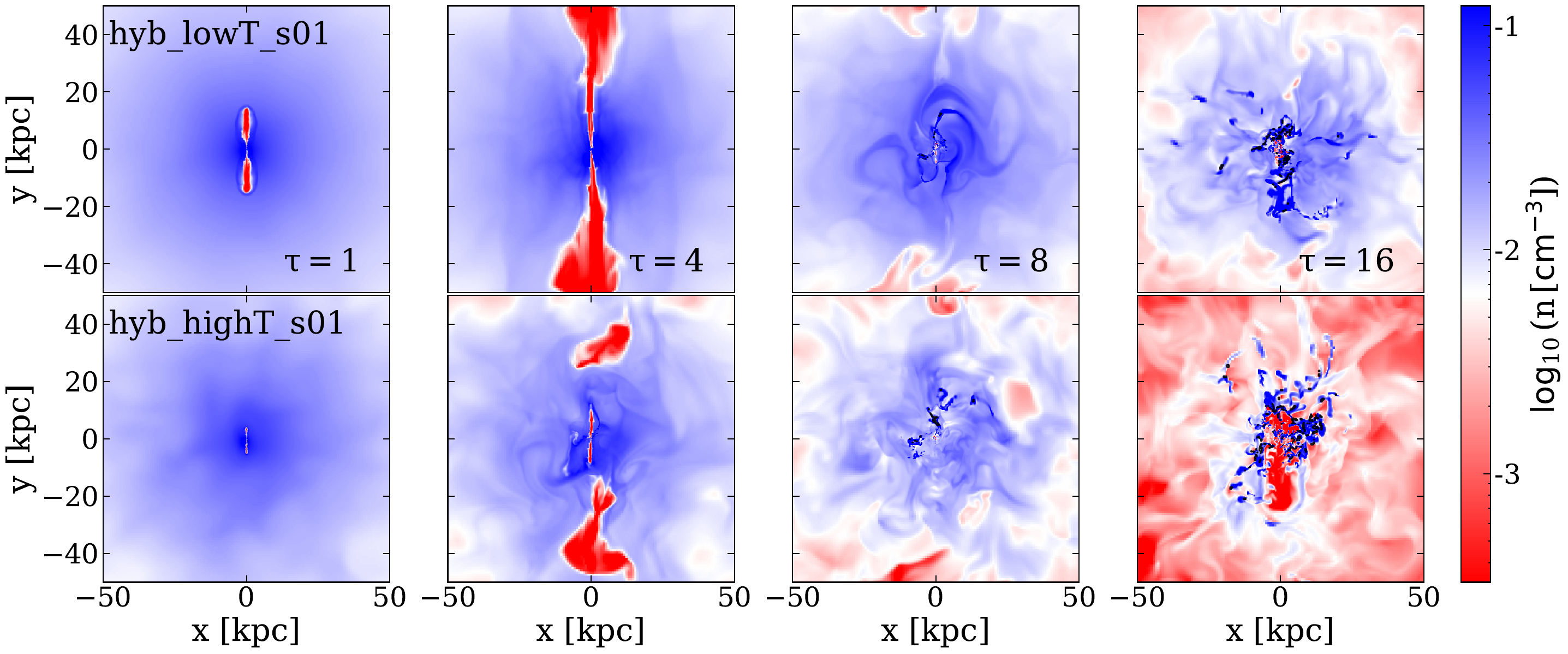}
\caption{Density slices for the low- and high-turbulence \texttt{Hybrid} runs at selected epochs (\(\tau=t/t_\mathrm{rain}=1,4,8,16\)) on larger spatial scales. The upper set shows the \(\sim 10\) kpc region (inner macro-scale), while the lower set extends to the \(\sim 50\) kpc (outer macro-scale).}
\label{slice_turb2}%
\end{figure}

Figure~\ref{slice_turb2} shows density slices for the low- and high-turbulence runs on progressively larger scales. The maps confirm that the absence or weakness of cold gas in the central \(100\) pc (micro-meso scale) at selected epochs does not correspond to a global shutdown of condensation. Instead, cold material can remain present at larger radii, with the low-turbulence run generally preserving more coherent structures, and the high-turbulence run showing more fragmented and mixed multiphase gas.

\subsection{Phase distributions}
\label{app:outer_pdfs}

Figure~\ref{pdf_turb2} shows the corresponding mass-weighted density phase distributions in radial shells outside the central \(100\) pc. These PDFs reinforce the picture inferred from the slices: multiphase gas can persist at \(0.1\)--\(1\) kpc and \(1\)--\(10\) kpc even when the nuclear region is temporarily hot-dominated.

\begin{figure}
\includegraphics[width=\textwidth]{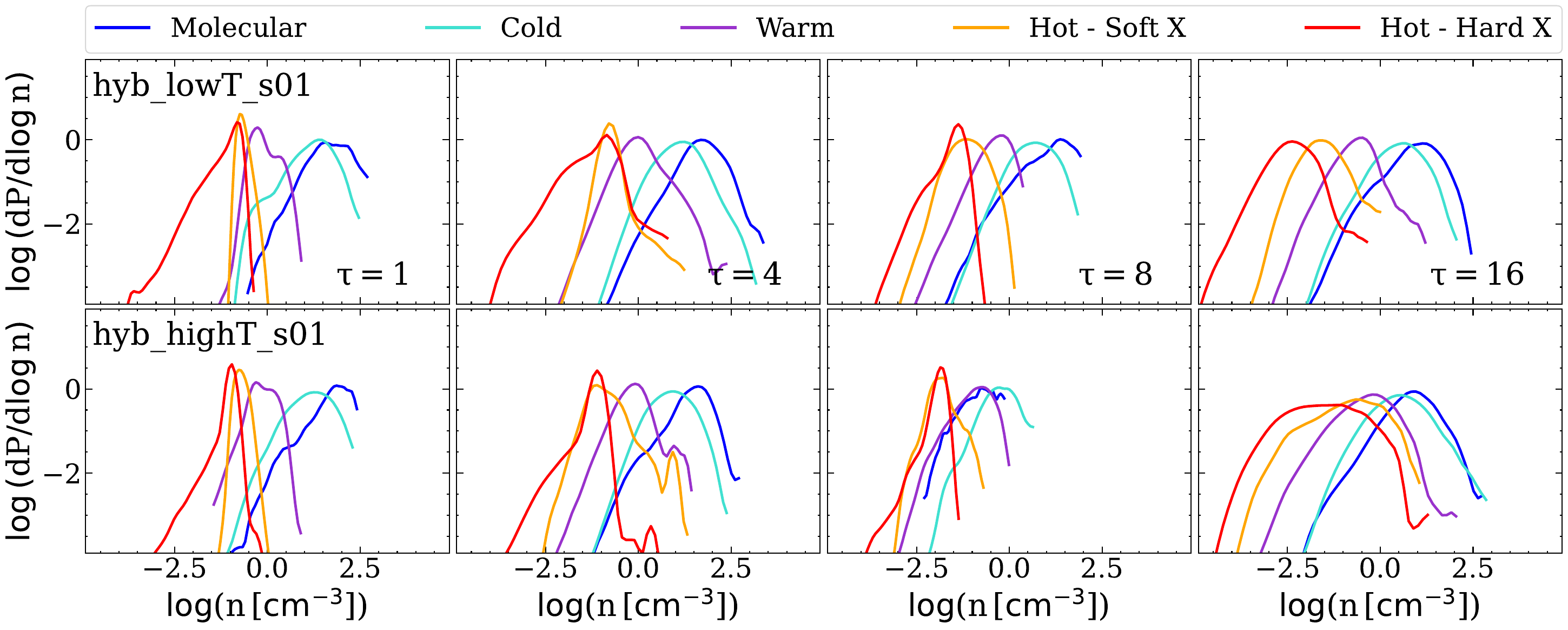}
\includegraphics[width=\textwidth]{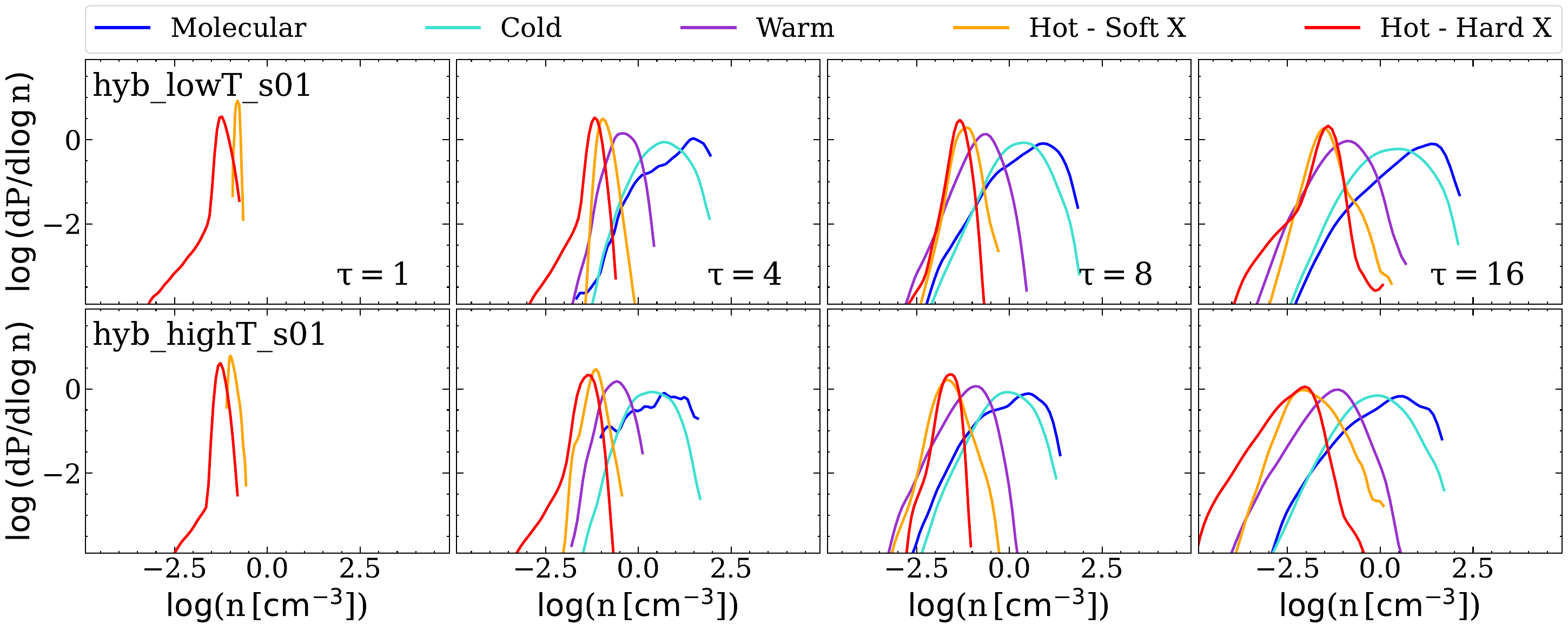}
\caption{Mass-weighted PDFs of gas density for the 5 phases for the low- and high-turbulence \texttt{Hybrid} runs at selected epochs (\(\tau=t/t_\mathrm{rain}=1,4,8,16\)) in two outer radial shells: \(0.1\)--\(1\) kpc (top) and \(1\)--\(10\) kpc (bottom).}
\label{pdf_turb2}
\end{figure}

\end{document}